\documentclass[twocolumn,aps,prl,showpacs,amsmath,superscriptaddress,longbibliography,notitlepage]{revtex4-2}
\usepackage{amsfonts}
\usepackage{amsmath}
\usepackage{amsthm}
\usepackage{mathtools}
\usepackage{longtable}
\usepackage{amssymb}
\usepackage{amsbsy}
\usepackage{graphicx}
\usepackage{bm}
\usepackage{xcolor}
\usepackage{mathrsfs}
\usepackage[colorlinks,linkcolor={red!75!black},citecolor={blue!75!black},urlcolor={blue!75!black}]{hyperref}
\usepackage{appendix}
\usepackage{booktabs}
\usepackage[export]{adjustbox}
\usepackage{braket}
\usepackage{multirow}

\usepackage{comment}

\usepackage{pdfpages}
\usepackage{pgffor}
\makeatletter
\AtBeginDocument{\let\LS@rot\@undefined}
\makeatother

\begin{document}
	\title{General Construction from Topological Loop States to Topological Invariants in Exactly Flat Bands}
	\author{Rui-Heng Liu}
	\affiliation{Beijing National Laboratory for Condensed Matter Physics and Institute of Physics, Chinese Academy of Sciences, Beijing 100190, China}
	\affiliation{University of Chinese Academy of Sciences, Beijing 100049, China}
	\author{Jiangping Hu}
	\affiliation{New Cornerstone Science Laboratory, Beijing National Laboratory for Condensed Matter Physics and Institute of Physics, Chinese Academy of Sciences, Beijing 100190, China}
	\author{Chen Fang}
	\email{cfang@iphy.ac.cn}
	\affiliation{Beijing National Laboratory for Condensed Matter Physics and Institute of Physics, Chinese Academy of Sciences, Beijing 100190, China}
	
	\begin{abstract}
		Electronic flat bands have localized Wannier-like orbitals as zero modes.
		In the Lieb or the kagome models, the localized orbitals satisfy a topological condition that entails two non-contractible loop eigenstates along $x/y$-axis in real space, and one topological band touching point with other bands in momentum space.
		In these topological flat bands, the Bloch state at the touching point is ill-defined, and so is any topological invariant for the entire band.
		We propose a new topological condition that the loop states in different directions be linearly dependent.
		Its satisfaction removes the singularity at the band touching point, and enforces nontrivial, well-defined topological invariants.
		Enforcing the new condition, we obtain topological-topological flat (top$^2$-flat) bands in 2D and 3D with nontrivial invariants including the Chern number, the $\mathbb{Z}_2$ invariants, and the topological-crystalline invariants.
		Under small, generic interactions, top$^2$-flat bands flow to correlated topological insulators with a symmetric mass term; and specially designed interacting models can have top$^2$-flat bands as exact zero modes.
	\end{abstract}
	\maketitle

	\emph{Introduction.---}
	In solids, an electronic Hamiltonian usually consists of the kinetic energy $\hat{H}_0$ and the interaction $\hat{U}$.
	Their interplay produces various correlation effects and complicates exact solutions.
	When $\hat{H}_0=0$, the kinetic energy is suppressed and one has a flat-band system \cite{sutherland1986localization,lieb1989two,vidal1998aharonov,liu2014exotic,leykam2018artificial,checkelsky2024flat} with the Landau level as a prototype.
	It proves more difficult than the non-interacting limit $\hat{U}=0$, but also more interesting as all dynamics derives from electron correlation \cite{lieb1989two,liu2014exotic,leykam2018artificial,checkelsky2024flat,shayegan2022wigner}. 
	Early studies in bipartite and line-graph lattice models proposed unusual magnetism in flat bands \cite{lieb1989two,mielke1991ferromagnetic,mielke1991ferromagnetism,mielke1992exact,tasaki1992ferromagnetism}. 
	Searching for exotic fractional states in flat bands has also been a longstanding focus \cite{kapit2010exact,tang2011high,sun2011nearly,neupert2011fractional,regnault2011fractional,qi2011generic,behrmann2016model}, with renewed interest particularly driven by studies of Moire systems \cite{tarnopolsky2019origin,ledwith2020fractional,repellin2020chern,wang2021exact,devakul2021magic,reddy2023fractional,wang2024fractional,jia2024moire,herzog2024moire,kwan2025moire,yu2025moire,liu2025theory}.
	More recently, experimental evidence of unconventional superconductivity \cite{cao2018unconventional,yankowitz2019tuning,balents2020superconductivity}, generalized Wigner crystal \cite{tang2020simulation,regan2020mott,shayegan2022wigner,li2024wigner}, fractional Chern insulator/quantum anomalous Hall effect \cite{cai2023signatures,park2023observation,xu2023observation,lu2024fractional} and various charge/spin orders \cite{lin2018flatbands,hase2018possibility,arachchige2022charge,teng2023magnetism,cao2023competing} are also attributed to flat-band physics.
	
	Vanishing kinetic energy implies that the flat band consists of local eigenstates known as compact localized states (CLSs) $b_{\mathbf{R}}$ \cite{wu2007flat,bergman2008band,morales2016simple,read2017compactly,maimaiti2017compact,rontgen2018compact,rhim2019classification,maimaiti2019universal,maimaiti2021flat,hwang2021general,hwang2021flat,graf2021designing,chen2023decoding,ara2025flat,liu2026symmetry}, with $\mathbf{R}$ labeling lattice sites and $\hat{H}_0 |b_{\mathbf{R}}\rangle = 0$.
	The CLSs and the Bloch states are both eigenstates of $\hat{H}_0$, but CLSs are (1) strictly local and (2) not (necessarily) orthogonal to each other.
	In the Lieb and kagome models \cite{bergman2008band,zong2016observation,xia2018unconventional,leykam2018artificial,rhim2019classification,rhim2020quantum,rhim2021singular}, adjacent CLSs overlap such that a Stokes-like theorem holds
	\begin{equation}\label{eq:1}
		\sum_{{V}}b_\mathbf{R}=\sum_{\partial{V}}a_\mathbf{R},
	\end{equation}
	with $\partial{V}$ the boundary of $V$ and $a_\mathbf{R}$ another local operator derived from $b_\mathbf{R}$.
	When the CLSs satisfy this topological condition, we call it a topological flat band, also known as the singular flat band \cite{rhim2019classification,rhim2020quantum,hwang2021general,rhim2021singular}.
	Under periodic boundary conditions, the system becomes a 2-torus without boundary, and Eq.~\eqref{eq:1} implies two dual facts \cite{bergman2008band}: (i) in real space, along the two non-contractible loops of the torus, two new loop states $\Theta_{x,y}$ can be built from the CLSs and the locality of $\hat{H}_0$; (ii) in momentum space, an unavoidable band touching between the flat band and other dispersive bands appears.
	At the touching point $\mathbf{k}_0$, the Bloch state is singular, usually marked by a nontrivial winding of the nearby spin/isospin structure \cite{rhim2019classification,rhim2020quantum,hwang2021general,rhim2021singular}.
	This singularity stimulates studies regarding its fate under perturbations such as electron interactions \cite{rhim2019classification,rhim2020quantum,rhim2021singular,ma2020spinorbit,sun2009topological,wen2010interaction,tsai2015interaction,zhu2016interaction,zeng2018tuning}, but precludes a well-defined projector and topological invariants for the flat band.
	
	In this Letter, we build topological flat bands from CLSs with well-defined, nontrivial topological invariants.
	We call them the topological-topological flat bands, or top$^2$-flat bands for short.
	The key is a new condition on top of Eq.~\eqref{eq:1}: loop states along different directions are linearly dependent; or, with multiple loops per direction, they span the same Hilbert subspace.
	We show that this criterion renders the Bloch state at $\mathbf{k}_0$ nonsingular despite the band touching: the directional loop states represent the same physical state and define a continuous projector. 
	This structure \emph{enforces} nonzero stable topological invariants for the flat band.
	Distinct from fragile topology in gapped flat bands \cite{chiu2020fragile,calugaru2022general}, the unavoidable touching places our construction outside no-go theorems for spectrally gapped exactly flat bands in translation-invariant, finite-range tight-binding models \cite{chen2014impossibility,dubail2015tensor,read2017compactly}.
	We obtain three elementary top$^2$-flat bands: a 2D Chern band \cite{thouless1982quantized,haldane1988model}, and $\mathbb{Z}_2$ topological bands in 2D \cite{kane2005quantum,kane2005z,bernevig2006quantum} and 3D \cite{fu2007topological,moore2007topological,roy2009topological}.
	Using layer construction \cite{isobe2015theory,fulga2016coupled,ezawa2016hourglass,song2017topological,huang2017building,song2018quantitative} and the topological-crystal method \cite{song2019topological,song2020real}, with the elementary top$^2$-flat $\mathbb{Z}_2$ band as the building block, we obtain top$^2$-flat bands for all topological crystalline bands in 3D space groups.
	Under a perturbative interaction of a proper sign (Hubbard attraction/repulsion for the Chern/$\mathbb{Z}_2$ case), a symmetric mass term is dynamically generated, so that under renormalization the top$^2$-flat bands flow to correlated topological insulators.
	We extend the theory to fully interacting Hamiltonians and show that CLSs in top$^2$-flat bands exist as exact zero modes in four-fermion interacting Hamiltonians.
	Our result is consistent with precedent constructions for chiral topological phases from projected-entangled-pair states \cite{wahl2013projected,wahl2014symmetries,yang2015chiral,dubail2015tensor}, and generalizes the construction of a flat-band Chern state on a dice lattice \cite{yang2025fractional}.

	\emph{First topological condition.---}
	CLSs are the real-space units of a flat band.
	Each CLS $b_\mathbf{R}$ is a single fermion operator with finite support near $\mathbf{R}$ as
	\begin{equation}\label{eq:2}
		b_\mathbf{R}=\sum_{\alpha,\mathbf{d}}\psi_{\mathbf{R}\alpha}(\mathbf{d})c_{\mathbf{R+d},\alpha}.
	\end{equation}
	Here $\alpha$ labels internal degrees of freedom in one unit cell and $\mathbf{d}$ is a lattice vector.
	$\psi_{\mathbf{R}\alpha}(\mathbf{d})$ is the wavefunction of the CLS, and finite support means $\psi_{\mathbf{R}\alpha}(\mathbf{d})=0$ if $|\mathbf{d}|>r>0$.
	Throughout, we assume an elementary CLS that is irreducible into smaller building blocks in the sense of Ref.~\cite{hwang2021flat}.
	In a flat-band system, each CLS is an exact zero mode of the non-interacting Hamiltonian $\hat{H}_0$ with
	\begin{equation}\label{eq:2.5}
		[b_\mathbf{R},\hat{H}_0]=0.
	\end{equation}
	
	For simplicity, we assume translation invariance in the main text so that $\psi_{\mathbf{R}\alpha}(\mathbf{d})$ is independent of $\mathbf{R}$. 
	Ref.~\cite{supp} shows that this assumption can be relaxed without affecting our main results except those involving space groups requiring translation symmetry. 
	The first topological condition characterizes whether $b_{\mathbf{R}}$ is a “total derivative”, \emph{i.e.}, the sum of $b_{\mathbf{R}}$ over a region equals the sum of $a_\mathbf{R}$ supported on the boundary. It is given by \cite{bergman2008band,rhim2019classification}
	\begin{equation}\label{eq:3}
		\sum_\mathbf{d}\psi_{\alpha}(\mathbf{d})=0.
	\end{equation}
	
	\begin{figure}
		\includegraphics[width=0.5\textwidth]{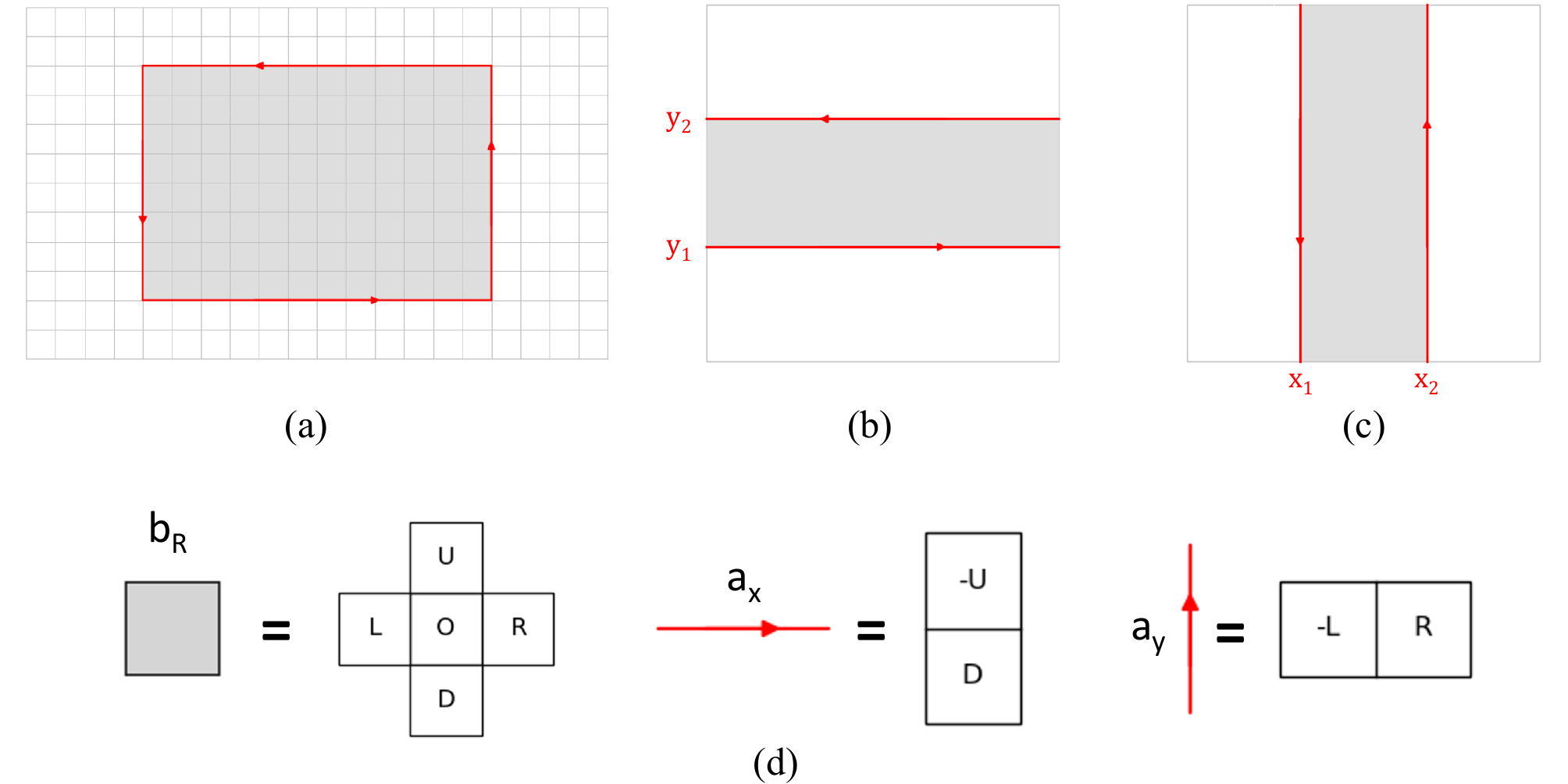}
		\caption{An explicit CLS satisfying the first topological condition. 
			(a) Summing $b_\mathbf{R}$ over the grey rectangle equals the sum of $a_{x/y}$ along the red loop.
			(b,c) If the rectangle spans the full lattice length/height, its boundaries form two separate loops at $y_{1,2}$/$x_{1,2}$.
			(d) Graphical expressions of $b_\mathbf{R}$ and $a_{x,y}$ on $x/y$ bonds.}
		\label{fig:first}
	\end{figure}
	
	Consider the explicit 2D example in Fig.~\ref{fig:first}: the CLS at $\mathbf{R}$ is a superposition of states at $\mathbf{R}$ and its four nearest neighbors as
	\begin{equation}
		b_\mathbf{R}=
		\sum_{\alpha}O_{\alpha}c_{\mathbf{R},\alpha}
		+L_{\alpha}c_{\mathbf{R}-\mathbf{x},\alpha}
		+R_{\alpha}c_{\mathbf{R}+\mathbf{x},\alpha}
		+U_{\alpha}c_{\mathbf{R}+\mathbf{y},\alpha}
		+D_{\alpha}c_{\mathbf{R}-\mathbf{y},\alpha}.
	\end{equation}
	The first topological condition becomes $O+L+R+U+D=0$, and when it is met, we have (Fig.~\ref{fig:first}(a))
	\begin{equation}\label{eq:4}
		\sum_{\blacksquare}b_\mathbf{R}=\oint_\square\mathbf{a}_{\mathbf{R}^\ast}\cdot\delta\mathbf{R}^\ast,
	\end{equation}
	where
	\begin{eqnarray}
		a^x_{\mathbf{R}^\ast}&=&\sum_{\alpha}-U_{\alpha}c_{\mathbf{R}^\ast+\mathbf{y}/2,\alpha}
		+D_{\alpha}c_{\mathbf{R}^\ast-\mathbf{y}/2,\alpha},\\
		\nonumber
		a^y_{\mathbf{R}^\ast}&=&\sum_{\alpha}-L_{\alpha}c_{\mathbf{R}^\ast-\mathbf{x}/2,\alpha}
		+R_{\alpha}c_{\mathbf{R}^\ast+\mathbf{x}/2,\alpha}.
	\end{eqnarray}
	Here $\mathbf{R}^\ast$ denotes the bond-center lattice in Fig.~\ref{fig:first}(d).
	
	Now we consider a periodic lattice and extend the rectangle to cover the full length of the lattice (Fig.~\ref{fig:first}(b)). 
	Applying Stokes' theorem [Eq.~\eqref{eq:4}] to Eq.~\eqref{eq:2.5} yields
	\begin{equation}\label{eq:5}
		\left[\sum_{x}a^x_{(x,y_1)}-\sum_xa^x_{(x,y_2)},\hat{H}_0\right]=0
	\end{equation}
	with $y_{1,2}$ the $y$-coordinates of the strip's lower and upper edges. 
	Since $|y_1-y_2|$ can be arbitrarily large, locality of $\hat{H}_0$ forces each sum in Eq.~\eqref{eq:5} to \emph{individually} commute with $\hat{H}_0$ \cite{bergman2008band}. 
	The same reasoning applies to a strip along the $y$-direction (Fig.~\ref{fig:first}(c)). 
	This leads to two loop states that are zero modes \emph{linearly independent} of the CLS
	\begin{eqnarray}
		\theta_y(y)=\sum_xa^x_{(x,y)},\quad
		\theta_x(x)&=&\sum_ya^y_{(x,y)}.
	\end{eqnarray}
	Using Eq.~\eqref{eq:4}, we see that $\theta_{y}$ at different $y$'s are equivalent up to superpositions of $b_\mathbf{R}$'s, and so are $\theta_x$ at different $x$'s. 
	This strip argument produces two extra non-contractible loop states as zero modes in two dimensions.
	
	Further, when we extend the rectangle in Fig.~\ref{fig:first}(a) to the entire lattice, Eq.~\eqref{eq:4} leads to $\sum{b}_\mathbf{R}=0$.
	Therefore, on a $N_x \times N_y$ lattice, we have $(N_xN_y-1)$ independent $b_\mathbf{R}$'s and 2 loops, making the total number of zero modes $(N_xN_y+1)$.
	This exceeds the number of Bloch states $N_xN_y$ in any single band, giving the counting argument for a degeneracy with another band \cite{bergman2008band}.
	In the sense of such real-space topology, we call a flat band satisfying Eq.~\eqref{eq:3} a topological flat (top-flat) band, and its enforced degeneracy a topological band touching.
	These touchings have been studied through the local geometry of Bloch states around them, including winding, Berry phase, and quantum distance \cite{rhim2019classification,rhim2020quantum,hwang2021general,rhim2021singular,hwang2021geometry}.
	While the band structure assumes translation symmetry, we remark that the counting argument for the degeneracy is completely general (Ref.~\cite{supp}).
	
	\begin{table*}[htbp]
		\centering
		\caption{Explicit top$^2$-CLSs. $O,R,A$ label components in Figs.~\ref{fig:second}(b) and \ref{fig:second}(c). The bases are $\{\uparrow,\downarrow\}$ and $\{A\uparrow,B\uparrow,A\downarrow,B\downarrow\}$ for two- and four-component wavefunctions, respectively, with $A,B$ orbitals in one unit cell. ``Not applicable'' means the (strong) Chern number is undefined in 3D.}   
		\label{table} 
		\begin{tabular}{|c|cc|cl|}
			\hline
			& \multicolumn{1}{c|}{2D square} & \multicolumn{1}{c|}{2D hexagonal} & \multicolumn{2}{c|}{3D} \\ \hline
			$C=1$                    & \multicolumn{1}{c|}{$O=[0,4],R=[i,-1]$}                  & $O=[0,6],A=[(\sqrt{3}+i)/2,-1]$          & \multicolumn{2}{c|}{Not applicable}                        \\ \hline
			\multirow{2}{*}{$\mathbb{Z}_2=1$} & \multicolumn{1}{c|}{$O=[0,4,0,0],R=[0,-1,i,0]$}          & $O=[0,6,0,0],A=[(\sqrt{3}+i)/2,-1,0,0]$  & \multicolumn{2}{c|}{$O=[0,6,0,0],A=[0,-1,i,0]$}  \\
			& \multicolumn{1}{c|}{$\text{T}O=[0,0,0,-4],\text{T}R=[-i,0,0,1]$} & $\text{T}O=[0,0,0,-6],\text{T}A=[0,0,(-\sqrt{3}+i)/2,1]$ & \multicolumn{2}{c|}{$\text{T}O=[0,0,0,-6],\text{T}A=[-i,0,0,1]$} \\ \hline
		\end{tabular}
	\end{table*}

\begin{figure}
		\includegraphics[width=0.5\textwidth]{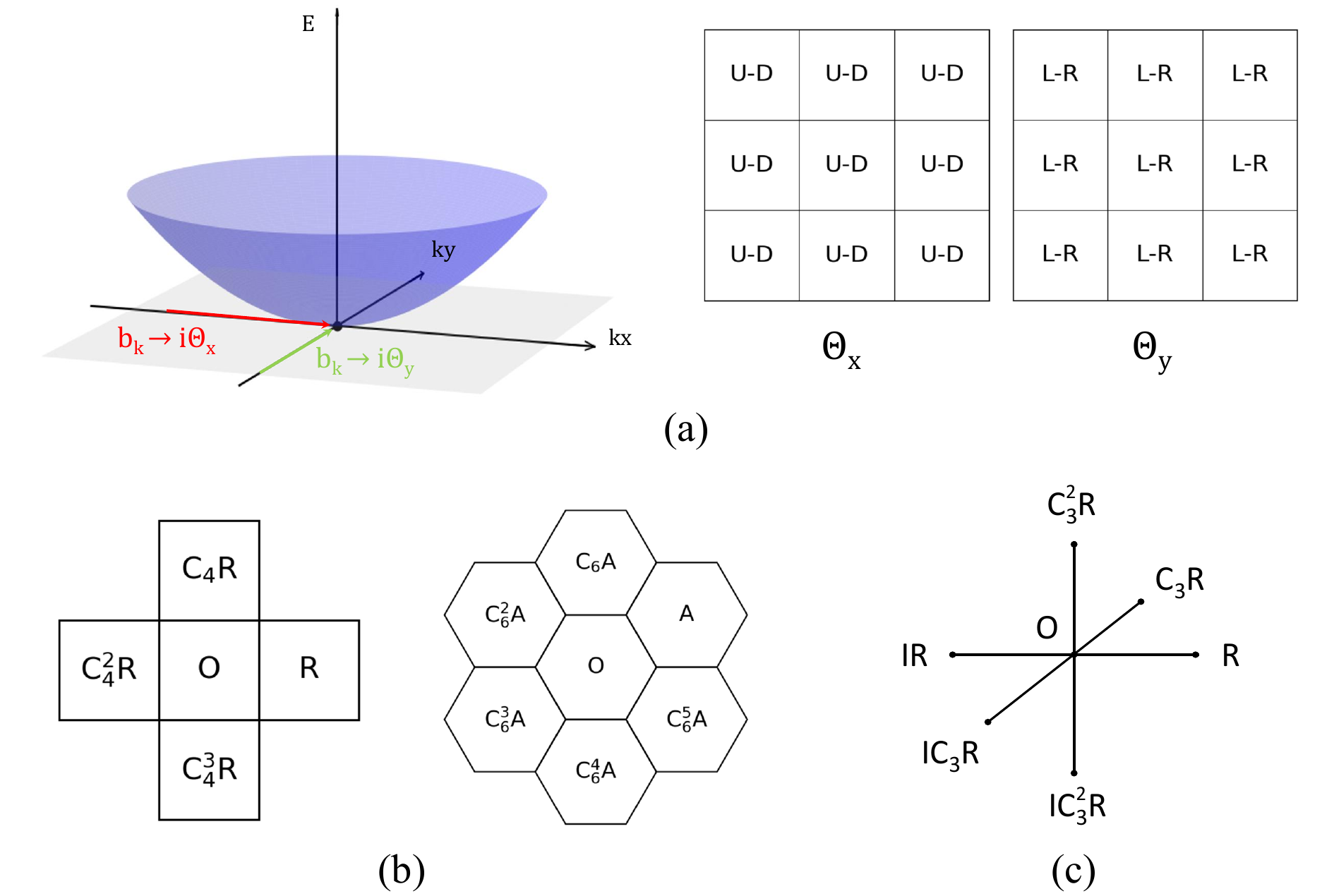}
		\caption{The second topological condition. (a) As $\mathbf{k}$ moves towards the origin along $x/y$-axis, the Bloch state converges to $\Theta_{x/y}$. (b) 2D CLSs satisfying the second topological condition with fourfold or sixfold rotation. (c) A 3D CLS satisfying the second topological condition with cubic symmetry. The spinor wavefunctions $O,R,A$ are given in Table~\ref{table}, and $\text{C}_4=\exp(-i\sigma_z\pi/4),\text{C}_3=\exp(-i\sigma_{111}\pi/3),\text{I}=\sigma_{0}\tau_{z}$ where $\tau_{z}=1(-1)$ for $A(B)$ orbitals.}
		\label{fig:second}
	\end{figure}

	\emph{Second topological condition.---}
	We now move from real to momentum space, where the second topological condition is most easily elucidated under translation symmetry, although the condition itself does not depend on it.
	Consider the Fourier transform of $b_\mathbf{R}$ as $b_\mathbf{k}=\sum_\mathbf{R}b_\mathbf{R}\exp(i\mathbf{k}\cdot\mathbf{R})$, with crystal momenta $(k_x,k_y)=2\pi(n_x/N_x,n_y/N_y)$, where $n_{x,y}=0,\dots,N_{x,y}-1$.
	For $\mathbf{k}\neq0$, $b_\mathbf{k}$ is the unnormalized Bloch state satisfying $[b_\mathbf{k},\hat{H}_0]=0$.
	But $b_\mathbf{k=0}=0$ due to the first topological condition and does not provide a physical state.
	It is then natural to speculate that the loop states from the previous strip argument must be related to the touching momentum. 
	To see this more clearly, note that since $\hat{H}_0 b_{\mathbf{k}}=0$ when $\mathbf{k}\neq 0$, its derivatives $\partial_{k_{x,y}} b_{\mathbf{k}=0}$ are also zero modes. 
	Hence, we have a correspondence between CLSs and loop states, as demonstrated in Ref.~\cite{udagawa2024pinch}:
	\begin{equation}\label{eq:8}
		-i\partial_{k_i}b_\mathbf{k=0}=\Theta_{i}\equiv\sum_{i}\theta_i(i),
	\end{equation}
	where $i=x,y$. Therefore, in translation-invariant systems, the loop states can also be viewed as derivatives of Bloch states $b_\mathbf{k}$ at $\mathbf{k}=0$. Eq.~\eqref{eq:8} shows why a band touching is necessary from a momentum-space perspective.
	As $\mathbf{k}$ approaches the origin along the $k_{x/y}$ axis, the state converges to two different limits $\Theta_{x/y}$ (Fig.~\ref{fig:second}(a)), thus making the flat band projector $P(\mathbf{k})$ discontinuous at $\mathbf{k}=0$.
	
	While this seems to be the complete story of the touching, let us consider what happens if $\Theta_{x,y}$ represent the same physical state (linearly dependent)?
	This is precisely the second topological condition
	\begin{equation}
		{\Theta}_y=\lambda\Theta_x,\quad \Im\lambda\neq 0,
	\end{equation}
	which translates into the model in Fig.~\ref{fig:first}(d) as $(L-R)+\lambda(U-D)=0$. 
	Now the first-order terms in Eq.~\eqref{eq:8} only yield one physical state $\Theta$, while the second orthogonal state $\Phi$ is invisible at this order. 
	For an elementary CLS, $\Phi$ can be obtained from higher-order terms in the expansion of $b_\mathbf{k}$, as detailed in Ref.~\cite{supp}; hence the original counting remains valid. 
	Under the second topological condition, we have a well-defined flat-band projector as $\mathbf{k}\rightarrow0$
	\begin{equation}
		P(\mathbf{k})\xrightarrow{\mathbf{k}\rightarrow0}\frac{|\Theta_x\rangle\langle{\Theta}_x|}{\langle{\Theta}_x|\Theta_x\rangle}=\frac{|\Theta_y\rangle\langle{\Theta}_y|}{\langle{\Theta}_y|\Theta_y\rangle}.
	\end{equation}
	The well-defined $P(\mathbf{k}\in\mathrm{BZ})$ allows all topological invariants, such as the Chern number, to be defined for the flat band.
	Also, the second topological condition \emph{enforces} a Chern number $C=\mathrm{sign}(\Im\lambda)=\pm1$ of the flat band: removing a disk $D_\epsilon$ around the touching leaves $b_{\mathbf{k}}$ smooth on the punctured Brillouin zone, with $C$ given by the phase winding on $\partial D_\epsilon$.
	There $b_{\mathbf{k}}\propto\Theta_x(k_x+\lambda k_y)$, where $\Im\lambda\neq0$ ensures that the scalar is nonzero in every real direction and has winding $2\pi\mathrm{sign}(\Im\lambda)$.
	Its phase cancels in $P(\mathbf{k})$, which remains continuous.
	We call this band topological-topological flat, or top$^2$-flat, where the first topological refers to the presence of topological loop states, and the second to the nonzero topological invariant.
	Figure~\ref{fig:second}(b) and Table~\ref{table} show an explicit CLS with $C=1$.
	The quantum-geometric implications of the two topological conditions are discussed in Ref.~\cite{supp}.
	
	A caveat in the explicit construction is that the Bloch state $b_{\mathbf{k}}$ may vanish at some $\mathbf{k}\neq 0$, causing additional crossings with dispersive bands that can affect the topology. 
	We discuss how to avoid this in Ref.~\cite{supp}.

	\emph{Elementary constructions.---}
	Having constructed the top$^2$-flat Chern bands ($C=1$), we turn to other topological invariants starting from the 2D time-reversal $\mathbb{Z}_2$ topological state. 
	Time-reversal symmetry enforces Kramers degeneracy, requiring at least two flat bands and a pair of CLSs $(b_{\mathbf{R}}, \mathrm{T}b_{\mathbf{R}})$ satisfying $b_{\mathbf{R}}\xrightarrow{\mathrm{T}}\mathrm{T}b_{\mathbf{R}}$ and $\mathrm{T}b_{\mathbf{R}}\xrightarrow{\mathrm{T}}-b_{\mathbf{R}}$.
	Following a similar process, one obtains a pair of loop states $\{\Theta_{x,y}, \mathrm{T}\Theta_{x,y}\}$.
	The continuity of the projector requires that $\{\Theta_x,\text{T}\Theta_x\}$ and $\{\Theta_y,\text{T}\Theta_y\}$ span the same linear subspace, \emph{i.e.}, $\Theta_y=\alpha\Theta_x+\beta{\text{T}}\Theta_x,(\alpha,\beta)\in\mathbb{C}^2$. 
	In Ref.~\cite{supp}, we show that this leads to a topological invariant $\mathbb{Z}_2=1$ so the degenerate flat band is a top$^2$-flat $\mathbb{Z}_2$ band.
	Here, we only present the simpler case with $\beta=0$, where the two bands near $\mathbf{k}=0$ have a smooth gauge as $b_{1\mathbf{k}}=i\Theta_x(k_x+\alpha{k}_y),b_{2\mathbf{k}}=i\text{T}\Theta_x(k_x+\alpha^\ast{k}_y)$. 
	If $\Im\alpha>0$, as $\mathbf{k}$ winds around the origin, the first/second band exhibits a $\pm 2\pi$ phase winding, corresponding to the Chern number $C=\pm1$, \emph{i.e.}, the quantum spin Hall limit of a $\mathbb{Z}_2$ topological state.
	Figure~\ref{fig:second}(b) and Table~\ref{table} show explicit time-reversal CLSs for top$^2$-flat bands with $D_{4}/D_{6}$ point-group symmetries.
	
	In 3D, the first topological condition [Eq.~\eqref{eq:3}] is unchanged.
	From this condition, one constructs three pairs of membrane states $\{\Theta_{x,y,z},\text{T}\Theta_{x,y,z}\}$ in 3D (analogous to loop states in 2D), one pair for each dimension.
	The second topological condition states that the three pairs span the \emph{same} linear subspace as
	\begin{eqnarray}\label{eq:10}
		\Theta_y=\alpha\Theta_x+\beta{\text{T}}\Theta_x,\quad
		\Theta_z=\mu\Theta_x+\nu{\text{T}}\Theta_x,
	\end{eqnarray}
	where $(\Im\alpha, \Re\beta, \Im\beta)$ and $(\Im\mu, \Re\nu, \Im\nu)$ are not colinear (Ref.~\cite{supp}). 
	An explicit CLS meeting this condition with cubic symmetry is given in Fig.~\ref{fig:second}(c) and Table~\ref{table}.
	In Ref.~\cite{supp}, we show that when Eq.~\eqref{eq:10} is met, the top$^2$-flat band has 3D $\mathbb{Z}_2=1$.
	
	\emph{Parent Hamiltonians.---}
	We have so far focused on constructing the CLS wavefunctions of flat bands.
	A natural question is: given CLSs satisfying one or both topological conditions, what parent Hamiltonians have these CLSs as zero modes?
	For a single set of CLSs, this question has been sufficiently addressed in Ref.~\cite{rhim2019classification}.
	In Ref.~\cite{supp}, we recapitulate and extend their method to cases with multiple time-reversal-symmetric CLSs.

	\emph{Topological crystalline constructions.---}
	We have obtained three elementary states in top$^2$-flat bands: a Chern band without symmetry, and 2D/3D $\mathbb{Z}_2$ bands protected by time-reversal symmetry.
	Spatial symmetries such as inversion, rotation, and mirror reflection further yield topological crystalline insulators \cite{fu2011topological} and higher-order topological insulators \cite{benalcazar2017quantized,benalcazar2017electric,Langbehn2017reflection,song2017d,schindler2018higher}. 
	These states can be built from elementary ones in real space \cite{isobe2015theory,fulga2016coupled,ezawa2016hourglass,song2017topological,huang2017building,song2018quantitative,song2019topological,song2020real}. 
	For example, with both inversion and time-reversal symmetry, the $\mathbb{Z}_2$ invariant is promoted to $\mathbb{Z}_4$, with $\mathbb{Z}_4=2$ corresponding to a second-order topological insulator jointly protected by both symmetries \cite{hughes2011inversion}.
	In real space, this state corresponds to a layer construction with 2D topological insulators placed at $z=n$ and $z=n+1/2$ ($n\in\mathbb{Z}$).
	The 3D topological crystalline state in this limit is a product state of all decoupled layers, hence the name ``layer construction''.
	
	The top$^2$-flat bands do not correspond to insulators because of the band touching, but have well-defined topological invariants since they have a projector $P(\mathbf{k})$ everywhere continuous.
	This implies that a 3D topological state from a layer construction of 2D top$^2$-flat bands also has a continuous projector, and moreover, well-defined topological invariants, while being dispersionless.
	These layer constructions from 2D top$^2$-flat bands are exactly the 3D top$^2$-flat bands having the same topological invariants as topological crystalline insulators.
	
	With time-reversal symmetry, every topological crystalline state in 218 of the 230 space groups admits a layer construction \cite{song2018quantitative,song2019topological}.
	For each space group, the lattice on a given layer belongs to one of the 17 wallpaper groups, which are all subgroups of $p4mm$ or $p6mm$; the $D_4$- and $D_6$-symmetric CLSs shown in Fig.~\ref{fig:second}(b) suffice to construct top$^2$-flat $\mathbb{Z}_2$ bands for any wallpaper group. 
	This concludes the construction of 3D top$^2$-flat bands for all topological crystalline states in 218 space groups.
	For the rest $230-218=12$ groups, topological states beyond layer construction can only be constructed in ``topological crystals'' \cite{song2019topological}.
	A topological crystal consists of small ``tiles'', called the 2-cells, arranged symmetrically with respect to the given space group, such that (i) the edges of each 2-cell are shared by another 2-cell, and (ii) each 2-cell is decorated by a 2D topological insulator.
	Inspired by this idea, we construct the top$^2$-flat bands for the topological crystalline states in the 12 groups by properly tiling each 2-cell with the 2D $\mathbb{Z}_2$-CLS (Ref.~\cite{supp}).
	For magnetic topological crystalline states, the corresponding construction with time-reversal-breaking Chern-state decorations \cite{peng2022topological}, rather than the present $\mathbb Z_2$ building blocks, is left for future work.

	\emph{Discussions on interacting phases.---}
	We focus on fully occupied top$^2$-flat bands and ask whether interactions gap the touching with dispersive bands. 
	In kagome and many 2D models, quadratic touchings jointly protected by rotation and time reversal require spontaneous symmetry breaking to gap \cite{sun2009topological,wen2010interaction,tsai2015interaction,zhu2016interaction,zeng2018tuning}.
	If time reversal is broken, the band becomes a Chern insulator, and if rotation is broken, a nematic state.
	
	By contrast, a top$^2$-flat band already has a well-defined topological invariant before gap opening, and its touching is not symmetry protected.
	We use the top$^2$-flat Chern band as an illustration.
	At $\mathbf{k}=0$, the two states are the first-order loop state $\Theta$ and the other degenerate state $\Phi$, as discussed above and detailed in Ref.~\cite{supp}.
	Their degeneracy arises purely from the CLS geometry and is not protected by any symmetry even when $D_{4,6}$ point groups are present.
	Therefore, a \emph{symmetric} mass term may be dynamically generated as the interaction is turned on.
	The mass term separates $\Theta$ and $\Phi$ in energy, and if $E(\Theta)<E(\Phi)$, the entire lower band is a Chern insulator; but if $E(\Theta)>E(\Phi)$, the system may instead become trivial or lie at a gapless critical point \cite{wahl2014symmetries}.
	
	\begin{figure}
		\includegraphics[width=0.5\textwidth]{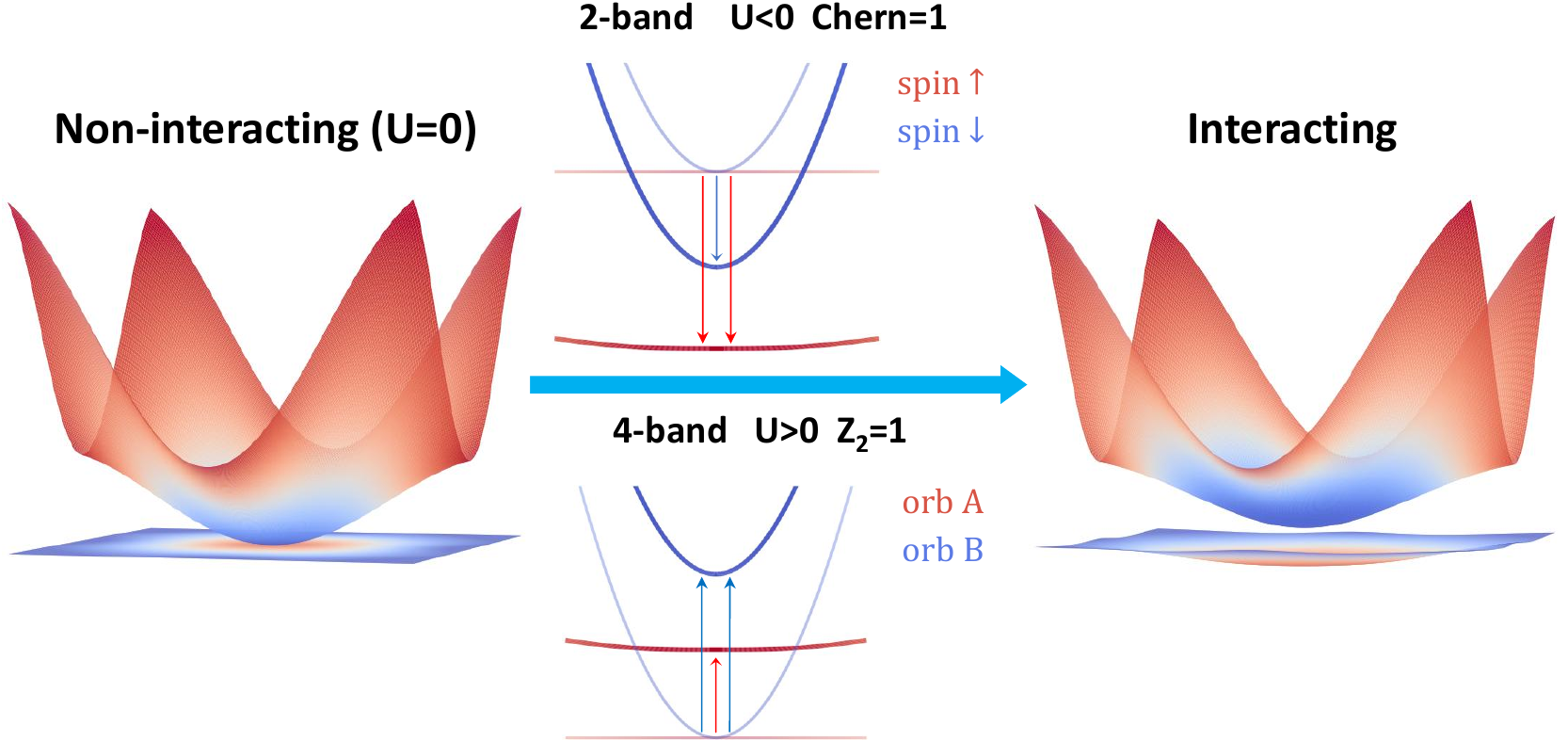}
		\caption{Hubbard interactions open a many-body gap in top$^2$-flat Chern/$\mathbb{Z}_2$ bands, with colors denoting spin or orbital polarization. In the middle, zoomed-in band structures near $\mathbf{k}=0$ are shown.}
		\label{fig:third}
	\end{figure}
	
	Whether $E(\Theta)<E(\Phi)$ depends on the details of the interaction; consider again the top$^2$-flat Chern band in Fig.~\ref{fig:second}(b) and Table~\ref{table}.
	In Fig.~\ref{fig:third}, calculation shows that in this model, the flat band is predominantly spin down, while near $\mathbf{k}=0$, $\Theta$ and $\Phi$ are spin-up and spin-down polarized, respectively.
	A Hubbard attraction favors occupation of opposite spins; we have $E(\Theta)<E(\Phi)$, and the many-body state is a correlated Chern insulator.
	The top$^2$-flat $\mathbb{Z}_2$ case is similar.
	At $\mathbf{k}=0$, the flat and dispersive bands correspond to $\{\Theta,\text{T}\Theta\}$ and $\{\Phi,\text{T}\Phi\}$, respectively.
	Here, $\{\Theta,\text{T}\Theta\}$ occupy the $A$ orbital and $\{\Phi,\text{T}\Phi\}$ occupy the $B$ orbital (Fig.~\ref{fig:third}), while the bands are predominantly $B$-orbital polarized.
	With an intra-orbital Hubbard repulsion, the predominantly occupied $B$-orbital states repel additional electrons on the same orbital.
	Therefore, $E(\Theta)<E(\Phi)$, and the many-body state is a correlated topological insulator.
	Since mass terms are relevant under the renormalization flow, small interactions of a proper sign drive the band to a correlated topological phase with the same invariant.
	The Hartree-Fock analysis in Ref.~\cite{supp} supports these polarization arguments and may yield a \emph{symmetric} mass gap $\Delta\propto |g|$, rather than the dynamically generated gap $\Delta\sim e^{-\mathrm{const}/g}$ for conventional quadratic touchings \cite{sun2009topological}.
	
	While generic interactions such as Hubbard terms generally spoil exact flatness, it is still possible to design fully interacting Hamiltonians that retain CLSs as zero modes.
	Specifically, for a given CLS $b_\mathbf{R}$, we look for a non-quadratic $\hat{H}$ such that $[b_{\mathbf{R}},\hat{H}]=0$, similar to Eq.~\eqref{eq:2.5}. Here we use a counting argument to show the existence of $\hat{H}$, and more details are given in Ref.~\cite{supp}.
	Suppose the support of each $b_\mathbf{R}$ has radius $r$, and each term in $\hat{H}$ has support of radius $\rho$.
	There are $n$ internal degrees of freedom in one unit cell, and $\hat{H}$ only contains up to four-fermion terms and has translation symmetry.
	The number of free parameters in $\hat{H}$ is $N_H\sim({\rho}^{d}n)^4/\rho^{d}=\rho^{3d}n^4$ for $\rho,n\gg1$, where the denominator comes from the translation invariance.
	In the commutator $[b_\mathbf{R},\hat{H}]$, only terms with support overlapping that of $b_\mathbf{R}$ contribute, so its support size is $(r+\rho)^{d}$.
	Each term in the commutator has up to three fermion operators, so the total number of terms in the full commutator is $N_C\sim(r+\rho)^{3d}n^3$ if $\rho,r,n\gg1$.
	Therefore, for $n>(1+r/\rho)^{3d}$, $N_H>N_C$ and a nonzero $\hat{H}$ always exists.
	
	We make no claim at fractional filling beyond the fully occupied case considered above. Nevertheless, Ref.~\cite{yang2025fractional} realizes a top$^2$-flat band and reports numerical evidence for an FQAH phase, implying potential relevance of this structure to correlated topological phases.

	\textit{Note added}:  After completing our major results, we became aware of independent research \cite{li2026stable} on perfectly flat bands with stable topology using a momentum-space framework with symmetry indicators; the results partly overlap and are consistent.
	
	\vspace{-0.6\baselineskip}
	\section{acknowledgments}
	\par R.-H.L. thanks Xin Shen for inspiring discussions on related topics. C.F. thanks Haijun Liao, Lei Wang and Hong-Hao Tu for fruitful discussions. R.-H.L. and C.F. are supported by the National Natural Science Foundation of China (NSFC) under grant numbers 12325404, 12547112, and 12188101; the National Key R\&D Program of China under grant numbers 2022YFA1403800 and 2023YFA1406704; and the Strategic Priority Research Program (B) of the Chinese Academy of Sciences (CAS) (Grant No.~XDB1720000). J.H. and C.F. are supported by the New Cornerstone Investigator Program.

\newpage


\section*{Supplementary material}

# Supplemental material for "The theory of topological-topological flat bands"

Rui-Heng Liu,[1,2] Jiangping Hu,[3] and Chen Fang[1,*]

[1]*Beijing National Laboratory for Condensed Matter Physics and Institute of Physics,
Chinese Academy of Sciences, Beijing 100190, China*
[2]*University of Chinese Academy of Sciences, Beijing 100049, China*
[3]*New Cornerstone Science Laboratory, Beijing National Laboratory for Condensed Matter Physics and Institute of Physics,
Chinese Academy of Sciences, Beijing 100190, China*

## CONTENTS

## Appendix A. Two topological conditions

### A.1. General definition

In this subsection, we define the two topological conditions in real space without assuming translational symmetry. We define the CLS as

$$b(x,y) = \sum_{\delta_x, \delta_y} b^{(\delta_x, \delta_y)}(x,y) \tag{S1}$$

Here, the superscript $(\delta_x, \delta_y)$ indicates the wavefunction is actually located at site $(x + \delta_x, y + \delta_y)$. Each $b^{(\delta_x,\delta_y)}(x,y)$ is a vector with the index of internal degree of freedom omitted. Summing wavefunctions over different sites is

understood as their direct sum as a single particle state. $b(x, y)$ could explicitly depend on the spatial coordinate so the translational symmetry is broken. The first topological condition is

$$\sum_{x,y} t_{xy} b(x, y) = 0 \quad \text{(PBC)} \quad \Rightarrow \quad \sum_{\delta_x, \delta_y} b^{(\delta_x, \delta_y)}(x_0 - \delta_x, y_0 - \delta_y) = 0 \quad \forall x_0, y_0 \tag{S2}$$

where we assume $t_{xy} = 1$ for $\Rightarrow$. This condition implies that superposing bulk CLSs yields nonzero amplitudes only at the boundary.

For simplicity, we focus on cross-shaped CLSs as

$$b(x, y) = b^O(x, y) + b^R(x, y) + b^U(x, y) + b^L(x, y) + b^D(x, y) \tag{S3}$$

Here, the superscript $O = (0, 0), R = (1, 0), U = (0, 1), L = (-1, 0), D = (0, -1)$. The explicit first topological condition for such CLS reads

$$\sum_{x,y} b(x, y) = 0 \quad \text{(PBC)} \iff b^O(x_0, y_0) + b^R(x_0 - 1, y_0) + b^U(x_0, y_0 - 1) + b^L(x_0 + 1, y_0) + b^D(x_0, y_0 + 1) = 0 \quad \forall x_0, y_0 \tag{S4}$$

As illustrated in the main text, under the first topological condition, we can further define loop states on the boundary of a strip as

$$L(x) = \sum_y \left[ b^R(x, y) - b^L(x + 1, y) \right] \qquad L(y) = \sum_x \left[ b^D(x, y) - b^U(x, y - 1) \right] \tag{S5}$$

The definition for general CLSs could be obtained but omitted here. We also only consider straight loops aligned with the axes here for simplicity; generic loops can be obtained by deforming a straight loop via superposing adjacent CLSs. Notably, the existence of loop states does not rely on translational symmetry, but only on the first topological condition.

The second topological condition is that all vertical loops $L(x)$ and horizontal loops $L(y)$ span the same Hilbert space. Formally

$$\sum_x c_x L(x) = \lambda \sum_y c_y L(y) \tag{S6}$$

For simplicity, we may restrict $\lambda = i$ and $c_x, c_y$ to be identical. Hence

$$b^R(x_0 - 1, y_0) - b^L(x_0 + 1, y_0) = i \left[ b^D(x_0, y_0 + 1) - b^U(x_0, y_0 - 1) \right] \qquad \forall x_0, y_0 \tag{S7}$$

We give an illustrative example of generating inhomogeneous CLSs while respecting the first and second topological conditions as follows

1. Define

$$b^U(x, y-1) = \begin{bmatrix} i \\ -1 \end{bmatrix} + \delta^U(x, y-1) \quad b^L(x+1, y) = \begin{bmatrix} -1 \\ -1 \end{bmatrix} + \delta^L(x+1, y) \quad b^D(x, y+1) = \begin{bmatrix} -i \\ -1 \end{bmatrix} + \delta^D(x, y+1) \tag{S8}$$

with $\delta^{U,L,D}$ randomly generated.

2. Define

$$b^R(x-1, y) = \begin{bmatrix} 1 \\ -1 \end{bmatrix} + \delta^R(x-1, y) \quad b^O(x, y) = \begin{bmatrix} 0 \\ 4 \end{bmatrix} + \delta^O(x, y) \tag{S9}$$

where

$$\delta^R(x-1, y) = i \left[ \delta^D(x, y+1) - \delta^U(x, y-1) \right] + \delta^L(x+1, y) \tag{S10}$$

$$\delta^O(x, y) = - \left[ \delta^R(x-1, y) + \delta^D(x, y+1) + \delta^U(x, y-1) + \delta^L(x+1, y) \right] \tag{S11}$$

When all $\delta = 0$, the translational symmetry is recovered and reduce to the case discussed in the main text, where we denote the wavefunctions as $O, R, U, L, D$ without ambiguity.

### A.2. The first topological condition with translational symmetry

In this subsection, we define the first topological condition in a periodic lattice. The core is to identify the singular point in $\boldsymbol{k}$-space. As in the main text, we first define the Fourier transform of the CLS $b_{\boldsymbol{R}}$ as

$$b_{\boldsymbol{k}} = \sum_{\boldsymbol{R}} b_{\boldsymbol{R}} e^{i\boldsymbol{k}\cdot\boldsymbol{R}} \tag{S12}$$

As conventional, we denote the periodic part of $b_{\boldsymbol{k}}$ as $u_{\boldsymbol{k}} = b_{\boldsymbol{k}} e^{-i\boldsymbol{k}\cdot\boldsymbol{r}}$. Due to compactness of $b_{\boldsymbol{R}}$, each component of $u_{\boldsymbol{k}}$ is a trigonometric polynomial, and $u_{\boldsymbol{k}}$ is typically not normalized. Generically, we can expand it near an arbitrary $\boldsymbol{k}'$ as

$$u_{\boldsymbol{k}} = u_0 + \nabla_{\boldsymbol{k}} u \cdot \boldsymbol{q} + \cdots \tag{S13}$$

where $u_0 = u_{\boldsymbol{k}=\boldsymbol{k}'}$ and $\boldsymbol{q}$ is measured from $\boldsymbol{k}'$. The Fourier transform is unitary and does not change the dimension of the Hilbert space. Since the first topological condition imposes a linear constraint on all CLSs in real space, there must exist a singular momentum such that $b_{\boldsymbol{k}}$ is not a physical state. Equivalently, we express the first topological condition as

$$\exists \boldsymbol{k_0} \quad u_0 = 0 \tag{S14}$$

We will mainly assume $\boldsymbol{k_0} = 0$ for simplicity but this is not necessary. Under the first topological condition, the Bloch state at $\boldsymbol{k_0}$ is absent in the Hilbert space spanned by the CLSs; instead, it is recovered by two loop states. As illustrated in the main text, these loop states are constructed from $\partial_{x,y} b_{\boldsymbol{k}}$, since $H_0(\partial_{x,y} b_{\boldsymbol{k}}) = \partial_{x,y}(H_0 b_{\boldsymbol{k}}) = 0$ validates $\partial_{x,y} b_{\boldsymbol{k}}$ as zero modes (as well as all higher derivatives). As a result, the flat-band subspace has dimension $N+1$, enforcing a band touching at $\boldsymbol{k_0}$.

### A.3. The second topological condition with translational symmetry

In this subsection, we provide a formal definition of the second topological condition in a periodic lattice, which is applicable to both time reversal symmetry broken/preserved systems. The core is to guarantee that the subspace spanned by the flat band wavefunction remains physically identical on any direction approaching the singular point, thus allowing for a well-defined projector and topological invariant. As $u_0 = 0$ by the first topological condition, it holds that

$$\frac{\partial b_{\boldsymbol{k}}}{\partial k_i}\bigg|_{\boldsymbol{k}=\boldsymbol{k_0}} = e^{i\boldsymbol{k_0}\cdot\boldsymbol{r}} \frac{\partial u_{\boldsymbol{k}}}{\partial k_i}\bigg|_{\boldsymbol{k}=\boldsymbol{k_0}} \tag{S15}$$

It is then natural to characterize the two loop states (essentially $\partial_{x,y} b_{\boldsymbol{k}}$) by $\nabla_{\boldsymbol{k}} u$, and ask what additional properties could be explored. We define

$$U_i = \frac{\partial \tilde{u}_{\boldsymbol{k}}}{\partial k_i}\bigg|_{\boldsymbol{k}=\boldsymbol{k_0}} \in \mathbb{C}^{n\times m} \tag{S16}$$

Here, for time reversal symmetry broken case $m=1$ and simply $\tilde{u}_{\boldsymbol{k}} = u_{\boldsymbol{k}}$; while for time reversal symmetric case $m=2$ and $\tilde{u}_{\boldsymbol{k}} = [u_{\boldsymbol{k}}, \mathrm{T} u_{-\boldsymbol{k}}]$ where T is the time reversal operator. $i=1,2,3$ corresponds to the spatial direction $x, y, z$. The second topological condition is defined upon these first-order derivatives as

$$\left|\det(U_i^\dagger U_j)\right|^2 = \det(U_i^\dagger U_i) \cdot \det(U_j^\dagger U_j) \qquad \forall i,j \tag{S17}$$

which guarantees that all $U_i$ span the same space. Also, we demand

$$S_{ij} = \mathrm{Re}\left[\mathrm{Tr}(U_i^\dagger U_j)\right] \qquad S > 0 \quad \text{(positive definite)} \tag{S18}$$

which guarantees that when approaching $\boldsymbol{k_0}$ in a general direction $\boldsymbol{q}$, the subspace spanned by $q_x U_x + q_y U_y + q_z U_z$ is identical and of rank $m$ (notice that for $m=2$, the two columns related by an antiunitary transformation could not

be of rank 1). This allows us to uniquely distinguish the degenerate states at the singular point, and hence define the projector of the flat band. For a 2D system without time reversal symmetry, the second topological condition could be simpflied as

$$U_y = \lambda U_x \qquad \text{Im}(\lambda) \neq 0 \tag{S19}$$

as claimed in the main text.

The following two lemmas may help understand the second topological condition mathematically.

*Lemma 1*: For $U_i \in \mathbb{C}^{n \times m}$ and $\text{rank}(U_i) = m$, they span the same space iff

$$\left| \det(U_i^\dagger U_j) \right|^2 = \det(U_i^\dagger U_i) \cdot \det(U_j^\dagger U_j) \tag{S20}$$

Proof: We set

$$A = U_i(U_i^\dagger U_i)^{-1/2}, \qquad B = U_j(U_j^\dagger U_j)^{-1/2}, \tag{S21}$$

so that $A^\dagger A = \mathbf{1}_m$ and $B^\dagger B = \mathbf{1}_m$. Then $M = A^\dagger B$ satisfies $M^\dagger M \leq \mathbf{1}_m$ (as a matrix inequality); hence all singular values of $M$ lie in $[0,1]$. Consequently, $|\det M| \leq 1$ with equality iff $M$ is unitary. Now compute

$$|\det M|^2 = \frac{|\det(U_i^\dagger U_j)|^2}{\det(U_i^\dagger U_i)\ \det(U_j^\dagger U_j)}. \tag{S22}$$

Thus the lemma is equivalent to $|\det M|^2 \leq 1$, and the equality corresponds to $|\det M| = 1$, i.e., $M$ unitary. When $M$ is unitary, we have $B = AM$ and

$$U_j = U_i\,(U_i^\dagger U_i)^{-1/2} M\,(U_j^\dagger U_j)^{1/2}, \tag{S23}$$

which expresses $U_j$ as an invertible linear transformation of $U_i$; hence $\text{Col}(U_j) = \text{Col}(U_i)$.

Conversely, if the column spaces coincide, we set $U_j = U_i C$ with $C \in \mathbb{C}^{m \times m}$ invertible. It follows that

$$|\det(U_i^\dagger U_j)|^2 = |\det(U_i^\dagger U_i C)|^2 = |\det(U_i^\dagger U_i)|^2 |\det C|^2, \tag{S24}$$

and

$$\det(U_i^\dagger U_i) \det(U_j^\dagger U_j) = \det(U_i^\dagger U_i) \det(C^\dagger U_i^\dagger U_i C) = \det(U_i^\dagger U_i)^2 |\det C|^2 \tag{S25}$$

so equality holds. This proves the equivalence.

*Lemma 2*: For $U_i \in \mathbb{C}^{n \times m}$, the equation $\sum_i q_i U_i = 0$ has real solutions iff

$$S_{ij} = \text{Re}\left[\text{Tr}(U_i^\dagger U_j)\right] \qquad \det(S) = 0 \tag{S26}$$

Proof: Introduce the real inner product

$$\langle X, Y \rangle_\mathbb{R} = \text{Re}\left[\text{Tr}(X^\dagger Y)\right] \tag{S27}$$

The matrix $S$ is the Gram matrix of $\{U_i\}$ with respect to $\langle \cdot, \cdot \rangle_\mathbb{R}$. If $\sum_i q_i U_i = 0$ for some real $q \neq 0$, then

$$0 = \left\langle \sum_i q_i U_i, U_j \right\rangle_\mathbb{R} = \sum_i q_i S_{ij} = (Sq)_j \qquad \forall j \tag{S28}$$

and hence $\det S = 0$.

Conversely, if $Sq = 0$ for some real $q \neq 0$, then $\langle \sum_i q_i U_i, U_j \rangle_{\mathbb{R}} = 0, \forall j$. Hence

$$\left\langle \sum_i q_i U_i, \sum_i q_i U_i \right\rangle_{\mathbb{R}} = q^{\mathrm{T}} S q = 0 \tag{S29}$$

so $\sum_i q_i U_i = 0$. This proves the equivalence.

Under the second topological condition, two loop states yield only one physical state, raising a question of how to obtain the additional $\boldsymbol{k}_0$ state in the flat-band subspace. We show that the missing state can be restored by following a designed path approaching $\boldsymbol{k}_0$ in the complexified momentum space. We first expand the Fourier-transformed CLS as

$$b_{\boldsymbol{k}} = b_0 + b_x q_x + b_y q_y + \frac{1}{2} \left( b_{xx} q_x^2 + 2 b_{xy} q_x q_y + b_{yy} q_y^2 \right) + \cdots \tag{S30}$$

where the subscript denotes partial derivatives in momentum space. We approach $\boldsymbol{k}_0$ on the parameterized curve

$$\begin{cases} k_x = t \\ k_y = \sum_n c_n t^n = c_1 t + c_2 t^2 + \cdots \end{cases} \tag{S31}$$

and it follows that

$$b_{\boldsymbol{k}} = b_0 + (b_x + c_1 b_y) t + \left( \frac{1}{2} b_{xx} + c_1 b_{xy} + \frac{c_1^2}{2} b_{yy} + c_2 b_y \right) t^2 + \cdots \tag{S32}$$

The two topological conditions lead to $b_0 = 0$ and $b_y = \lambda b_x, \mathrm{Im}(\lambda) \neq 0$. As we have illustrated, the loop states at $\boldsymbol{k}_0$ are exactly given by the first-order derivatives. To obtain the missing state, one must choose appropriate parameters to approach a state orthogonal to the loop state. In most generic cases, this could be achieved by choosing $c_1 = -1/\lambda, c_2 = 0$. The leading term is in the second order as

$$\frac{t^2}{2\lambda^2} \left( \lambda^2 b_{xx} - 2\lambda b_{xy} + b_{yy} \right) \tag{S33}$$

Noticed that

$$b_{k_x, k_y} = \sum_{\boldsymbol{R}} b_{\boldsymbol{R}} \exp(i\boldsymbol{k} \cdot \boldsymbol{R}) = \sum_{X,Y} b(X,Y) \exp(ik_x X) \exp(ik_y Y) \tag{S34}$$

Thus, the missing wavefunction could be constructed as

$$\psi \sim \sum_{X,Y} (\lambda^2 X^2 - 2\lambda XY + Y^2) b(X,Y) \tag{S35}$$

For concreteness, we still consider a cross-shaped CLS with components denoted as $O, R, U, L, D$. The two topological conditions demand that

$$O + L + R + U + D = 0 \qquad (D - U) = \lambda(L - R) \tag{S36}$$

It follows that

$$\sum_{X,Y} X^2 b(X,Y) = L + R + 2(L-R)X \qquad \sum_{X,Y} XY b(X,Y) = (D-U)X + (L-R)Y \qquad \sum_{X,Y} Y^2 b(X,Y) = D + U + 2(D-U)Y \tag{S37}$$

and

$$\psi \sim \sum_{X,Y} \lambda^2 (L+R) + D + U \tag{S38}$$

Here, the summed term is independent of spatial position, consistent with its correspondence to a state at $\boldsymbol{k}_0 = 0$.

More generally, the resulting state is ensured to be homogeneous with an properly chosen parameterized curve, and the choice of coefficients $c_n$ can be directly determined from $u_{\boldsymbol{k}}$. To see this, write $\psi(k_x, k_y) = u(k_x, k_y)e^{i(k_x X + k_y Y)}$ as a component of $b_{\boldsymbol{k}}$, and restrict to a curve $\psi(t) \equiv \psi(t, f(t))$. Expanding in $t$, the zeroth-order term of $u$ vanishes by the first topological condition, so the leading contribution comes from the $t^1$ term in $u$ and the $t^0$ term in the phase, with no $X, Y$ dependence. If the first-order term of $u$ is set to zero, the second-order contribution arises solely from the $t^2$ term in $u$. Iterating this argument, if all terms in $u$ below order $n$ vanish, then the order-$n$ contribution comes entirely from $u$, with no inhomogeneous contribution from the phase. Therefore, it suffices to expand $u_{\boldsymbol{k}}$ in $t$ and choose parameters so that its first $n$ orders vanish, where $n = 1$ typically but could depends on the model, thereby obtaining the missing state.

When it is unfortunate that the remaining second-order term cannot provide a state orthogonal to the loop state, we can always find the missing state by working on higher-order contributions. For example, consider a Fourier-transformed CLS with $u_{\boldsymbol{k}} = [\sin k_x + i \sin k_y, 2 - \cos k_x - \cos k_y]^{\mathrm{T}}$, as previously argued, we expand it on the parameterized curve as

$$
\begin{bmatrix}
(1 + ic_1)t + ic_2 t^2 + \left(-\dfrac{1}{6} - \dfrac{ic_1^3}{6} + ic_3\right)t^3 + \left(-\dfrac{i}{2}c_1^2 c_2 + ic_4\right)t^4 \\
\left(-\dfrac{1}{2} + \dfrac{c_1^2}{2}\right)t^2 + c_1 c_2 t^3 + \left(-\dfrac{1}{24} + c_1 c_3 + \dfrac{c_2^2}{2} - \dfrac{c_1^4}{24}\right)t^4
\end{bmatrix}
\tag{S39}
$$

Since the loop states are polarized on the first component, we need to choose $c_1 = i, c_2 = 0, c_3 = -i/3, c_4 = 0$ to finally get the missing $\boldsymbol{k}_0$ state, and it will be translated back to real space as a uniform wavefunction. While the result is obvious in this two-band model, this general method will be useful in more complicate cases.

### A.4. Caveat on additional singularities and a practical solution

In this subsection, we identify a caveat in the construction of top$^2$-flat band models and provide a detailed recipe to resolve it. The second topological condition is defined in terms of local derivatives, and thus does not rule out additional singularities (degeneracies) in the band structure. For example, one can readily check that the following wavefunction satisfies both topological conditions at $\boldsymbol{k} = 0$

$$
u_{\boldsymbol{k}} = \begin{bmatrix} \sin k_x + i \sin k_y \\ \cos k_x - \cos k_y \end{bmatrix}
\tag{S40}
$$

However, this wavefunction also satisfies the first topological condition at $(\pi, \pi)$, leading to a degeneracy there. As a result, the band topology cannot be determined solely from its winding behavior $\sin k_x + i \sin k_y \sim k_x + i k_y$ around the origin.

To remedy this, we provide a general recipe for avoiding such additional singularities. For a 2D system without time reversal symmetry, we propose the following parameterization

$$
u_{\boldsymbol{k}} = A \sin k_x + \lambda A \sin k_y + D(1 - \cos k_x) + E(1 - \cos k_y)
\tag{S41}
$$

Here, we first specify $A$ as a nonzero $n \times 1$ vector and $\lambda$ with $\mathrm{Im}(\lambda) \neq 0$. $D, E$ should be properly chosen to guarantee $\boldsymbol{k} = 0$ is the only singular point. Formally, we define $Q$ as the projector to project to the complement of $\mathrm{Span}\{A\}$, and the following condition is sufficient

$$
Q\left[D(1 - \cos k_x) + E(1 - \cos k_y)\right] = 0 \Rightarrow k_x = k_y = 0
\tag{S42}
$$

Practically, this could be easily satisfied by including a $(2 - \cos k_x - \cos k_y)$ component. In the simplest $n = 2$ case, the criterion could be reduced to

$$
(A \times D)(A \times E) > 0
\tag{S43}
$$

A simple example is $A = [1, 0]^{\mathrm{T}}, D = E = [0, 1]^{\mathrm{T}}$ and $\lambda = i$ as

$$
u_{\boldsymbol{k}} = \begin{bmatrix} \sin k_x + i \sin k_y \\ 2 - \cos k_x - \cos k_y \end{bmatrix}
\tag{S44}
$$

For a time reversal symmetric system, we provide a general parameterization as

$$u_{\boldsymbol{k},1} = A\sin k_x + (\alpha A + \beta \mathrm{T}A)\sin k_y + (\mu A + \nu \mathrm{T}A)\sin k_z + D(1-\cos k_x) + E(1-\cos k_y) + F(1-\cos k_z) \quad \text{(S45)}$$

with $X = [X_\uparrow, X_\downarrow]^{\mathrm{T}}$ has $2n$ components and $\mathrm{T}X = [X_\downarrow^*, -X_\uparrow^*]^{\mathrm{T}}$ with $\mathrm{T} = i\sigma_y \mathcal{K}$, $\mathrm{T}^2 = -1$. Its time reversal partner is

$$u_{\boldsymbol{k},2} = \mathrm{T}b_{-\boldsymbol{k},1} = -\mathrm{T}A\sin k_x + (\beta^* A - \alpha^* \mathrm{T}A)\sin k_y + (\nu^* A - \mu^* \mathrm{T}A)\sin k_z + \mathrm{T}D(1-\cos k_x) + \mathrm{T}E(1-\cos k_y) + \mathrm{T}F(1-\cos k_z) \quad \text{(S46)}$$

Here, we first specify $A$ as a nonzero $2n \times 1$ vector and choose $\alpha, \beta, \mu, \nu$ such that

$$\left[\mathrm{Im}(\alpha)^2 + |\beta|^2\right]\left[\mathrm{Im}(\mu)^2 + |\nu|^2\right] > \left[\mathrm{Im}(\alpha)\mathrm{Im}(\mu) + \mathrm{Re}(\beta^*\nu)\right]^2 \quad \text{(S47)}$$

or equivalently $(\mathrm{Im}(\alpha), \mathrm{Re}(\beta), \mathrm{Im}(\beta))$ and $(\mathrm{Im}(\mu), \mathrm{Re}(\nu), \mathrm{Im}(\nu))$ are not colinear to guarantee $S > 0$ (the second topological condition). $D, E, F$ should be properly chosen to guarantee $\boldsymbol{k} = 0$ is the only singular point. Formally, we define $Q$ as the projector to project to the complement of $\mathrm{Span}\{A, \mathrm{T}A\}$, and the following condition is sufficient

$$Q\left[D(1-\cos k_x) + E(1-\cos k_y) + F(1-\cos k_z)\right] = 0 \Rightarrow k_x = k_y = k_z = 0 \quad \text{(S48)}$$

Practically, this could be easily satisfied by including a $(3 - \cos k_x - \cos k_y - \cos k_z)$ component.

This parameterization naturally reduces to 2D systems by removing all terms involving $k_z$. A simple example is $A = [1,0,0,0]^{\mathrm{T}}, D = E = [0,1,0,0]^{\mathrm{T}}$ and $\alpha = i$ as

$$u_{\boldsymbol{k},1} = \begin{bmatrix} \sin k_x + i\sin k_y \\ 2 - \cos k_x - \cos k_y \\ 0 \\ 0 \end{bmatrix} \qquad u_{\boldsymbol{k},2} = \begin{bmatrix} 0 \\ 0 \\ \sin k_x - i\sin k_y \\ -2 + \cos k_x + \cos k_y \end{bmatrix} \quad \text{(S49)}$$

In 3D, a simple example is $A = [0,0,1,0]^{\mathrm{T}}, D = E = F = [0,1,0,0]^{\mathrm{T}}$ and $\alpha = i, \nu = 1$ as

$$u_{\boldsymbol{k},1} = \begin{bmatrix} \sin k_z \\ 3 - \cos k_x - \cos k_y - \cos k_z \\ \sin k_x + i\sin k_y \\ 0 \end{bmatrix} \qquad u_{\boldsymbol{k},2} = \begin{bmatrix} -\sin k_x + i\sin k_y \\ 0 \\ \sin k_z \\ -3 + \cos k_x + \cos k_y + \cos k_z \end{bmatrix} \quad \text{(S50)}$$

### Appendix B. $\mathbb{Z}_2$ invariant of top$^2$-flat bands

In this section, we verify our top$^2$-flat 3D TI has a nontrivial $\mathbb{Z}_2$ index by computing its Wilson loop. The occupied states in this model are

$$|u_1(\boldsymbol{k})\rangle = \frac{1}{\mathcal{N}}\begin{bmatrix} \sin k_z \\ 3 - \cos k_x - \cos k_y - \cos k_z \\ \sin k_x + i\sin k_y \\ 0 \end{bmatrix} \qquad |u_2(\boldsymbol{k})\rangle = \frac{1}{\mathcal{N}}\begin{bmatrix} -\sin k_x + i\sin k_y \\ 0 \\ \sin k_z \\ -3 + \cos k_x + \cos k_y + \cos k_z \end{bmatrix} \quad \text{(S51)}$$

with $\mathcal{N}$ the normalization factor.

On the $k_z = 0$ plane, it reduces to

$$|u_1(k_x, k_y, k_z = 0)\rangle = \frac{1}{\mathcal{N}}\begin{bmatrix} 0 \\ 2 - \cos k_x - \cos k_y \\ \sin k_x + i\sin k_y \\ 0 \end{bmatrix} \qquad |u_2(k_x, k_y, k_z = 0)\rangle = \frac{1}{\mathcal{N}}\begin{bmatrix} -\sin k_x + i\sin k_y \\ 0 \\ 0 \\ -2 + \cos k_x + \cos k_y \end{bmatrix} \quad \text{(S52)}$$

On the $k_z = \pi$ plane, it reduces to

$$|u_1(k_x, k_y, k_z = \pi)\rangle = \frac{1}{\mathcal{N}}\begin{bmatrix} 0 \\ 4 - \cos k_x - \cos k_y \\ \sin k_x + i\sin k_y \\ 0 \end{bmatrix} \qquad |u_2(k_x, k_y, k_z = \pi)\rangle = \frac{1}{\mathcal{N}}\begin{bmatrix} -\sin k_x + i\sin k_y \\ 0 \\ 0 \\ -4 + \cos k_x + \cos k_y \end{bmatrix} \quad \text{(S53)}$$

We compute the Wilson loops of the two occupied bands on each plane, as shown in Fig. S1. The winding is nontrivial on the $k_z = 0$ plane but trivial on the $k_z = \pi$ plane, leading to strong $\mathbb{Z}_2 = 1$ index.

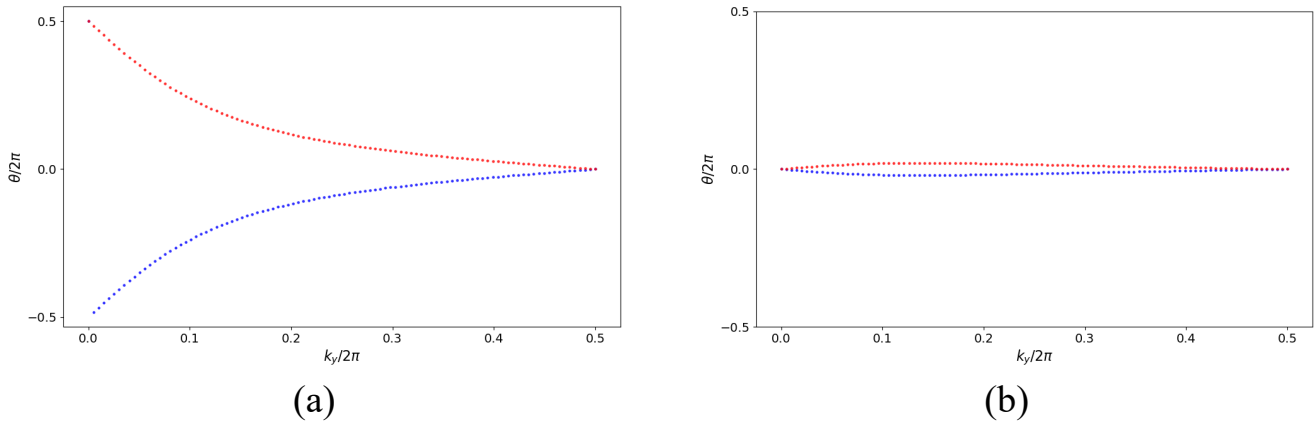

<div align="center">(a)</div> <div align="center">(b)</div>

FIG. S1. (a) Wilson loop calculation (evolution of Wannier center) on the $k_z = 0$ plane, showing a nontrivial winding. (b) Same on the $k_z = \pi$ plane, showing a trivial winding.

The nontrivial $\mathbb{Z}_2$ index for the 2D counterpart could be similarly verified by repeating the wilson loop calculation, which is almost identical to the calculation for the 3D model restriced to the $k_z = 0$ plane.

### Appendix C. Parent Hamiltonian

In this section, we first briefly review the discussion of the parent Hamiltonian in Ref. [1], and then reformulate the construction of the parent Hamiltonian as a syzygy problem and discuss its solution; in particular, we provide a simple explicit solution applicable to both time reversal symmetry broken/preserved top$^2$-flat bands. From our construction, the Fourier transformed CLS, i.e., $u_{\boldsymbol{k}} = [v_{\boldsymbol{k},1}, \cdots, v_{\boldsymbol{k},n}]^{\mathrm{T}}$, is an unnormalized Bloch wavefunction with elements a trigonometric polynomial. A natural question is whether there exists a corresponding parent Hamiltonian with from strictly local hoppings, and how to find them.

Ref. [1] provides a solution as follows: Given $u_{\boldsymbol{k}}$, one can obtain the remaining $(n-1)$ orthonormal states via Gram-Schmidt orthogonalization, denoted as $d_{\boldsymbol{k},\alpha}$ with $\alpha = 1, \cdots, n-1$, and then define

$$\mathcal{U}(\boldsymbol{k}) = [u_{\boldsymbol{k}}, d_{\boldsymbol{k},1}, \cdots, d_{\boldsymbol{k},n-1}] \qquad \mathcal{V}(\boldsymbol{k}) = [0, E_{\boldsymbol{k},1} d_{\boldsymbol{k},1}, \cdots, E_{\boldsymbol{k},n-1} d_{\boldsymbol{k},n-1}] \tag{S54}$$

where $E_{\boldsymbol{k},\alpha}$ are the dispersions of the other $(n-1)$ bands. The constructed Hamiltonian should satisfy $H(\boldsymbol{k})\mathcal{U}(\boldsymbol{k}) = \mathcal{V}(\boldsymbol{k})$ and the elements are given by

$$[H(\boldsymbol{k})]_{ij} = \sum_{\alpha=1}^{n-1} E_{\boldsymbol{k}}^\alpha d_{\boldsymbol{k},i}^\alpha d_{\boldsymbol{k},j}^{\alpha*} \tag{S55}$$

It is required, and in principle possible, to properly choose $d_{\boldsymbol{k},\alpha}$ and $E_{\boldsymbol{k}}^\alpha$ such that $[H(\boldsymbol{k})]_{ij}$ are also trigonometric polynomials.

We provide a more formal viewpoint in the following. Mathematically, $v_{\boldsymbol{k}}$ is a Laurent polynomial of $e^{i\boldsymbol{k}}$, i.e., $v_{\boldsymbol{k}} \in \mathbb{C}[e^{\pm i\boldsymbol{k}_1}, \cdots, e^{\pm i\boldsymbol{k}_d}]$. Also, we demand that the Hamiltonian $H(\boldsymbol{k})$ should annihilate $u_{\boldsymbol{k}}$, with matrix elements given by trigonometric polynomials as well (so that the corresponding real space hopping is of finite range). Mathematically, this task is equivalent to requiring that each row of $H(\boldsymbol{k})$ be a syzygy, with an additional physical constraint that the full matrix is Hermitian. We do not attempt a detailed and rigorous introduction to polynomial rings and modules. Intuitively, a syzygy can be viewed as solving a linear system, with the unknowns promoted from numbers to polynomials (elements of a ring). For Laurent polynomials we are working on, all solutions are finitely generated (known as the syzygy module). In practice, once we analytically or computationally work out several (at least one,

which is always possible) syzygies such that

$$\sum_i s_{\boldsymbol{k},i} v_{\boldsymbol{k},i} = 0 \qquad s_{\boldsymbol{k}} \in \mathbb{C}[e^{\pm i\boldsymbol{k}_1}, \cdots, e^{\pm i\boldsymbol{k}_d}] \tag{S56}$$

we can arrange $s_{\boldsymbol{k}}$ as the row of a matrix $S(\boldsymbol{k})$, then $H(\boldsymbol{k}) = S^\dagger(\boldsymbol{k})S(\boldsymbol{k})$ is always a valid Hamiltonian.

Actually, for our setup there exists an elegant construction of the Hamiltonian in a projector-like form as

$$H(\boldsymbol{k}) = u_{\boldsymbol{k}}^\dagger u_{\boldsymbol{k}} \mathbf{1} - u_{\boldsymbol{k}} u_{\boldsymbol{k}}^\dagger \tag{S57}$$

which has several advantages:

1. It requires only the flat-band wavefunction, without any additional band information;

2. It is positive semidefinite, so the resulting flat band lies at the bottom of the spectrum, which is potentially useful for studying interacting many-body states;

3. It can be directly generalized to the time-reversal-symmetric case: noting that $u_{\boldsymbol{k},1}$ and $u_{\boldsymbol{k},2} = \mathrm{T} u_{-\boldsymbol{k},1}$ have equal norms and can be made orthogonal, we can directly construct their corresponding parent Hamiltonian

$$H(\boldsymbol{k}) = u_{\boldsymbol{k},1}^\dagger u_{\boldsymbol{k},1} \mathbf{1} - u_{\boldsymbol{k},1} u_{\boldsymbol{k},1}^\dagger - u_{\boldsymbol{k},2} u_{\boldsymbol{k},2}^\dagger \tag{S58}$$

This is not obvious from the general syzygy setup, since it is not trivial to determine whether two sets of generators share the same syzygy.

Using our construction, the parent Hamiltonian of our top$^2$-flat Chern band is

$$H(\boldsymbol{k}) = \begin{bmatrix} (2 - \cos k_x - \cos k_y)^2 & (-2 + \cos k_x + \cos k_y)(\sin k_x + i \sin k_y) \\ (-2 + \cos k_x + \cos k_y)(\sin k_x - i \sin k_y) & \sin^2 k_x + \sin^2 k_y \end{bmatrix} \tag{S59}$$

with $u_{\boldsymbol{k}} = [\sin k_x + i \sin k_y, 2 - \cos k_x - \cos k_y]^{\mathrm{T}}$.

The parent Hamiltonian of our top$^2$-flat 3D TI band is

$$H(\boldsymbol{k}) = \begin{bmatrix} (3 - C_x - C_y - C_z)^2 & (-3 + C_x + C_y + C_z)S_z & 0 & (-3 + C_x + C_y + C_z)(S_z - iS_y) \\ (-3 + C_x + C_y + C_z)S_z & S_x^2 + S_y^2 + S_z^2 & (-3 + C_x + C_y + C_z)(S_x - iS_y) & 0 \\ 0 & (-3 + C_x + C_y + C_z)(S_x + iS_y) & (3 - C_x - C_y - C_z)^2 & (3 - C_x - C_y - C_z)S_z \\ (-3 + C_x + C_y + C_z)(S_z + iS_y) & 0 & (3 - C_x - C_y - C_z)S_z & S_x^2 + S_y^2 + S_z^2 \end{bmatrix} \tag{S60}$$

with $u_{\boldsymbol{k},1} = [\sin k_z, 3 - \cos k_x - \cos k_y - \cos k_z, \sin k_x + i \sin k_y, 0]^{\mathrm{T}}$, $u_{\boldsymbol{k},2} = [-\sin k_x + i \sin k_y, 0, \sin k_z, -3 + \cos k_x + \cos k_y + \cos k_z]^{\mathrm{T}}$. Here, $S, C$ are short hand notations for sin, cos.

### Appendix D. Construction of top$^2$-flat bands for topological crystals

#### D.1. Review of real space construction of TCI

In this subsection, we review the idea and main steps of real space construction of TCI. The main point of the real-space construction is to use intrinsically symmetry-protected lower-dimensional topological states as building blocks, and arrange them in real space according to crystal symmetry to obtain crystalline-symmetry-protected topological phases. To systematically understand the real-space construction, it is useful to introduce the concept of a cell complex, which is formed by assembling geometrical entities in different dimensions. The cell complex is defined according to a given space group. The first step is to identify an asymmetric unit, or called a 3-cell, which is the interior of a region of space that is as large as possible, such that any two distinct points in the region can not be related by a crystalline symmetry. Practically, a 3-cell is constructed by first determining the primitive unit cell according to translational symmetry, and then further subdivides the unit cell using the crystal symmetries beyond translation. This subdivision produces several regions of equal volume and each of the region is a 3-cell uniquely associated with a group element. Further, we define a 2-cell on the surface shared by two neighboring 3-cells, a 1-cell

on the edge where several 2-cells meet and a 0-cell as a vertex where multiple 1-cells intersect. In the original setup, all low-dimensional cells are also defined such that no two distinct points in the cell can be related under symmetries, which will be slightly relaxed later. There could be more than one type of low-dimensional cells. Within each type, all the cells could transform to each other by a symmetry element; while the cells belong to distinct type could not. A cell complex could be simply understood as gluing all cells of different dimensions together.

To apply the real-space construction to three-dimensional topological crystalline insulators (TCIs), we need to properly decorate the cell complex by some topological states as building blocks. It is proven that the only relevant cases are

1. Decorating two-dimensional mirror Chern insulators (MCIs) on the 2-cell which is a mirror plane. In this case, the MCIs are actually placed on decoupled two-dimensional planes, refereed to as layer constructions.

2. Decorating two-dimensional time-reversal-invariant topological insulators (2DTIs) on the 2-cell which is not a mirror plane. In this case, the decorated 2-cells satisfy the gluing condition that each 1-cell is the edge of an even number of these 2DTI blocks. Most of such constructions are still decoupled into layers, but there are exceptions that the 2DTI should be decorated on an intricate connected surface without boundary, instead of a plane, and hence called nonlayer constructions, or topological crysyals. There are 12 of them which live in the following space groups: Pnn2 (#34,A), Pnnn (#48,B), $P4_2$ (#77,C), $P4_2/n$ (#86,D), $P4_22$ (#93,E), $P4_22_1^2$ (#94,F), $P4_2$cm (#102,G), $P\bar{4}n2$ (#118,H), $P4_2/nnm$ (#134,I), $Pn\bar{3}$ (#201,J), $P4_232$ (#208,K), $Pn\bar{3}m$ (#224,L). For intuitive pictures of these topological crystals, see Ref. [2].

### D.2. Top²-flat bands in topological crystals

In this subsection, we employ a graph theory perspective to demonstrate how to obtain top²-flat bands in all topological crystals. In the main text, we construct a two-dimensional top² Chern band. Consequently, the MCI and 2DTI states follow directly. Thus, all TCIs attainable via the layer construction are exhausted. We use a pair of time-reversal partners with $D_4$ symmetry as building blocks to construct the remaining 12 topological crystals. Concretely, the wavefunction of one CLS is given by

$$O = [0, 4, 0, 0]^{\mathrm{T}} \qquad R = [0, -1, 1, 0]^{\mathrm{T}} \qquad U = [0, -1, i, 0]^{\mathrm{T}} \qquad L = [0, -1, -1, 0]^{\mathrm{T}} \qquad D = [0, -1, -i, 0]^{\mathrm{T}} \tag{S61}$$

To proceed, we first introduce a slight modification of the cell complex. In the original definition, each 2-cell is either a parallelogram or a triangle, and none carries any spatial symmetry (more precisely, no nontrivial group element leaves an individual 2-cell invariant). For our purpose, whenever a 2-cell is triangular, we merge it with a neighboring triangle to form a parallelogram. This can always be done for the 12 topological crystals under consideration. The resulting "composite" 2-cell then has a nontrivial stabilizer subgroup of order two, physically realized by a twofold rotation or a mirror. Since our CLS has higher symmetry, these additional symmetries will not hinder our following construction. With this change, we obtain a fully connected, boundaryless two-dimensional surface tiled by parallelograms sharing edges. In the spirit of the real-space construction, resolving the remaining 12 topological crystals reduces to the following task: place the $O$ component of the CLS at each (possibly "composite") 2-cell of the topological crystal, and self-consistently position the four legs $(R, U, L, D)$ on adjacent 2-cells, such that every 2-cell is covered exactly once by the set $\{O, R, U, L, D\}$. Once this is achieved, we have decorated the entire topological crystal with a 2DTI, thereby yielding the corresponding TCI.

The solution will become apparent once we employ a graph theory perspective. To facilitate our presentation, we introduce some useful definitions and conclusions, which could be skipped for readers who are familiar with graph theory.

1. **Graph.** A graph $G$ is an ordered pair $(V(G), E(G))$ where $V(G) = \{v_1, \ldots, v_n\}$ is a finite vertex set and $E(G)$ is a set of edges. Each edge is an unordered pair of distinct vertices, i.e., $E(G) \subseteq \{\{v_i, v_j\} | v_i, v_j \in V(G), \ i \neq j\}$. This defines a simple undirected graph (no loops or multiple edges).

2. **Adjacent and degree.** Two vertices $u, v \in V(G)$ are called adjacent (or neighbors) if $\{u, v\} \in E(G)$. The degree of a vertex $v$, denoted $\deg(v)$, is the number of edges incident to $v$. A graph is called $k$-regular if $\deg(v) = k, \forall v \in V(G)$.

3. **Adjacency matrix.** Given an ordering $V(G) = \{v_1, \ldots, v_n\}$, the adjacency matrix $A = [a_{ij}]$ is an $n \times n$ matrix defined by

$$a_{ij} = \begin{cases} 1, & \{v_i, v_j\} \in E(G), \\ 0, & \text{otherwise.} \end{cases} \tag{S62}$$

For a simple undirected graph, $A$ is symmetric and $a_{ii} = 0$.

4. **Complete graph.** The complete graph $K_n$ has $n$ vertices and contains every possible edge between distinct vertices. The adjacency matrix of $K_n$ is simply $a_{ij} = 1 - \delta_{ij}$.

5. **Subgraph and spanning subgraph.** A graph $G'$ is a subgraph of $G$ if $V(G') \subseteq V(G), E(G') \subseteq E(G)$, and each edge in $E(G')$ has both endpoints in $V(G')$. If $V(G') = V(G)$, then $G'$ is called a spanning subgraph of $G$.

6. **$n$-factor and $n$-factorization.** A $n$-factor of $G$ is a spanning $n$-regular subgraph, and a $n$-factorization partitions the edges of the graph into disjoint $n$-factors. In particular, a 2-factor is a collection of vertex-disjoint cycles whose union covers all vertices of $G$. A 1-factor is also called a perfect matching. The complete graph $K_{2n}$ always admit a 1-factorization by equally dividing all $2n$ vertices and matching them.

7. **Petersen's 2-factor theorem.** Let $G$ be a $2n$-regular graph, then the edge set $E(G)$ can be partitioned into $n$ edge-disjoint 2-factors. Equivalently, a $2n$-regular graph can be decomposed into $n$ spanning subgraphs, each of which is a union of vertex-disjoint cycles covering all vertices of $G$. The proof can be found in standard graph theory textbooks, see as an example.

8. **Corollary of the 2-factor theorem.** Let $G$ be a $2n$-regular graph with adjacency matrix $A$, then $A$ could be written as

$$A = \sum_{i=1}^{n} \left( P_i + P_i^{\mathrm{T}} \right) \tag{S63}$$

with $P_i$ permutation matrices.

Proof: Using Petersen's 2-factor theorem, we partition $E(G)$ as edge-disjoint 2-factors $H_{1,\ldots,n}$. Consequently the adjacency matrix $A$ is the sum of the adjacency matrices of these 2-factors as $A = \sum_{i=1}^{n} A_i$. Therefore, we only need to prove that the adjacency matrix of a 2-factor could be written as $P + P^{\mathrm{T}}$. Noticed that a 2-factor consists of vertex-disjoint cycles, we orientate each cycle in one of the two directions to define a permutation $\sigma$ on the vertex set (each vertex maps to its successor). Let $P$ be the permutation matrix corresponding to $\sigma$, i.e. $P_{ij} = 1$ iff $\sigma(i) = j$. Then the edge set of $H$ is exactly $\{\{i, j\} \mid \sigma(i) = j \text{ or } \sigma(j) = i\}$, hence $A_H = P + P^{\mathrm{T}}$. This completes the proof.

We now interpret each topological crystal as a graph. We first assign a vertex to each composite 2-cell (parallelogram), located at its center where the $\{O, R, U, L, D\}$ wavefunction is placed. To define the edge set, a natural way is to assign an edge between two vertices whenever the corresponding parallelograms are adjacent (sharing a 1-cell). Noticed that according to the gluing condition, each 1-cell is shared by even number of 2-cells, therefore such an assignment locally creates a complete graph $K_{2n}$. However, we only choose a 1-factor of this complete graph for our purpose. Intuitively, this corresponds to virtually separating some originally adjacent 2-cells so that each 1-cell is now shared by exactly two 2-cells. In fact, among all 12 topological crystals, only B and H exhibit the situation where a single 1-cell is shared by four 2-cells. In all other cases, no such modification is required. After this step, the original topological crystal can be interpreted as a 4-regular graph, where each edge represents the adjacency between two (possibly "composite") 2-cells. Every vertex is of degree 4, reflecting the fact that a topological crystal has no boundary.

With the graph interpretation, the desired decoration is straightforward. We decompose the adjacency matrix of the resulting 4-regular graph as

$$A = P_R + P_L + P_U + P_D \qquad P_L = P_R^{\mathrm{T}} \quad P_D = P_U^{\mathrm{T}} \tag{S64}$$

with $P_n, n = \{R, U, L, D\}$ permutation matrices. We then define

$$(P_n)_{ij} = 1 \Rightarrow f_n(v_j) = v_i, \tag{S65}$$

These four mappings completely determine the desired decoration: if the $j$-th 2-cell is assigned $O$, then the 2-cell $f_\alpha(v_j)$ is assigned $n = \{R, U, L, D\}$. Since each $P_n$ is a permutation, it is guaranteed that when every 2-cell is decorated once by $O$, each 2-cell is also decorated exactly once by $R, U, L, D$.

One may notice a subtle issue in the above construction. When a given topological crystal contains more than one type of 2-cell (which indeed occurs in 10 cases excluding J and L), these 2-cells can not be related by any symmetry of the corresponding space group. If we decorate all 2-cells using identical CLSs, the resulting state, while still realizing a TCI protected by the given space group, may accidentally exhibit an enlarged symmetry. This loophole can be resolved with some complication. We divide the remaining 10 cases into two classes.

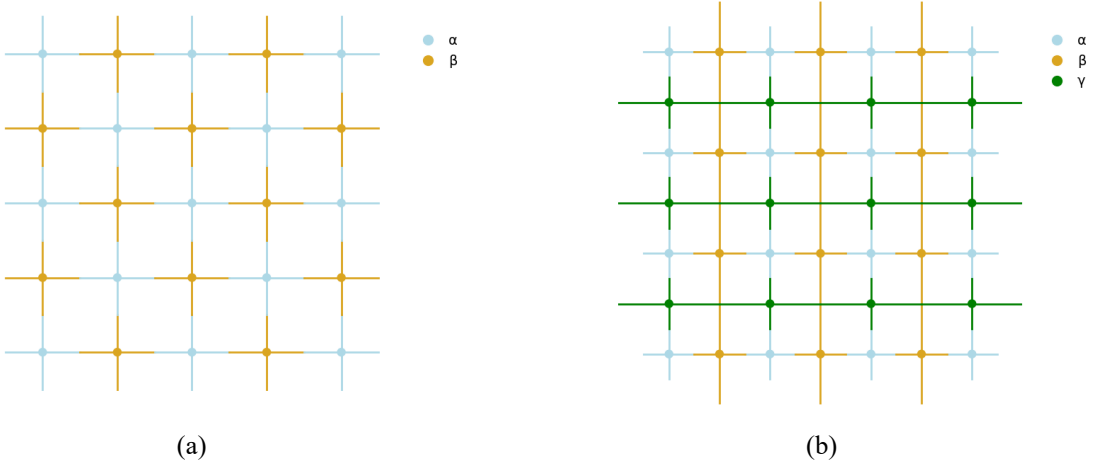

FIG. S2. Connectivity of topological crystals. (a) D, G and K with a staggered square lattice structure. (b) A, B, C, E, F, H and I with a modified Lieb lattice structure.

1. D, G and K. These 3 topological crystals contain two types of 2-cells, whose connectivity is equivalent to that of a square lattice with staggered $\alpha$ and $\beta$ sublattices (Fig. S2). In these circumstances, we introduce two set of distinct $D_4$-symmetric CLSs, with wavefunctions denoted by $O_{\alpha,\beta}, R_{\alpha,\beta}, U_{\alpha,\beta}, L_{\alpha,\beta}, D_{\alpha,\beta}$. The two topological conditions then become

$$\text{First:} \begin{cases} O_\alpha + R_\beta + U_\beta + L_\beta + D_\beta = 0 \\ O_\beta + R_\alpha + U_\alpha + L_\alpha + D_\alpha = 0 \end{cases} \qquad \text{Second:} \begin{cases} R_\alpha - L_\alpha = \lambda(U_\alpha - D_\alpha) \\ R_\beta - L_\beta = \lambda(U_\beta - D_\beta) \end{cases} \tag{S66}$$

with $\text{Im}(\lambda) \neq 0$. It is straightforward to see that these conditions can always be satisfied as the number of orbitals per site becomes larger. The previous construction corresponds to the special case where the two type of CLSs are identical.

2. A, B, C, E, F, H and I. These 7 topological crystals contain two or three types of 2-cells, whose connectivity is equivalent to that a modified Lieb lattice (Fig. S2). In these circumstances, we first introduce three set of distinct $D_4$-symmetric CLSs, with wavefunctions denoted by $O_{\alpha,\beta,\gamma}, R_{\alpha,\beta,\gamma}, U_{\alpha,\beta,\gamma}, L_{\alpha,\beta,\gamma}, D_{\alpha,\beta,\gamma}$. The two topological conditions then become

$$\text{First:} \begin{cases} O_\alpha + R_\beta + U_\gamma + L_\beta + D_\gamma = 0 \\ O_\beta + R_\alpha + U_\beta + L_\alpha + D_\beta = 0 \\ O_\gamma + R_\gamma + U_\alpha + L_\gamma + D_\alpha = 0 \end{cases} \qquad \text{Second:} \begin{cases} R_\beta - L_\beta = \lambda(U_\gamma - D_\gamma) \\ R_\alpha - L_\alpha = \lambda(U_\beta - D_\beta) \\ R_\gamma - L_\gamma = \lambda(U_\alpha - D_\alpha) \end{cases} \tag{S67}$$

with $\text{Im}(\lambda) \neq 0$. In case that there are only two distinct types of 2-cells, just simply take the corresponding CLSs to be identical. Similarly, one can see that this is always solvable.

## Appendix E. Perturbative interaction on top²-flat bands

### E.1. A $k \cdot p$ argument

In this subsection, we construct a low-energy $\boldsymbol{k} \cdot \boldsymbol{p}$ model for our two-band top²-flat band model near the touching point. We study how the Chern number depends on the perturbative parameters. Although interactions are not explicitly considered, we provide it as complementary evidence that opening the gap in the top²-flat band may yield a nontrivial topology. We consider the top²-flat Chern band model

$$H(\boldsymbol{k}) = \begin{bmatrix} (2 - \cos k_x - \cos k_y)^2 & (-2 + \cos k_x + \cos k_y)(\sin k_x + i \sin k_y) \\ (-2 + \cos k_x + \cos k_y)(\sin k_x - i \sin k_y) & \sin^2 k_x + \sin^2 k_y \end{bmatrix} \tag{S68}$$

with the zero mode

$$u_{\boldsymbol{k}} = \begin{bmatrix} \sin k_x + i \sin k_y \\ 2 - \cos k_x - \cos k_y \end{bmatrix} \tag{S69}$$

The $\boldsymbol{k} \cdot \boldsymbol{p}$ expansion near $\boldsymbol{k} = (0,0)$ is

$$H(\boldsymbol{k}) = \begin{bmatrix} \frac{1}{4}(k_x^2 + k_y^2)^2 & -\frac{1}{2}(k_x^2 + k_y^2)(k_x + ik_y) \\ -\frac{1}{2}(k_x^2 + k_y^2)(k_x - ik_y) & (k_x^2 + k_y^2) - \frac{1}{3}(k_x^4 + k_y^4) \end{bmatrix} \tag{S70}$$

This is equivalent to $\boldsymbol{d} \cdot \boldsymbol{\sigma}$ with

$$\begin{cases} d_x = -\frac{1}{2}(k_x^2 + k_y^2)k_x = -\frac{1}{2}k^3 \cos\varphi \\ d_y = \frac{1}{2}(k_x^2 + k_y^2)k_y = \frac{1}{2}k^3 \sin\varphi \\ d_z = -\frac{1}{2}(k_x^2 + k_y^2) + \frac{1}{8}(k_x^2 + k_y^2)^2 + \frac{1}{6}(k_x^4 + k_y^4) = -\frac{1}{2}k^2 + \frac{7}{24}k^4 - \frac{1}{3}k^4 \cos^2\varphi \sin^2\varphi \end{cases} \tag{S71}$$

Notice that the original lattice model is gapless at $(0,0)$, we introduce a small perturbation $\boldsymbol{t} \cdot \boldsymbol{\sigma}$ to open the gap. Thus, we consider

$$\begin{cases} d'_x = -\frac{1}{2}(k_x^2 + k_y^2)k_x + t_x = 0 \\ d'_y = \frac{1}{2}(k_x^2 + k_y^2)k_y + t_y = 0 \\ d'_z = -\frac{1}{2}(k_x^2 + k_y^2) + \frac{1}{8}(k_x^2 + k_y^2)^2 + \frac{1}{6}(k_x^4 + k_y^4) + t_z = 0 \end{cases} \qquad \begin{cases} t_x = \frac{1}{2}(k_x^2 + k_y^2)k_x \\ t_y = -\frac{1}{2}(k_x^2 + k_y^2)k_y \\ t_z = \frac{1}{2}(k_x^2 + k_y^2) - \frac{1}{8}(k_x^2 + k_y^2)^2 - \frac{1}{6}(k_x^4 + k_y^4) \end{cases} \tag{S72}$$

This parameterized the gapless surface in phase space $(t_x, t_y, t_z)$, which explicitly reads

$$t_z = 2^{-1/3}(t_x^2 + t_y^2)^{1/3} - \frac{7}{12}2^{1/3}(t_x^2 + t_y^2)^{2/3} + \frac{2}{3}2^{1/3}\frac{t_x^2 t_y^2}{(t_x^2 + t_y^2)^{4/3}} \tag{S73}$$

The maximum value is $t_z = 3/10$ when $k_x = k_y = \pm\sqrt{3/5}$ and $t_x = t_y = \pm(3/5)^{3/2}$. When $|t_{x,y}|$ are very small, the leading term is $t_z \sim (t_x^2 + t_y^2)^{1/3}$ and for a fixed $t_z$ we have $t_x^2 + t_y^2 \approx C$. As $|t_{x,y}|$ gets larger, the curve deviates from a circle and becomes anisotropic due to the third term but retain a four-fold rotation.

The $\boldsymbol{k} \cdot \boldsymbol{p}$ model is an effective low-energy description. To have a correct topological analysis, we should compactify the parameter space as $\mathbb{R}^2 \cup \{\infty\} \cong S^2$ to define a mapping $S^2 \mapsto S^2$. This is achieved by introducing $\hat{\boldsymbol{d}}'(k \to \infty) = (0,0,1)$. If $t_x = t_y = 0, t_z < 0$, $\hat{\boldsymbol{d}}'(k = 0) = (0,0,-1)$ and the image of the mapping should cover the sphere and the mapping degree is 1 (noticed the winding from $d_x - id_y$ which covers the azimuthal direction), which is consistent with the lattice model (since $b_{\boldsymbol{k}} \to [1,0]^{\mathrm{T}}$ near the touching point and $t_z < 0$ lowers its energy). As long as the parameters do not cross the gapless surface, $C = 1$ remains.

One may understand $\boldsymbol{t} \cdot \boldsymbol{\sigma}$ as the effective correction near the touching point from interaction. Hence, the result here indicates the possibility of a topological gap.

## E.2. Hartree-Fock formalism

In this subsection, we provide a general Hartree-Fock formalism for later perturbative analysis of Hubbard interactions on top$^2$-flat bands. Consider a $n$ band system with Hamiltonian $H_0(\boldsymbol{k})$, the transformation from the original basis to the band basis is given by a unitary matrix $U(\boldsymbol{k})$, whose columns are the eigenvectors of $H_0(\boldsymbol{k})$

$$H_0(\boldsymbol{k})U(\boldsymbol{k}) = U(\boldsymbol{k}) \cdot \text{diag}[E_1(\boldsymbol{k}), \cdots, E_n(\boldsymbol{k})] \tag{S74}$$

with $E_n(\boldsymbol{k})$ the band dispersions in ascending order. The eigenstates are created by

$$\psi_{\boldsymbol{k},n}^\dagger = \sum_\alpha c_{\boldsymbol{k},\alpha}^\dagger u_{\boldsymbol{k},\alpha n} \qquad \psi_{\boldsymbol{k}}^\dagger = c_{\boldsymbol{k}}^\dagger U(\boldsymbol{k}) = \begin{bmatrix} \psi_{\boldsymbol{k},1}^\dagger & \cdots & \psi_{\boldsymbol{k},n}^\dagger \end{bmatrix} \tag{S75}$$

with $u_{\boldsymbol{k},\alpha n}$ the element of $U(\boldsymbol{k})$.

We use Hartree-Fock formalism to study interaction $H'$ as a perturbation. To start, we define the density matrix $\rho(\boldsymbol{k})$ at each $\boldsymbol{k}$ with $\rho(\boldsymbol{k}) = \rho^\dagger(\boldsymbol{k}), 0 \le \rho \le \mathbb{1}$ and $\sum_{\boldsymbol{k}} \text{Tr}\rho(\boldsymbol{k}) = N_{\text{particle}}$. We also define

$$D_{pq} = \frac{1}{N} \sum_{\boldsymbol{k}} \sum_{mn} \rho_{mn}(\boldsymbol{k}) u_{\boldsymbol{k},pm} u_{\boldsymbol{k},qn}^* \tag{S76}$$

Hence, the total energy functional is

$$E[\rho] = \sum_{\boldsymbol{k}} \text{Tr}(\rho(\boldsymbol{k})\varepsilon(\boldsymbol{k})) + F(D(\rho(\boldsymbol{k}))) \tag{S77}$$

Here, $\varepsilon(\boldsymbol{k}) = E_m(\boldsymbol{k})\delta_{mn}$ is a diagonal matrix with elements the noninteracting band dispersion. $F(D) = \langle H' \rangle$ is a functional depending on the interaction. When

To compute the variation, notice that

$$\delta F = \sum_{pq} \frac{\partial F}{\partial D_{pq}} \delta D_{pq} = \text{Tr}(h \delta D) \qquad h_{pq} = \frac{\partial F}{\partial D_{qp}} \tag{S78}$$

and

$$\delta D_{pq} = \frac{1}{N} \sum_{\boldsymbol{k}} \sum_{mn} \delta \rho_{mn}(\boldsymbol{k}) u_{\boldsymbol{k},pm} u_{\boldsymbol{k},qn}^* \tag{S79}$$

We find

$$\frac{\delta(NF)}{\delta \rho_{mn}(\boldsymbol{k})} = \sum_{pq} u_{\boldsymbol{k},pm} h_{qp} u_{\boldsymbol{k},qn}^* \tag{S80}$$

and the total variation is

$$\frac{\delta E[f]}{\delta \rho_{mn}(\boldsymbol{k})} = E_m(\boldsymbol{k})\delta_{mn} + \sum_{pq} h_{pq} \frac{\delta D_{qp}}{\delta \rho_{mn}(\boldsymbol{k})} = E_m(\boldsymbol{k})\delta_{mn} + \langle u_{\boldsymbol{k},m}| h |u_{\boldsymbol{k},n}\rangle \tag{S81}$$

We hence define the Hartree-Fock Hamiltonian as

$$H_{mn}^{\text{HF}} = E_m(\boldsymbol{k})\delta_{mn} + \langle u_{\boldsymbol{k},m}| h |u_{\boldsymbol{k},n}\rangle \tag{S82}$$

The one-shot Hartree-Fock band is obtained by diagonalizing this Hamiltonian.

### E.3. Application to the two-band Chern model

In this subsection, we apply Hartree-Fock formalism to our top²-flat Chern band with the Hamiltonian

$$H(\boldsymbol{k}) = \begin{bmatrix} (2-\cos k_x - \cos k_y)^2 & (-2+\cos k_x + \cos k_y)(\sin k_x + i\sin k_y) \\ (-2+\cos k_x + \cos k_y)(\sin k_x - i\sin k_y) & \sin^2 k_x + \sin^2 k_y \end{bmatrix} \tag{S83}$$

with the bands $E_1 = 0, E_2 = 6 - 4(\cos k_x + \cos k_y) + 2\cos k_x \cos k_y \geq 0$. The transformation to band basis is given by

$$U(\boldsymbol{k}) = \frac{1}{\mathcal{N}} \begin{bmatrix} \sin k_x + i\sin k_y & -2 + \cos k_x + \cos k_y \\ 2 - \cos k_x - \cos k_y & \sin k_x - i\sin k_y \end{bmatrix} \tag{S84}$$

with each column the band Bloch states and $\mathcal{N}^2 = 6 - 4(\cos k_x + \cos k_y) + 2\cos k_x \cos k_y$. The projector of band 1 is

$$P(\boldsymbol{k}) = \frac{1}{\mathcal{N}^2} \begin{bmatrix} \sin^2 k_x + \sin^2 k_y & (2-\cos k_x - \cos k_y)(\sin k_x + i\sin k_y) \\ (2-\cos k_x - \cos k_y)(\sin k_x - i\sin k_y) & (2-\cos k_x - \cos k_y)^2 \end{bmatrix} \tag{S85}$$

Here, we denote the basis as $\{A, B\}$ (one could also call it spin) and we introduce a Hubbard interaction

$$H' = U\sum_j n_{jA} n_{jB} \tag{S86}$$

We fix half filling and evaluate $\rho(\boldsymbol{k})$ and $D$ in the noninteracting ground state which is a Slater determinant as

$$D = \frac{1}{N}\sum_{\boldsymbol{k}} P(\boldsymbol{k}) = \frac{1}{(2\pi)^2}\int_{\text{FBZ}} 2k \cdot P(\boldsymbol{k}) \tag{S87}$$

Thus, form the Wick theorem, the energy functional of this interaction is

$$F(D) = U\left(D_{11} D_{22} - |D_{12}|^2\right) \tag{S88}$$

with the Hartree term and Fock term. Then

$$h_{pq} = \frac{\partial F}{\partial D_{qp}} = \begin{bmatrix} UD_{22} & -UD_{12} \\ -UD_{12}^* & UD_{11} \end{bmatrix} \tag{S89}$$

Since only diagonal elements of $P$ are even, $D = \text{diag}[D_A, D_B]$ with

$$D_A = \frac{1}{(2\pi)^2}\int_{\text{FBZ}} 2k \frac{\sin^2 k_x + \sin^2 k_y}{6 - 4(\cos k_x + \cos k_y) + 2\cos k_x \cos k_y} \approx 0.301 \tag{S90}$$

$$D_B = \frac{1}{(2\pi)^2}\int_{\text{FBZ}} 2k \frac{(2-\cos k_x - \cos k_y)^2}{6 - 4(\cos k_x + \cos k_y) + 2\cos k_x \cos k_y} \approx 0.699 \tag{S91}$$

and

$$h = U \cdot \text{diag}[D_B, D_A] \tag{S92}$$

Explicitly, the Hartree-Fock correction is

$$h_{11} = U \cdot \frac{1}{\mathcal{N}^2}\left[D_A\left(2-\cos k_x - \cos k_y\right)^2 + D_B\left(\sin^2 k_x + \sin^2 k_y\right)\right] \tag{S93}$$

$$h_{12} = h_{21}^* = U \cdot \frac{1}{\mathcal{N}^2}\left[(D_A - D_B)\left(2-\cos k_x - \cos k_y\right)\left(\sin k_x - i\sin k_y\right)\right] \tag{S94}$$

$$h_{22} = U \cdot \frac{1}{\mathcal{N}^2}\left[D_A\left(\sin^2 k_x + \sin^2 k_y\right) + D_B\left(2-\cos k_x - \cos k_y\right)^2\right] \tag{S95}$$

The bands along high symmetry line are plotted in Fig. S3.

Notice that the noninteracting system has a finite gap $\Delta$ away from $\boldsymbol{k} = (0,0)$, but $\Delta \to 0$ as $\boldsymbol{k} \to (0,0)$, and it is exactly the $\sim k_x + ik_y$ behavior around that contributes to the nontrivial topology. Near $\boldsymbol{k} = (0,0)$, the projector $P \to \text{diag}[1,0]$ and hence the Hartree-Fock Hamiltonian remain diagonal. When $U < 0$, the gap is opened and the original Bloch states in band 1 has lower energy. The physical picture is as follows: near $\boldsymbol{k} = (0,0)$, $|\psi_{\boldsymbol{k},1}\rangle$ are polarized on $A$ while $|\psi_{\boldsymbol{k},2}\rangle$ are polarized on $B$, but the occupied band are mainly distributed on $B$ after integrating over BZ ($D_B > D_A$). Hence, if an electron near $\boldsymbol{k} = (0,0)$ hop from the lower band to the upper band, we effectively has less double occupation. Hence, we must choose $U < 0$ to suppress this procedure so that the original state tends to be stable.

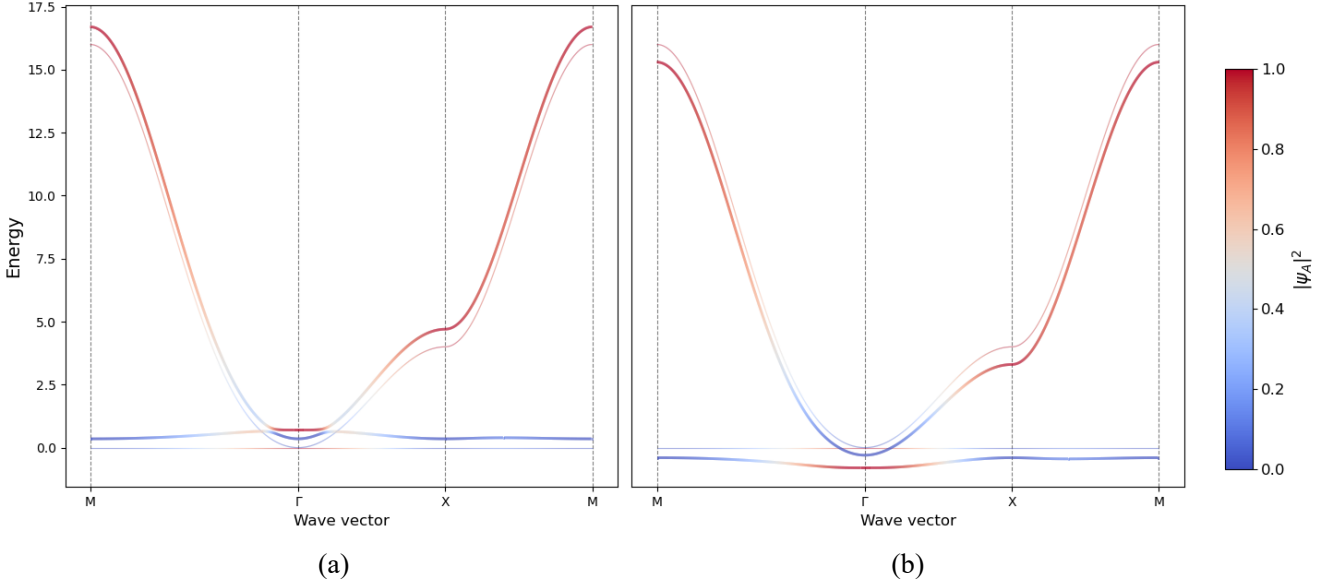

FIG. S3. Band structure of the top²-flat Chern band after one-shot Hartree-Fock correction. Thin (thick) lines show the noninteracting (Hartree-Fock) bands; color denotes the weight of the $A$ component. Parameter is (a) $U = 1$. (b) $U = -1$.

### E.4. Application to the four-band 3DTI model

In this subsection, we apply Hartree-Fock formalism to our top²-flat 3DTI band with the Hamiltonian

$$H(\boldsymbol{k}) = \begin{bmatrix} (3-C_x-C_y-C_z)^2 & (-3+C_x+C_y+C_z)S_z & 0 & (-3+C_x+C_y+C_z)(S_x-iS_y) \\ (-3+C_x+C_y+C_z)S_z & S_x^2+S_y^2+S_z^2 & (-3+C_x+C_y+C_z)(S_x-iS_y) & 0 \\ 0 & (-3+C_x+C_y+C_z)(S_x+iS_y) & (3-C_x-C_y-C_z)^2 & (3-C_x-C_y-C_z)S_z \\ (-3+C_x+C_y+C_z)(S_x+iS_y) & 0 & (3-C_x-C_y-C_z)S_z & S_x^2+S_y^2+S_z^2 \end{bmatrix} \quad (S96)$$

with the bands $E_1 = E_2 = 0, E_3 = E_4 = 12 - 6(\cos k_x + \cos k_y + \cos k_z) + 2(\cos k_x \cos k_y + \cos k_x \cos k_z + \cos k_y \cos k_z) \geq 0$. The transformation to band basis is given by

$$U(\boldsymbol{k}) = \frac{1}{\mathcal{N}} \begin{bmatrix} S_z & -S_x+iS_y & 3-C_x-C_y-C_z & 0 \\ 3-C_x-C_y-C_z & 0 & -S_z & S_x-iS_y \\ S_x+iS_y & S_z & 0 & -3+C_x+C_y+C_z \\ 0 & -3+C_x+C_y+C_z & -S_x-iS_y & -S_z \end{bmatrix} \quad (S97)$$

with each column the band Bloch states and $\mathcal{N}^2 = 12 - 6(\cos k_x + \cos k_y + \cos k_z) + 2(\cos k_x \cos k_y + \cos k_x \cos k_z + \cos k_y \cos k_z)$. The projector of band 1,2 is

$$P(\boldsymbol{k}) = \frac{1}{\mathcal{N}^2} \begin{bmatrix} S_x^2+S_y^2+S_z^2 & (3-C_x-C_y-C_z)S_z & 0 & (3-C_x-C_y-C_z)(S_x-iS_y) \\ (3-C_x-C_y-C_z)S_z & (3-C_x-C_y-C_z)^2 & (3-C_x-C_y-C_z)(S_x-iS_y) & 0 \\ 0 & (3-C_x-C_y-C_z)(S_x+iS_y) & S_x^2+S_y^2+S_z^2 & (-3+C_x+C_y+C_z)S_z \\ (3-C_x-C_y-C_z)(S_x+iS_y) & 0 & (-3+C_x+C_y+C_z)S_z & (3-C_x-C_y-C_z)^2 \end{bmatrix} \quad (S98)$$

Here, we denote the basis as $\{A \uparrow, B \uparrow, A \downarrow, B \downarrow\}$ and we introduce a Hubbard interaction

$$H' = \sum_j \left( U_A n_{j,1} n_{j,3} + U_B n_{j,2} n_{j,4} + U' \sum_j \sum_{p \in \{1,3\}, q \in \{2,4\}} n_{j,p} n_{j,q} \right) \quad (S99)$$

We fix half filling and evaluate $\rho(\boldsymbol{k})$ and $D$ in the noninteracting ground state which is a Slater determinant as

$$D = \frac{1}{N} \sum_{\boldsymbol{k}} P(\boldsymbol{k}) = \frac{1}{(2\pi)^3} \int_{\text{FBZ}} 3k \cdot P(\boldsymbol{k}) \quad (S100)$$

Thus, from the Wick theorem, the energy functional of this interaction is

$$F(D) = \left[ U_A \left( D_{11}D_{33} - |D_{13}|^2 \right) + U_B \left( D_{22}D_{44} - |D_{24}|^2 \right) \right.$$
$$\left. + U' \left( D_{11}D_{22} - |D_{12}|^2 + D_{11}D_{44} - |D_{14}|^2 + D_{22}D_{33} - |D_{23}|^2 + D_{33}D_{44} - |D_{34}|^2 \right) \right] \tag{S101}$$

with the Hartree term and Fock term. Then

$$h_{pq} = \frac{\partial F}{\partial D_{qp}} = \begin{bmatrix} U_A D_{33} + U'(D_{22} + D_{44}) & -U'D_{12} & -U_A D_{13} & -U'D_{14} \\ -U'D_{12}^* & U_B D_{44} + U'(D_{11} + D_{33}) & -U'D_{23} & -U_B D_{24} \\ -U_A D_{13}^* & -U'D_{23}^* & U_A D_{11} + U'(D_{22} + D_{44}) & -U'D_{34} \\ -U'D_{14}^* & -U_B D_{24}^* & -U'D_{34}^* & U_B D_{22} + U'(D_{11} + D_{33}) \end{bmatrix} \tag{S102}$$

Since only diagonal elements of P are even, $D = \mathrm{diag}[D_A, D_B, D_A, D_B]$ with

$$D_A = \frac{1}{(2\pi)^3} \int_{\mathrm{FBZ}} 3k \frac{\sin^2 k_x + \sin^2 k_y + \sin^2 k_z}{12 - 6(\cos k_x + \cos k_y + \cos k_z) + 2(\cos k_x \cos k_y + \cos k_x \cos k_z + \cos k_y \cos k_z)} \approx 0.205 \tag{S103}$$

$$D_B = \frac{1}{(2\pi)^3} \int_{\mathrm{FBZ}} 3k \frac{(3 - \cos k_x - \cos k_y - \cos k_z)^2}{12 - 6(\cos k_x + \cos k_y + \cos k_z) + 2(\cos k_x \cos k_y + \cos k_x \cos k_z + \cos k_y \cos k_z)} \approx 0.795 \tag{S104}$$

and

$$h = \mathrm{diag}[h_A, h_B, h_A, h_B] \qquad \begin{cases} h_A = U_A D_A + 2U'D_B \\ h_B = U_B D_B + 2U'D_A \end{cases} \tag{S105}$$

Explicitly, the Hartree-Fock correction is

$$h_{11} = \frac{1}{\mathcal{N}^2} \left[ h_A \left( \sin^2 k_x + \sin^2 k_y + \sin^2 k_z \right) + h_B \left( 3 - \cos k_x - \cos k_y - \cos k_z \right)^2 \right] \tag{S106}$$

$$h_{13} = h_{31}^* = \frac{1}{\mathcal{N}^2} \left[ (h_A - h_B) \left( 3 - \cos k_x - \cos k_y - \cos k_z \right) \sin k_z \right] \tag{S107}$$

$$h_{14} = h_{41}^* = -\frac{1}{\mathcal{N}^2} \left[ (h_A - h_B) \left( 3 - \cos k_x - \cos k_y - \cos k_z \right) \left( \sin k_x - i \sin k_y \right) \right] \tag{S108}$$

$$h_{22} = \frac{1}{\mathcal{N}^2} \left[ h_A \left( \sin^2 k_x + \sin^2 k_y + \sin^2 k_z \right) + h_B \left( 3 - \cos k_x - \cos k_y - \cos k_z \right)^2 \right] \tag{S109}$$

$$h_{23} = h_{32}^* = -\frac{1}{\mathcal{N}^2} \left[ (h_A - h_B) \left( 3 - \cos k_x - \cos k_y - \cos k_z \right) \left( \sin k_x + i \sin k_y \right) \right] \tag{S110}$$

$$h_{24} = h_{42}^* = -\frac{1}{\mathcal{N}^2} \left[ (h_A - h_B) \left( 3 - \cos k_x - \cos k_y - \cos k_z \right) \sin k_z \right] \tag{S111}$$

$$h_{33} = \frac{1}{\mathcal{N}^2} \left[ h_A \left( 3 - \cos k_x - \cos k_y - \cos k_z \right)^2 + h_B \left( \sin^2 k_x + \sin^2 k_y + \sin^2 k_z \right) \right] \tag{S112}$$

$$h_{44} = \frac{1}{\mathcal{N}^2} \left[ h_A \left( 3 - \cos k_x - \cos k_y - \cos k_z \right)^2 + h_B \left( \sin^2 k_x + \sin^2 k_y + \sin^2 k_z \right) \right] \tag{S113}$$

The bands along high symmetry line are plotted in Fig. S4.

We notice that $\Delta \to 0, P \to \mathrm{diag}[1, 0, 1, 0], Q \to \mathrm{diag}[0, 1, 0, 1]$ as $\boldsymbol{k} \to (0, 0, 0)$, and follow a similar logic as in the two band model, we should set $(U_A - 2U')D_A < (U_B - 2U')D_B$ to open a gap. If we take $U_A = U_B = U$, then $U > 2U'$ should hold. When $U = 0$, a negative $U'$ recovers the two band model.

## Appendix F. Detailed counting

In this section, we provide a more detailed counting to validate the statement that a CLS can serve as an exact zero mode of some four-operator interacting Hamiltonians. We denote the CLS creation operator as $d^\dagger = \sum c^\dagger$ hereafter, which is a superposition of onsite $c^\dagger$. We assume that $d^\dagger$ creates a local state supported within a $(L+1) \times (L+1)$ square region. The commutation $[H, d^\dagger] \sim \sum c^\dagger c^\dagger c = 0$ could be viewed as a set of linear equations, where the number of parameters is determined by the interaction and the number of equations is determined by the three-operator terms.

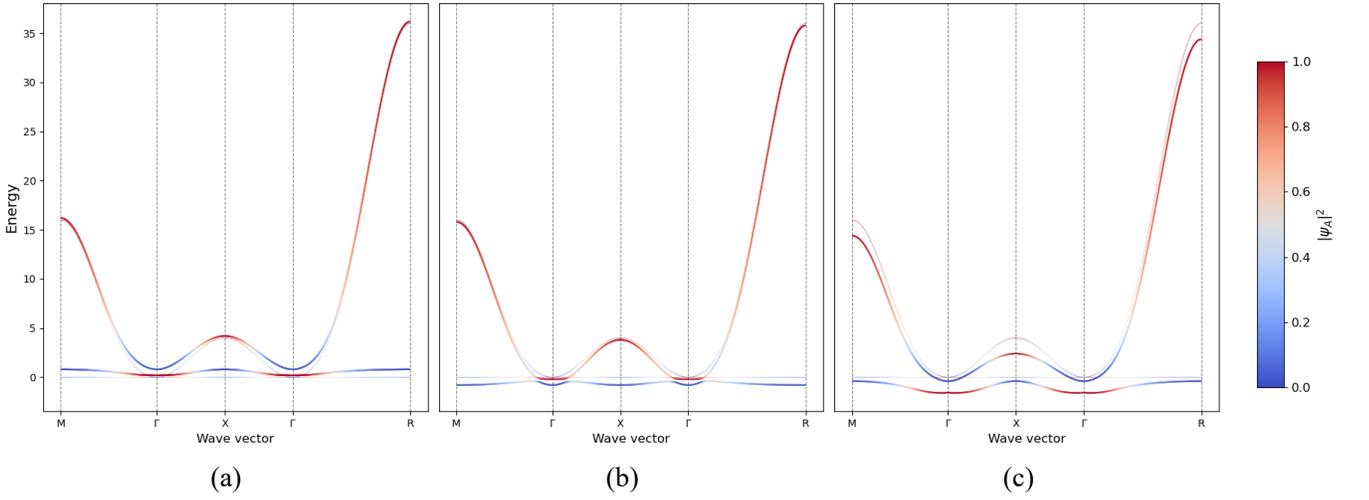

FIG. S4. Band structure of the top²-flat 3DTI band after one-shot Hartree-Fock correction. Thin (thick) lines show the noninteracting (Hartree-Fock) bands; color denotes the weight of the $A$ component. Parameters are (a) $U_A = U_B = 1, U' = 0$. (b) $U_A = U_B = -1, U' = 0$. (c) $U_A = U_B = 0, U' = -1$

### F.1. Counting of parameters

In this subsection, we present a detailed counting on the parameters of a local interacting Hamiltonian. The model is set up as follows: given a square lattice with $n$ flavors of fermions per site, we consider the following interaction term

$$U_{\alpha\beta\gamma\delta} c_\alpha^\dagger c_\beta^\dagger c_\gamma c_\delta \tag{S114}$$

where $\alpha, \beta, \gamma, \delta$ are multiple indices of site and flavor. The coordinates involved in the interaction, i.e., $\boldsymbol{r}_{\alpha,\beta,\gamma,\delta}$, must lie within a $(R+1) \times (R+1)$ square (consisting of lattice sites), where $R$ is the interaction range. The goal is to count the number of independent parameters $U_{\alpha\beta\gamma\delta}$. It must be noticed that two configurations $\boldsymbol{r}_{\alpha,\beta,\gamma,\delta}$ that only differ by a lattice translation are equivalent and correspond to the same parameter.

We divide the derivation into two parts. First, consider a $(R+1) \times (R+1)$ square, we ask how many inequivalent $p$ point configurations are allowed. We use the following trick to avoid multiple counting: noticed that for each configuration, we can always push it until it reaches the left and the bottom edge of the square. If two configurations are identical after this procedure, they must be equivalent. Thus, we should only count configurations in which the leftmost column and the downmost row are occupied. From the inclusion-exclusion principle, the number of inequivalent $p$ point configurations

$$f_p(R) = \binom{(R+1)^2}{p} - 2\binom{R(R+1)}{p} + \binom{R^2}{p} \sim R^{2(p-1)} \tag{S115}$$

Explicitly

$$f_1(R) = 1 \quad f_2(R) = 2R^2 + 2R \quad f_3(R) = \frac{3}{2}R^4 + 3R^3 + \frac{R^2}{2} - R \quad f_4(R) = \frac{2R^6}{3} + 2R^5 + \frac{7R^4}{12} - \frac{13R^3}{6} - \frac{3R^2}{4} + \frac{2R}{3} \tag{S116}$$

Second, assuming $\boldsymbol{r}_{\alpha,\beta,\gamma,\delta}$ already occupy $p$ distinct sites, we ask what the number of parameters is. We allow $U$ to take complex values, and use the fact that a $N \times N$ Hermitian matrix has $N^2$ real parameters. For $p = 1$, we have $C_2^n$ pair operators $c^\dagger c^\dagger (cc)$ and the number of parameters is

$$S_1(n) = \binom{n}{2}^2 \sim n^4 \tag{S117}$$

For $p > 2$, we have $C_2^{pn}$ pair operators but to get a correct counting, we should subtract the contribution from the

case that $r_{\alpha,\beta,\gamma,\delta}$ actually only occupy $q < p$ distinct sites. Therefore

$$S_p(n) = \binom{pm}{2}^2 - \sum_{q=1}^{p-1} \binom{p}{q} S_q(n) \sim n^4 \tag{S118}$$

Explicitly

$$S_1(n) = \frac{1}{4}n^4 - \frac{1}{2}n^3 + \frac{1}{4}n^2 \quad S_2(n) = \frac{7}{2}n^4 - 3n^3 + \frac{1}{2}n^2 \quad S_3(n) = 9n^4 - 3n^3 \quad S_4(n) = 6n^4 \tag{S119}$$

Combining the two, the total counting is

$$\begin{aligned}
\sum_{p=1}^{4} f_p(R)S_p(n) &= \frac{n^2}{4}\left[n^2(4R^3 + 6R^2 + 4R + 1)^2 - 2n(3R^2 + 3R + 1)^2 + (2R + 1)^2\right] \\
&= \frac{n^2}{4}(1 + 4R + 4R^2) - \frac{n^3}{2}(1 + 6R + 15R^2 + 18R^3 + 9R^4) \\
&\quad + \frac{n^4}{4}(1 + 8R + 28R^2 + 56R^3 + 68R^4 + 48R^5 + 16R^6)
\end{aligned} \tag{S120}$$

### F.2. Counting of three-operator terms

In this subsection, we analyze the commutator $[H, d^\dagger] \sim c^\dagger c^\dagger c$ and count the number of distinct three-operator terms. Here $d^\dagger$ creates a local state supported within a $(L + 1) \times (L + 1)$ square region. Since the Hamiltonian $H$ has range $R$, the operators appearing in the commutator may spread over a larger region. In particular, the resulting operators can lie within an $M \times M$ square with $M = 2R + L + 1$. However, each individual operator product $c^\dagger c^\dagger c$ must still be contained within a $(R + 1) \times (R + 1)$ square. We adopt a counting strategy similar to the previous. First, consider the full $M \times M$ region and ask how many distinct configurations of $p' = \{1, 2, 3\}$ points are possible under the constraint that their separations along both lattice directions do not exceed $R$. To facilitate the counting, we define

$$r_x = \text{Max}\left(|\boldsymbol{r_i} - \boldsymbol{r_j}|_x\right) \qquad r_y = \text{Max}\left(|\boldsymbol{r_i} - \boldsymbol{r_j}|_y\right) \tag{S121}$$

where $i, j$ label the three operators. Then locality constraint becomes $r_x \leq R, r_y \leq R$, and our goal is to count the possibilities of embedding a $r_x \times r_y$ rectangle in the $M \times M$ square.

For $p' = 1$, the counting is simply the number of sites in the square as

$$g_1(L, R) = M^2 \tag{S122}$$

For $p' = 2$, the counting is classified as $(0 < r_x \leq R, r_y = 0), (r_x = 0, 0 < r_y \leq R)$ and $(0 < r_x \leq R, 0 < r_y \leq R)$. The formula is

$$g_2(L, R) = \sum_{r_x=1}^{R}(M - r_x) \cdot M + M \cdot \sum_{r_y=1}^{R}(M - r_y) + 2\sum_{r_x=1}^{R}\sum_{r_y=1}^{R}(M - r_x)(M - r_y) \tag{S123}$$

Here, the factor 2 comes from the two distinct ways of placing the points on opposite corners of the $r_x \times r_y$ rectangle.

For $p' = 3$, we use the same classification and the formula is

$$g_3(L, R) = \sum_{r_x=1}^{R}(r_x - 1)(M - r_x) \cdot M + M \cdot \sum_{r_y=1}^{R}(r_y - 1)(M - r_y) + \sum_{r_x=1}^{R}\sum_{r_y=1}^{R}(6r_x r_y - 2)(M - r_x)(M - r_y) \tag{S124}$$

Here, the factor $(6r_x r_y - 2)$ receives contributions from several cases: (a) 4 configurations when the three points occupy three corners; (b) $2(r_x - 1) + 2(r_y - 1)$ configurations when the three points occupy two adjacent corners; (c) $2[(r_x + 1)(r_y + 1) - 4]$ configurations when the three points occupy two opposite corners; (d) $4(r_x - 1)(r_y - 1)$

configurations when the three points occupy a single corner.

Explicitly

$$g_1(L, R) = (2R + L + 1)^2 \qquad g_2(L, R) = \frac{1}{2}R(R+1)(3R + 2L + 1)(3R + 2L + 2) \tag{S125}$$

$$g_3(L, R) = \frac{1}{18}R(R+1)\left[(3R^2 + 3R - 2)(4R + 3L + 2)^2 - (R^2 + R - 2)\right] \tag{S126}$$

Second, we assume the three operators occupy $p'$ distinct sites and count the number of $c^\dagger c^\dagger c$ terms. Noticing the Pauli exclusion, the result is

$$T_1(n) = \binom{n}{2} \cdot n = \frac{1}{2}n^3 - \frac{1}{2}n^2 \quad T_2(n) = 2n^3 + 2 \cdot \binom{n}{2} \cdot n = 3n^3 - n^2 \quad T_3(n) = 3n^3 \tag{S127}$$

Combining the two, the total counting is

$$\begin{aligned}
\sum_{p'=1}^{3} g_{p'}(R, L)T_{p'}(n) = &\frac{1}{2}n^2(-1 - 2L - L^2 + n + 2Ln + L^2n) + \frac{1}{2}n^2(-6 - 10L - 4L^2 + 8n + 14Ln + 6L^2n)R \\
&+ \frac{1}{2}n^2(-15 - 18L - 4L^2 + 28n + 42Ln + 15L^2n)R^2 + \frac{1}{2}n^2(-18 - 12L + 56n + 68Ln + 18L^2n)R^3 \\
&+ \frac{1}{2}n^2(-9 + 68n + 60Ln + 9L^2n)R^4 + \frac{1}{2}n^2(48n + 24Ln)R^5 + 8n^3R^6
\end{aligned} \tag{S128}$$

### F.3. Conclusion

Noticing that each three-operator term could correspond to a complex coefficient in principle, we want the following condition to hold

$$\text{number of parameters} > 2 \times \text{number of three-operator terms} \geq \text{number of equations} \tag{S129}$$

In the parameter counting, the leading term is $4n^4R^6$. In the operator counting, the leading term is $8n^3R^6$. Hence, when $R$ becomes large but still finite, and these two terms with $R^6$ scaling dominate, this general condition could hold under proper interaction when $n > 4$.

Numerically, we first fix $L = 2$ for simplicity, corresponding to a CLS with 9 occupied sites at most. The counting becomes

$$2\left[n^2\left(-\frac{9}{2} - 21R - \frac{67}{2}R^2 - 21R^3 - \frac{9}{2}R^4\right) + n^3\left(\frac{9}{2} + 30R + 86R^2 + 132R^3 + 112R^4 + 48R^5 + 8R^6\right)\right] \tag{S130}$$

The condition is satisfied for $L = 2$, $(n, R) = (5, 12)(6, 7)(7, 5)$ or larger $R$.

Such a rough estimation is likely to be more stringent than reality for the following reasons:

1. We simplify the support of the CLS as a square and when it actually occupies a smaller region, the number of three-operator terms must be reduced.

2. The coefficient of each three-operator term may be restricted as a real number for a specific $d^\dagger$. Also, each three-operator term may not correspond to an independent equation. Introducing the factor 2 could only be an overestimation.

As a final remark, we notice that even the counting grows only polynomially rather than exponentially, the system of equations from $[H, d^\dagger] = 0$ is still quite large. One may exploit symmetries to reduce the size of the system of equations and simplify the problem, but finding a concrete example remains challenging.




\begin{thebibliography}{112}%
		\makeatletter
		\providecommand \@ifxundefined [1]{%
			\@ifx{#1\undefined}
		}%
		\providecommand \@ifnum [1]{%
			\ifnum #1\expandafter \@firstoftwo
			\else \expandafter \@secondoftwo
			\fi
		}%
		\providecommand \@ifx [1]{%
			\ifx #1\expandafter \@firstoftwo
			\else \expandafter \@secondoftwo
			\fi
		}%
		\providecommand \natexlab [1]{#1}%
		\providecommand \enquote  [1]{``#1''}%
		\providecommand \bibnamefont  [1]{#1}%
		\providecommand \bibfnamefont [1]{#1}%
		\providecommand \citenamefont [1]{#1}%
		\providecommand \href@noop [0]{\@secondoftwo}%
		\providecommand \href [0]{\begingroup \@sanitize@url \@href}%
		\providecommand \@href[1]{\@@startlink{#1}\@@href}%
		\providecommand \@@href[1]{\endgroup#1\@@endlink}%
		\providecommand \@sanitize@url [0]{\catcode `\\12\catcode `\$12\catcode `\&12\catcode `\#12\catcode `\^12\catcode `\_12\catcode `\%12\relax}%
		\providecommand \@@startlink[1]{}%
		\providecommand \@@endlink[0]{}%
		\providecommand \url  [0]{\begingroup\@sanitize@url \@url }%
		\providecommand \@url [1]{\endgroup\@href {#1}{\urlprefix }}%
		\providecommand \urlprefix  [0]{URL }%
		\providecommand \Eprint [0]{\href }%
		\providecommand \doibase [0]{https://doi.org/}%
		\providecommand \selectlanguage [0]{\@gobble}%
		\providecommand \bibinfo  [0]{\@secondoftwo}%
		\providecommand \bibfield  [0]{\@secondoftwo}%
		\providecommand \translation [1]{[#1]}%
		\providecommand \BibitemOpen [0]{}%
		\providecommand \bibitemStop [0]{}%
		\providecommand \bibitemNoStop [0]{.\EOS\space}%
		\providecommand \EOS [0]{\spacefactor3000\relax}%
		\providecommand \BibitemShut  [1]{\csname bibitem#1\endcsname}%
		\let\auto@bib@innerbib\@empty
		\bibitem [{\citenamefont {Sutherland}(1986)}]{sutherland1986localization}%
		\BibitemOpen
		\bibfield  {author} {\bibinfo {author} {\bibfnamefont {B.}~\bibnamefont {Sutherland}},\ }\bibfield  {title} {\bibinfo {title} {Localization of electronic wave functions due to local topology},\ }\href@noop {} {\bibfield  {journal} {\bibinfo  {journal} {Physical Review B}\ }\textbf {\bibinfo {volume} {34}},\ \bibinfo {pages} {5208} (\bibinfo {year} {1986})}\BibitemShut {NoStop}%
		\bibitem [{\citenamefont {Lieb}(1989)}]{lieb1989two}%
		\BibitemOpen
		\bibfield  {author} {\bibinfo {author} {\bibfnamefont {E.~H.}\ \bibnamefont {Lieb}},\ }\bibfield  {title} {\bibinfo {title} {Two theorems on the hubbard model},\ }\href@noop {} {\bibfield  {journal} {\bibinfo  {journal} {Physical review letters}\ }\textbf {\bibinfo {volume} {62}},\ \bibinfo {pages} {1201} (\bibinfo {year} {1989})}\BibitemShut {NoStop}%
		\bibitem [{\citenamefont {Vidal}\ \emph {et~al.}(1998)\citenamefont {Vidal}, \citenamefont {Mosseri},\ and\ \citenamefont {Dou{\c{c}}ot}}]{vidal1998aharonov}%
		\BibitemOpen
		\bibfield  {author} {\bibinfo {author} {\bibfnamefont {J.}~\bibnamefont {Vidal}}, \bibinfo {author} {\bibfnamefont {R.}~\bibnamefont {Mosseri}},\ and\ \bibinfo {author} {\bibfnamefont {B.}~\bibnamefont {Dou{\c{c}}ot}},\ }\bibfield  {title} {\bibinfo {title} {Aharonov-bohm cages in two-dimensional structures},\ }\href@noop {} {\bibfield  {journal} {\bibinfo  {journal} {Physical review letters}\ }\textbf {\bibinfo {volume} {81}},\ \bibinfo {pages} {5888} (\bibinfo {year} {1998})}\BibitemShut {NoStop}%
		\bibitem [{\citenamefont {Liu}\ \emph {et~al.}(2014)\citenamefont {Liu}, \citenamefont {Liu},\ and\ \citenamefont {Wu}}]{liu2014exotic}%
		\BibitemOpen
		\bibfield  {author} {\bibinfo {author} {\bibfnamefont {Z.}~\bibnamefont {Liu}}, \bibinfo {author} {\bibfnamefont {F.}~\bibnamefont {Liu}},\ and\ \bibinfo {author} {\bibfnamefont {Y.-S.}\ \bibnamefont {Wu}},\ }\bibfield  {title} {\bibinfo {title} {Exotic electronic states in the world of flat bands: From theory to material},\ }\href@noop {} {\bibfield  {journal} {\bibinfo  {journal} {Chinese Physics B}\ }\textbf {\bibinfo {volume} {23}},\ \bibinfo {pages} {077308} (\bibinfo {year} {2014})}\BibitemShut {NoStop}%
		\bibitem [{\citenamefont {Leykam}\ \emph {et~al.}(2018)\citenamefont {Leykam}, \citenamefont {Andreanov},\ and\ \citenamefont {Flach}}]{leykam2018artificial}%
		\BibitemOpen
		\bibfield  {author} {\bibinfo {author} {\bibfnamefont {D.}~\bibnamefont {Leykam}}, \bibinfo {author} {\bibfnamefont {A.}~\bibnamefont {Andreanov}},\ and\ \bibinfo {author} {\bibfnamefont {S.}~\bibnamefont {Flach}},\ }\bibfield  {title} {\bibinfo {title} {Artificial flat band systems: from lattice models to experiments},\ }\href@noop {} {\bibfield  {journal} {\bibinfo  {journal} {Advances in Physics: X}\ }\textbf {\bibinfo {volume} {3}},\ \bibinfo {pages} {1473052} (\bibinfo {year} {2018})}\BibitemShut {NoStop}%
		\bibitem [{\citenamefont {Checkelsky}\ \emph {et~al.}(2024)\citenamefont {Checkelsky}, \citenamefont {Bernevig}, \citenamefont {Coleman}, \citenamefont {Si},\ and\ \citenamefont {Paschen}}]{checkelsky2024flat}%
		\BibitemOpen
		\bibfield  {author} {\bibinfo {author} {\bibfnamefont {J.~G.}\ \bibnamefont {Checkelsky}}, \bibinfo {author} {\bibfnamefont {B.~A.}\ \bibnamefont {Bernevig}}, \bibinfo {author} {\bibfnamefont {P.}~\bibnamefont {Coleman}}, \bibinfo {author} {\bibfnamefont {Q.}~\bibnamefont {Si}},\ and\ \bibinfo {author} {\bibfnamefont {S.}~\bibnamefont {Paschen}},\ }\bibfield  {title} {\bibinfo {title} {Flat bands, strange metals and the kondo effect},\ }\href@noop {} {\bibfield  {journal} {\bibinfo  {journal} {Nature Reviews Materials}\ }\textbf {\bibinfo {volume} {9}},\ \bibinfo {pages} {509} (\bibinfo {year} {2024})}\BibitemShut {NoStop}%
		\bibitem [{\citenamefont {Shayegan}(2022)}]{shayegan2022wigner}%
		\BibitemOpen
		\bibfield  {author} {\bibinfo {author} {\bibfnamefont {M.}~\bibnamefont {Shayegan}},\ }\bibfield  {title} {\bibinfo {title} {Wigner crystals in flat band 2d electron systems},\ }\href@noop {} {\bibfield  {journal} {\bibinfo  {journal} {Nature Reviews Physics}\ }\textbf {\bibinfo {volume} {4}},\ \bibinfo {pages} {212} (\bibinfo {year} {2022})}\BibitemShut {NoStop}%
		\bibitem [{\citenamefont {Mielke}(1991{\natexlab{a}})}]{mielke1991ferromagnetic}%
		\BibitemOpen
		\bibfield  {author} {\bibinfo {author} {\bibfnamefont {A.}~\bibnamefont {Mielke}},\ }\bibfield  {title} {\bibinfo {title} {Ferromagnetic ground states for the hubbard model on line graphs},\ }\href@noop {} {\bibfield  {journal} {\bibinfo  {journal} {Journal of Physics A: Mathematical and General}\ }\textbf {\bibinfo {volume} {24}},\ \bibinfo {pages} {L73} (\bibinfo {year} {1991}{\natexlab{a}})}\BibitemShut {NoStop}%
		\bibitem [{\citenamefont {Mielke}(1991{\natexlab{b}})}]{mielke1991ferromagnetism}%
		\BibitemOpen
		\bibfield  {author} {\bibinfo {author} {\bibfnamefont {A.}~\bibnamefont {Mielke}},\ }\bibfield  {title} {\bibinfo {title} {Ferromagnetism in the hubbard model on line graphs and further considerations},\ }\href@noop {} {\bibfield  {journal} {\bibinfo  {journal} {Journal of Physics A: Mathematical and General}\ }\textbf {\bibinfo {volume} {24}},\ \bibinfo {pages} {3311} (\bibinfo {year} {1991}{\natexlab{b}})}\BibitemShut {NoStop}%
		\bibitem [{\citenamefont {Mielke}(1992)}]{mielke1992exact}%
		\BibitemOpen
		\bibfield  {author} {\bibinfo {author} {\bibfnamefont {A.}~\bibnamefont {Mielke}},\ }\bibfield  {title} {\bibinfo {title} {Exact ground states for the hubbard model on the kagome lattice},\ }\href@noop {} {\bibfield  {journal} {\bibinfo  {journal} {Journal of Physics A: Mathematical and General}\ }\textbf {\bibinfo {volume} {25}},\ \bibinfo {pages} {4335} (\bibinfo {year} {1992})}\BibitemShut {NoStop}%
		\bibitem [{\citenamefont {Tasaki}(1992)}]{tasaki1992ferromagnetism}%
		\BibitemOpen
		\bibfield  {author} {\bibinfo {author} {\bibfnamefont {H.}~\bibnamefont {Tasaki}},\ }\bibfield  {title} {\bibinfo {title} {Ferromagnetism in the hubbard models with degenerate single-electron ground states},\ }\href@noop {} {\bibfield  {journal} {\bibinfo  {journal} {Physical review letters}\ }\textbf {\bibinfo {volume} {69}},\ \bibinfo {pages} {1608} (\bibinfo {year} {1992})}\BibitemShut {NoStop}%
		\bibitem [{\citenamefont {Kapit}\ and\ \citenamefont {Mueller}(2010)}]{kapit2010exact}%
		\BibitemOpen
		\bibfield  {author} {\bibinfo {author} {\bibfnamefont {E.}~\bibnamefont {Kapit}}\ and\ \bibinfo {author} {\bibfnamefont {E.}~\bibnamefont {Mueller}},\ }\bibfield  {title} {\bibinfo {title} {Exact parent hamiltonian for the quantum hall states in a lattice},\ }\href@noop {} {\bibfield  {journal} {\bibinfo  {journal} {Physical review letters}\ }\textbf {\bibinfo {volume} {105}},\ \bibinfo {pages} {215303} (\bibinfo {year} {2010})}\BibitemShut {NoStop}%
		\bibitem [{\citenamefont {Tang}\ \emph {et~al.}(2011)\citenamefont {Tang}, \citenamefont {Mei},\ and\ \citenamefont {Wen}}]{tang2011high}%
		\BibitemOpen
		\bibfield  {author} {\bibinfo {author} {\bibfnamefont {E.}~\bibnamefont {Tang}}, \bibinfo {author} {\bibfnamefont {J.-W.}\ \bibnamefont {Mei}},\ and\ \bibinfo {author} {\bibfnamefont {X.-G.}\ \bibnamefont {Wen}},\ }\bibfield  {title} {\bibinfo {title} {High-temperature fractional quantum hall states},\ }\href@noop {} {\bibfield  {journal} {\bibinfo  {journal} {Physical review letters}\ }\textbf {\bibinfo {volume} {106}},\ \bibinfo {pages} {236802} (\bibinfo {year} {2011})}\BibitemShut {NoStop}%
		\bibitem [{\citenamefont {Sun}\ \emph {et~al.}(2011)\citenamefont {Sun}, \citenamefont {Gu}, \citenamefont {Katsura},\ and\ \citenamefont {Das~Sarma}}]{sun2011nearly}%
		\BibitemOpen
		\bibfield  {author} {\bibinfo {author} {\bibfnamefont {K.}~\bibnamefont {Sun}}, \bibinfo {author} {\bibfnamefont {Z.}~\bibnamefont {Gu}}, \bibinfo {author} {\bibfnamefont {H.}~\bibnamefont {Katsura}},\ and\ \bibinfo {author} {\bibfnamefont {S.}~\bibnamefont {Das~Sarma}},\ }\bibfield  {title} {\bibinfo {title} {Nearly flatbands with nontrivial topology},\ }\href@noop {} {\bibfield  {journal} {\bibinfo  {journal} {Physical review letters}\ }\textbf {\bibinfo {volume} {106}},\ \bibinfo {pages} {236803} (\bibinfo {year} {2011})}\BibitemShut {NoStop}%
		\bibitem [{\citenamefont {Neupert}\ \emph {et~al.}(2011)\citenamefont {Neupert}, \citenamefont {Santos}, \citenamefont {Chamon},\ and\ \citenamefont {Mudry}}]{neupert2011fractional}%
		\BibitemOpen
		\bibfield  {author} {\bibinfo {author} {\bibfnamefont {T.}~\bibnamefont {Neupert}}, \bibinfo {author} {\bibfnamefont {L.}~\bibnamefont {Santos}}, \bibinfo {author} {\bibfnamefont {C.}~\bibnamefont {Chamon}},\ and\ \bibinfo {author} {\bibfnamefont {C.}~\bibnamefont {Mudry}},\ }\bibfield  {title} {\bibinfo {title} {Fractional quantum hall states at zero magnetic field},\ }\href@noop {} {\bibfield  {journal} {\bibinfo  {journal} {Physical review letters}\ }\textbf {\bibinfo {volume} {106}},\ \bibinfo {pages} {236804} (\bibinfo {year} {2011})}\BibitemShut {NoStop}%
		\bibitem [{\citenamefont {Regnault}\ and\ \citenamefont {Bernevig}(2011)}]{regnault2011fractional}%
		\BibitemOpen
		\bibfield  {author} {\bibinfo {author} {\bibfnamefont {N.}~\bibnamefont {Regnault}}\ and\ \bibinfo {author} {\bibfnamefont {B.~A.}\ \bibnamefont {Bernevig}},\ }\bibfield  {title} {\bibinfo {title} {Fractional chern insulator},\ }\href@noop {} {\bibfield  {journal} {\bibinfo  {journal} {Physical Review X}\ }\textbf {\bibinfo {volume} {1}},\ \bibinfo {pages} {021014} (\bibinfo {year} {2011})}\BibitemShut {NoStop}%
		\bibitem [{\citenamefont {Qi}(2011)}]{qi2011generic}%
		\BibitemOpen
		\bibfield  {author} {\bibinfo {author} {\bibfnamefont {X.-L.}\ \bibnamefont {Qi}},\ }\bibfield  {title} {\bibinfo {title} {Generic wave-function description of fractional quantum anomalous hall states and fractional topological insulators},\ }\href@noop {} {\bibfield  {journal} {\bibinfo  {journal} {Physical review letters}\ }\textbf {\bibinfo {volume} {107}},\ \bibinfo {pages} {126803} (\bibinfo {year} {2011})}\BibitemShut {NoStop}%
		\bibitem [{\citenamefont {Behrmann}\ \emph {et~al.}(2016)\citenamefont {Behrmann}, \citenamefont {Liu},\ and\ \citenamefont {Bergholtz}}]{behrmann2016model}%
		\BibitemOpen
		\bibfield  {author} {\bibinfo {author} {\bibfnamefont {J.}~\bibnamefont {Behrmann}}, \bibinfo {author} {\bibfnamefont {Z.}~\bibnamefont {Liu}},\ and\ \bibinfo {author} {\bibfnamefont {E.~J.}\ \bibnamefont {Bergholtz}},\ }\bibfield  {title} {\bibinfo {title} {Model fractional chern insulators},\ }\href@noop {} {\bibfield  {journal} {\bibinfo  {journal} {Physical Review Letters}\ }\textbf {\bibinfo {volume} {116}},\ \bibinfo {pages} {216802} (\bibinfo {year} {2016})}\BibitemShut {NoStop}%
		\bibitem [{\citenamefont {Tarnopolsky}\ \emph {et~al.}(2019)\citenamefont {Tarnopolsky}, \citenamefont {Kruchkov},\ and\ \citenamefont {Vishwanath}}]{tarnopolsky2019origin}%
		\BibitemOpen
		\bibfield  {author} {\bibinfo {author} {\bibfnamefont {G.}~\bibnamefont {Tarnopolsky}}, \bibinfo {author} {\bibfnamefont {A.~J.}\ \bibnamefont {Kruchkov}},\ and\ \bibinfo {author} {\bibfnamefont {A.}~\bibnamefont {Vishwanath}},\ }\bibfield  {title} {\bibinfo {title} {Origin of magic angles in twisted bilayer graphene},\ }\href@noop {} {\bibfield  {journal} {\bibinfo  {journal} {Physical review letters}\ }\textbf {\bibinfo {volume} {122}},\ \bibinfo {pages} {106405} (\bibinfo {year} {2019})}\BibitemShut {NoStop}%
		\bibitem [{\citenamefont {Ledwith}\ \emph {et~al.}(2020)\citenamefont {Ledwith}, \citenamefont {Tarnopolsky}, \citenamefont {Khalaf},\ and\ \citenamefont {Vishwanath}}]{ledwith2020fractional}%
		\BibitemOpen
		\bibfield  {author} {\bibinfo {author} {\bibfnamefont {P.~J.}\ \bibnamefont {Ledwith}}, \bibinfo {author} {\bibfnamefont {G.}~\bibnamefont {Tarnopolsky}}, \bibinfo {author} {\bibfnamefont {E.}~\bibnamefont {Khalaf}},\ and\ \bibinfo {author} {\bibfnamefont {A.}~\bibnamefont {Vishwanath}},\ }\bibfield  {title} {\bibinfo {title} {Fractional chern insulator states in twisted bilayer graphene: An analytical approach},\ }\href@noop {} {\bibfield  {journal} {\bibinfo  {journal} {Physical Review Research}\ }\textbf {\bibinfo {volume} {2}},\ \bibinfo {pages} {023237} (\bibinfo {year} {2020})}\BibitemShut {NoStop}%
		\bibitem [{\citenamefont {Repellin}\ and\ \citenamefont {Senthil}(2020)}]{repellin2020chern}%
		\BibitemOpen
		\bibfield  {author} {\bibinfo {author} {\bibfnamefont {C.}~\bibnamefont {Repellin}}\ and\ \bibinfo {author} {\bibfnamefont {T.}~\bibnamefont {Senthil}},\ }\bibfield  {title} {\bibinfo {title} {Chern bands of twisted bilayer graphene: Fractional chern insulators and spin phase transition},\ }\href@noop {} {\bibfield  {journal} {\bibinfo  {journal} {Physical Review Research}\ }\textbf {\bibinfo {volume} {2}},\ \bibinfo {pages} {023238} (\bibinfo {year} {2020})}\BibitemShut {NoStop}%
		\bibitem [{\citenamefont {Wang}\ \emph {et~al.}(2021)\citenamefont {Wang}, \citenamefont {Cano}, \citenamefont {Millis}, \citenamefont {Liu},\ and\ \citenamefont {Yang}}]{wang2021exact}%
		\BibitemOpen
		\bibfield  {author} {\bibinfo {author} {\bibfnamefont {J.}~\bibnamefont {Wang}}, \bibinfo {author} {\bibfnamefont {J.}~\bibnamefont {Cano}}, \bibinfo {author} {\bibfnamefont {A.~J.}\ \bibnamefont {Millis}}, \bibinfo {author} {\bibfnamefont {Z.}~\bibnamefont {Liu}},\ and\ \bibinfo {author} {\bibfnamefont {B.}~\bibnamefont {Yang}},\ }\bibfield  {title} {\bibinfo {title} {Exact landau level description of geometry and interaction in a flatband},\ }\href@noop {} {\bibfield  {journal} {\bibinfo  {journal} {Physical review letters}\ }\textbf {\bibinfo {volume} {127}},\ \bibinfo {pages} {246403} (\bibinfo {year} {2021})}\BibitemShut {NoStop}%
		\bibitem [{\citenamefont {Devakul}\ \emph {et~al.}(2021)\citenamefont {Devakul}, \citenamefont {Cr{\'e}pel}, \citenamefont {Zhang},\ and\ \citenamefont {Fu}}]{devakul2021magic}%
		\BibitemOpen
		\bibfield  {author} {\bibinfo {author} {\bibfnamefont {T.}~\bibnamefont {Devakul}}, \bibinfo {author} {\bibfnamefont {V.}~\bibnamefont {Cr{\'e}pel}}, \bibinfo {author} {\bibfnamefont {Y.}~\bibnamefont {Zhang}},\ and\ \bibinfo {author} {\bibfnamefont {L.}~\bibnamefont {Fu}},\ }\bibfield  {title} {\bibinfo {title} {Magic in twisted transition metal dichalcogenide bilayers},\ }\href@noop {} {\bibfield  {journal} {\bibinfo  {journal} {Nature communications}\ }\textbf {\bibinfo {volume} {12}},\ \bibinfo {pages} {6730} (\bibinfo {year} {2021})}\BibitemShut {NoStop}%
		\bibitem [{\citenamefont {Reddy}\ \emph {et~al.}(2023)\citenamefont {Reddy}, \citenamefont {Alsallom}, \citenamefont {Zhang}, \citenamefont {Devakul},\ and\ \citenamefont {Fu}}]{reddy2023fractional}%
		\BibitemOpen
		\bibfield  {author} {\bibinfo {author} {\bibfnamefont {A.~P.}\ \bibnamefont {Reddy}}, \bibinfo {author} {\bibfnamefont {F.}~\bibnamefont {Alsallom}}, \bibinfo {author} {\bibfnamefont {Y.}~\bibnamefont {Zhang}}, \bibinfo {author} {\bibfnamefont {T.}~\bibnamefont {Devakul}},\ and\ \bibinfo {author} {\bibfnamefont {L.}~\bibnamefont {Fu}},\ }\bibfield  {title} {\bibinfo {title} {Fractional quantum anomalous hall states in twisted bilayer mote 2 and wse 2},\ }\href@noop {} {\bibfield  {journal} {\bibinfo  {journal} {Physical Review B}\ }\textbf {\bibinfo {volume} {108}},\ \bibinfo {pages} {085117} (\bibinfo {year} {2023})}\BibitemShut {NoStop}%
		\bibitem [{\citenamefont {Wang}\ \emph {et~al.}(2024)\citenamefont {Wang}, \citenamefont {Zhang}, \citenamefont {Liu}, \citenamefont {He}, \citenamefont {Xu}, \citenamefont {Ran}, \citenamefont {Cao},\ and\ \citenamefont {Xiao}}]{wang2024fractional}%
		\BibitemOpen
		\bibfield  {author} {\bibinfo {author} {\bibfnamefont {C.}~\bibnamefont {Wang}}, \bibinfo {author} {\bibfnamefont {X.-W.}\ \bibnamefont {Zhang}}, \bibinfo {author} {\bibfnamefont {X.}~\bibnamefont {Liu}}, \bibinfo {author} {\bibfnamefont {Y.}~\bibnamefont {He}}, \bibinfo {author} {\bibfnamefont {X.}~\bibnamefont {Xu}}, \bibinfo {author} {\bibfnamefont {Y.}~\bibnamefont {Ran}}, \bibinfo {author} {\bibfnamefont {T.}~\bibnamefont {Cao}},\ and\ \bibinfo {author} {\bibfnamefont {D.}~\bibnamefont {Xiao}},\ }\bibfield  {title} {\bibinfo {title} {Fractional chern insulator in twisted bilayer mote 2},\ }\href@noop {} {\bibfield  {journal} {\bibinfo  {journal} {Physical Review Letters}\ }\textbf {\bibinfo {volume} {132}},\ \bibinfo {pages} {036501} (\bibinfo {year} {2024})}\BibitemShut {NoStop}%
		\bibitem [{\citenamefont {Jia}\ \emph {et~al.}(2024)\citenamefont {Jia}, \citenamefont {Yu}, \citenamefont {Liu}, \citenamefont {Herzog-Arbeitman}, \citenamefont {Qi}, \citenamefont {Pi}, \citenamefont {Regnault}, \citenamefont {Weng}, \citenamefont {Bernevig},\ and\ \citenamefont {Wu}}]{jia2024moire}%
		\BibitemOpen
		\bibfield  {author} {\bibinfo {author} {\bibfnamefont {Y.}~\bibnamefont {Jia}}, \bibinfo {author} {\bibfnamefont {J.}~\bibnamefont {Yu}}, \bibinfo {author} {\bibfnamefont {J.}~\bibnamefont {Liu}}, \bibinfo {author} {\bibfnamefont {J.}~\bibnamefont {Herzog-Arbeitman}}, \bibinfo {author} {\bibfnamefont {Z.}~\bibnamefont {Qi}}, \bibinfo {author} {\bibfnamefont {H.}~\bibnamefont {Pi}}, \bibinfo {author} {\bibfnamefont {N.}~\bibnamefont {Regnault}}, \bibinfo {author} {\bibfnamefont {H.}~\bibnamefont {Weng}}, \bibinfo {author} {\bibfnamefont {B.~A.}\ \bibnamefont {Bernevig}},\ and\ \bibinfo {author} {\bibfnamefont {Q.}~\bibnamefont {Wu}},\ }\bibfield  {title} {\bibinfo {title} {Moir{\'e} fractional chern insulators. i. first-principles calculations and continuum models of twisted bilayer mote 2},\ }\href@noop {} {\bibfield  {journal} {\bibinfo  {journal} {Physical Review B}\ }\textbf {\bibinfo {volume} {109}},\ \bibinfo {pages} {205121} (\bibinfo {year} {2024})}\BibitemShut {NoStop}%
		\bibitem [{\citenamefont {Herzog-Arbeitman}\ \emph {et~al.}(2024)\citenamefont {Herzog-Arbeitman}, \citenamefont {Wang}, \citenamefont {Liu}, \citenamefont {Tam}, \citenamefont {Qi}, \citenamefont {Jia}, \citenamefont {Efetov}, \citenamefont {Vafek}, \citenamefont {Regnault}, \citenamefont {Weng} \emph {et~al.}}]{herzog2024moire}%
		\BibitemOpen
		\bibfield  {author} {\bibinfo {author} {\bibfnamefont {J.}~\bibnamefont {Herzog-Arbeitman}}, \bibinfo {author} {\bibfnamefont {Y.}~\bibnamefont {Wang}}, \bibinfo {author} {\bibfnamefont {J.}~\bibnamefont {Liu}}, \bibinfo {author} {\bibfnamefont {P.~M.}\ \bibnamefont {Tam}}, \bibinfo {author} {\bibfnamefont {Z.}~\bibnamefont {Qi}}, \bibinfo {author} {\bibfnamefont {Y.}~\bibnamefont {Jia}}, \bibinfo {author} {\bibfnamefont {D.~K.}\ \bibnamefont {Efetov}}, \bibinfo {author} {\bibfnamefont {O.}~\bibnamefont {Vafek}}, \bibinfo {author} {\bibfnamefont {N.}~\bibnamefont {Regnault}}, \bibinfo {author} {\bibfnamefont {H.}~\bibnamefont {Weng}}, \emph {et~al.},\ }\bibfield  {title} {\bibinfo {title} {Moir{\'e} fractional chern insulators. ii. first-principles calculations and continuum models of rhombohedral graphene superlattices},\ }\href@noop {} {\bibfield  {journal} {\bibinfo  {journal} {Physical Review B}\ }\textbf {\bibinfo {volume} {109}},\ \bibinfo {pages} {205122} (\bibinfo {year} {2024})}\BibitemShut
		{NoStop}%
		\bibitem [{\citenamefont {Kwan}\ \emph {et~al.}(2025)\citenamefont {Kwan}, \citenamefont {Yu}, \citenamefont {Herzog-Arbeitman}, \citenamefont {Efetov}, \citenamefont {Regnault},\ and\ \citenamefont {Bernevig}}]{kwan2025moire}%
		\BibitemOpen
		\bibfield  {author} {\bibinfo {author} {\bibfnamefont {Y.~H.}\ \bibnamefont {Kwan}}, \bibinfo {author} {\bibfnamefont {J.}~\bibnamefont {Yu}}, \bibinfo {author} {\bibfnamefont {J.}~\bibnamefont {Herzog-Arbeitman}}, \bibinfo {author} {\bibfnamefont {D.~K.}\ \bibnamefont {Efetov}}, \bibinfo {author} {\bibfnamefont {N.}~\bibnamefont {Regnault}},\ and\ \bibinfo {author} {\bibfnamefont {B.~A.}\ \bibnamefont {Bernevig}},\ }\bibfield  {title} {\bibinfo {title} {Moir{\'e} fractional chern insulators. iii. hartree-fock phase diagram, magic angle regime for chern insulator states, role of moir{\'e} potential, and goldstone gaps in rhombohedral graphene superlattices},\ }\href@noop {} {\bibfield  {journal} {\bibinfo  {journal} {Physical Review B}\ }\textbf {\bibinfo {volume} {112}},\ \bibinfo {pages} {075109} (\bibinfo {year} {2025})}\BibitemShut {NoStop}%
		\bibitem [{\citenamefont {Yu}\ \emph {et~al.}(2025)\citenamefont {Yu}, \citenamefont {Herzog-Arbeitman}, \citenamefont {Kwan}, \citenamefont {Regnault},\ and\ \citenamefont {Bernevig}}]{yu2025moire}%
		\BibitemOpen
		\bibfield  {author} {\bibinfo {author} {\bibfnamefont {J.}~\bibnamefont {Yu}}, \bibinfo {author} {\bibfnamefont {J.}~\bibnamefont {Herzog-Arbeitman}}, \bibinfo {author} {\bibfnamefont {Y.~H.}\ \bibnamefont {Kwan}}, \bibinfo {author} {\bibfnamefont {N.}~\bibnamefont {Regnault}},\ and\ \bibinfo {author} {\bibfnamefont {B.~A.}\ \bibnamefont {Bernevig}},\ }\bibfield  {title} {\bibinfo {title} {Moir{\'e} fractional chern insulators. iv. fluctuation-driven collapse in multiband exact diagonalization calculations on rhombohedral graphene},\ }\href@noop {} {\bibfield  {journal} {\bibinfo  {journal} {Physical Review B}\ }\textbf {\bibinfo {volume} {112}},\ \bibinfo {pages} {075110} (\bibinfo {year} {2025})}\BibitemShut {NoStop}%
		\bibitem [{\citenamefont {Liu}\ \emph {et~al.}(2025)\citenamefont {Liu}, \citenamefont {Mera}, \citenamefont {Fujimoto}, \citenamefont {Ozawa},\ and\ \citenamefont {Wang}}]{liu2025theory}%
		\BibitemOpen
		\bibfield  {author} {\bibinfo {author} {\bibfnamefont {Z.}~\bibnamefont {Liu}}, \bibinfo {author} {\bibfnamefont {B.}~\bibnamefont {Mera}}, \bibinfo {author} {\bibfnamefont {M.}~\bibnamefont {Fujimoto}}, \bibinfo {author} {\bibfnamefont {T.}~\bibnamefont {Ozawa}},\ and\ \bibinfo {author} {\bibfnamefont {J.}~\bibnamefont {Wang}},\ }\bibfield  {title} {\bibinfo {title} {Theory of generalized landau levels and its implications for non-abelian states},\ }\href@noop {} {\bibfield  {journal} {\bibinfo  {journal} {Physical Review X}\ }\textbf {\bibinfo {volume} {15}},\ \bibinfo {pages} {031019} (\bibinfo {year} {2025})}\BibitemShut {NoStop}%
		\bibitem [{\citenamefont {Cao}\ \emph {et~al.}(2018)\citenamefont {Cao}, \citenamefont {Fatemi}, \citenamefont {Fang}, \citenamefont {Watanabe}, \citenamefont {Taniguchi}, \citenamefont {Kaxiras},\ and\ \citenamefont {Jarillo-Herrero}}]{cao2018unconventional}%
		\BibitemOpen
		\bibfield  {author} {\bibinfo {author} {\bibfnamefont {Y.}~\bibnamefont {Cao}}, \bibinfo {author} {\bibfnamefont {V.}~\bibnamefont {Fatemi}}, \bibinfo {author} {\bibfnamefont {S.}~\bibnamefont {Fang}}, \bibinfo {author} {\bibfnamefont {K.}~\bibnamefont {Watanabe}}, \bibinfo {author} {\bibfnamefont {T.}~\bibnamefont {Taniguchi}}, \bibinfo {author} {\bibfnamefont {E.}~\bibnamefont {Kaxiras}},\ and\ \bibinfo {author} {\bibfnamefont {P.}~\bibnamefont {Jarillo-Herrero}},\ }\bibfield  {title} {\bibinfo {title} {Unconventional superconductivity in magic-angle graphene superlattices},\ }\href@noop {} {\bibfield  {journal} {\bibinfo  {journal} {Nature}\ }\textbf {\bibinfo {volume} {556}},\ \bibinfo {pages} {43} (\bibinfo {year} {2018})}\BibitemShut {NoStop}%
		\bibitem [{\citenamefont {Yankowitz}\ \emph {et~al.}(2019)\citenamefont {Yankowitz}, \citenamefont {Chen}, \citenamefont {Polshyn}, \citenamefont {Zhang}, \citenamefont {Watanabe}, \citenamefont {Taniguchi}, \citenamefont {Graf}, \citenamefont {Young},\ and\ \citenamefont {Dean}}]{yankowitz2019tuning}%
		\BibitemOpen
		\bibfield  {author} {\bibinfo {author} {\bibfnamefont {M.}~\bibnamefont {Yankowitz}}, \bibinfo {author} {\bibfnamefont {S.}~\bibnamefont {Chen}}, \bibinfo {author} {\bibfnamefont {H.}~\bibnamefont {Polshyn}}, \bibinfo {author} {\bibfnamefont {Y.}~\bibnamefont {Zhang}}, \bibinfo {author} {\bibfnamefont {K.}~\bibnamefont {Watanabe}}, \bibinfo {author} {\bibfnamefont {T.}~\bibnamefont {Taniguchi}}, \bibinfo {author} {\bibfnamefont {D.}~\bibnamefont {Graf}}, \bibinfo {author} {\bibfnamefont {A.~F.}\ \bibnamefont {Young}},\ and\ \bibinfo {author} {\bibfnamefont {C.~R.}\ \bibnamefont {Dean}},\ }\bibfield  {title} {\bibinfo {title} {Tuning superconductivity in twisted bilayer graphene},\ }\href@noop {} {\bibfield  {journal} {\bibinfo  {journal} {Science}\ }\textbf {\bibinfo {volume} {363}},\ \bibinfo {pages} {1059} (\bibinfo {year} {2019})}\BibitemShut {NoStop}%
		\bibitem [{\citenamefont {Balents}\ \emph {et~al.}(2020)\citenamefont {Balents}, \citenamefont {Dean}, \citenamefont {Efetov},\ and\ \citenamefont {Young}}]{balents2020superconductivity}%
		\BibitemOpen
		\bibfield  {author} {\bibinfo {author} {\bibfnamefont {L.}~\bibnamefont {Balents}}, \bibinfo {author} {\bibfnamefont {C.~R.}\ \bibnamefont {Dean}}, \bibinfo {author} {\bibfnamefont {D.~K.}\ \bibnamefont {Efetov}},\ and\ \bibinfo {author} {\bibfnamefont {A.~F.}\ \bibnamefont {Young}},\ }\bibfield  {title} {\bibinfo {title} {Superconductivity and strong correlations in moir{\'e} flat bands},\ }\href@noop {} {\bibfield  {journal} {\bibinfo  {journal} {Nature Physics}\ }\textbf {\bibinfo {volume} {16}},\ \bibinfo {pages} {725} (\bibinfo {year} {2020})}\BibitemShut {NoStop}%
		\bibitem [{\citenamefont {Tang}\ \emph {et~al.}(2020)\citenamefont {Tang}, \citenamefont {Li}, \citenamefont {Li}, \citenamefont {Xu}, \citenamefont {Liu}, \citenamefont {Barmak}, \citenamefont {Watanabe}, \citenamefont {Taniguchi}, \citenamefont {MacDonald}, \citenamefont {Shan} \emph {et~al.}}]{tang2020simulation}%
		\BibitemOpen
		\bibfield  {author} {\bibinfo {author} {\bibfnamefont {Y.}~\bibnamefont {Tang}}, \bibinfo {author} {\bibfnamefont {L.}~\bibnamefont {Li}}, \bibinfo {author} {\bibfnamefont {T.}~\bibnamefont {Li}}, \bibinfo {author} {\bibfnamefont {Y.}~\bibnamefont {Xu}}, \bibinfo {author} {\bibfnamefont {S.}~\bibnamefont {Liu}}, \bibinfo {author} {\bibfnamefont {K.}~\bibnamefont {Barmak}}, \bibinfo {author} {\bibfnamefont {K.}~\bibnamefont {Watanabe}}, \bibinfo {author} {\bibfnamefont {T.}~\bibnamefont {Taniguchi}}, \bibinfo {author} {\bibfnamefont {A.~H.}\ \bibnamefont {MacDonald}}, \bibinfo {author} {\bibfnamefont {J.}~\bibnamefont {Shan}}, \emph {et~al.},\ }\bibfield  {title} {\bibinfo {title} {Simulation of hubbard model physics in wse2/ws2 moir{\'e} superlattices},\ }\href@noop {} {\bibfield  {journal} {\bibinfo  {journal} {Nature}\ }\textbf {\bibinfo {volume} {579}},\ \bibinfo {pages} {353} (\bibinfo {year} {2020})}\BibitemShut {NoStop}%
		\bibitem [{\citenamefont {Regan}\ \emph {et~al.}(2020)\citenamefont {Regan}, \citenamefont {Wang}, \citenamefont {Jin}, \citenamefont {Bakti~Utama}, \citenamefont {Gao}, \citenamefont {Wei}, \citenamefont {Zhao}, \citenamefont {Zhao}, \citenamefont {Zhang}, \citenamefont {Yumigeta} \emph {et~al.}}]{regan2020mott}%
		\BibitemOpen
		\bibfield  {author} {\bibinfo {author} {\bibfnamefont {E.~C.}\ \bibnamefont {Regan}}, \bibinfo {author} {\bibfnamefont {D.}~\bibnamefont {Wang}}, \bibinfo {author} {\bibfnamefont {C.}~\bibnamefont {Jin}}, \bibinfo {author} {\bibfnamefont {M.~I.}\ \bibnamefont {Bakti~Utama}}, \bibinfo {author} {\bibfnamefont {B.}~\bibnamefont {Gao}}, \bibinfo {author} {\bibfnamefont {X.}~\bibnamefont {Wei}}, \bibinfo {author} {\bibfnamefont {S.}~\bibnamefont {Zhao}}, \bibinfo {author} {\bibfnamefont {W.}~\bibnamefont {Zhao}}, \bibinfo {author} {\bibfnamefont {Z.}~\bibnamefont {Zhang}}, \bibinfo {author} {\bibfnamefont {K.}~\bibnamefont {Yumigeta}}, \emph {et~al.},\ }\bibfield  {title} {\bibinfo {title} {Mott and generalized wigner crystal states in wse2/ws2 moir{\'e} superlattices},\ }\href@noop {} {\bibfield  {journal} {\bibinfo  {journal} {Nature}\ }\textbf {\bibinfo {volume} {579}},\ \bibinfo {pages} {359} (\bibinfo {year} {2020})}\BibitemShut {NoStop}%
		\bibitem [{\citenamefont {Li}\ \emph {et~al.}(2024)\citenamefont {Li}, \citenamefont {Xiang}, \citenamefont {Reddy}, \citenamefont {Devakul}, \citenamefont {Sailus}, \citenamefont {Banerjee}, \citenamefont {Taniguchi}, \citenamefont {Watanabe}, \citenamefont {Tongay}, \citenamefont {Zettl} \emph {et~al.}}]{li2024wigner}%
		\BibitemOpen
		\bibfield  {author} {\bibinfo {author} {\bibfnamefont {H.}~\bibnamefont {Li}}, \bibinfo {author} {\bibfnamefont {Z.}~\bibnamefont {Xiang}}, \bibinfo {author} {\bibfnamefont {A.~P.}\ \bibnamefont {Reddy}}, \bibinfo {author} {\bibfnamefont {T.}~\bibnamefont {Devakul}}, \bibinfo {author} {\bibfnamefont {R.}~\bibnamefont {Sailus}}, \bibinfo {author} {\bibfnamefont {R.}~\bibnamefont {Banerjee}}, \bibinfo {author} {\bibfnamefont {T.}~\bibnamefont {Taniguchi}}, \bibinfo {author} {\bibfnamefont {K.}~\bibnamefont {Watanabe}}, \bibinfo {author} {\bibfnamefont {S.}~\bibnamefont {Tongay}}, \bibinfo {author} {\bibfnamefont {A.}~\bibnamefont {Zettl}}, \emph {et~al.},\ }\bibfield  {title} {\bibinfo {title} {Wigner molecular crystals from multielectron moir{\'e} artificial atoms},\ }\href@noop {} {\bibfield  {journal} {\bibinfo  {journal} {Science}\ }\textbf {\bibinfo {volume} {385}},\ \bibinfo {pages} {86} (\bibinfo {year} {2024})}\BibitemShut {NoStop}%
		\bibitem [{\citenamefont {Cai}\ \emph {et~al.}(2023)\citenamefont {Cai}, \citenamefont {Anderson}, \citenamefont {Wang}, \citenamefont {Zhang}, \citenamefont {Liu}, \citenamefont {Holtzmann}, \citenamefont {Zhang}, \citenamefont {Fan}, \citenamefont {Taniguchi}, \citenamefont {Watanabe} \emph {et~al.}}]{cai2023signatures}%
		\BibitemOpen
		\bibfield  {author} {\bibinfo {author} {\bibfnamefont {J.}~\bibnamefont {Cai}}, \bibinfo {author} {\bibfnamefont {E.}~\bibnamefont {Anderson}}, \bibinfo {author} {\bibfnamefont {C.}~\bibnamefont {Wang}}, \bibinfo {author} {\bibfnamefont {X.}~\bibnamefont {Zhang}}, \bibinfo {author} {\bibfnamefont {X.}~\bibnamefont {Liu}}, \bibinfo {author} {\bibfnamefont {W.}~\bibnamefont {Holtzmann}}, \bibinfo {author} {\bibfnamefont {Y.}~\bibnamefont {Zhang}}, \bibinfo {author} {\bibfnamefont {F.}~\bibnamefont {Fan}}, \bibinfo {author} {\bibfnamefont {T.}~\bibnamefont {Taniguchi}}, \bibinfo {author} {\bibfnamefont {K.}~\bibnamefont {Watanabe}}, \emph {et~al.},\ }\bibfield  {title} {\bibinfo {title} {Signatures of fractional quantum anomalous hall states in twisted mote2},\ }\href@noop {} {\bibfield  {journal} {\bibinfo  {journal} {Nature}\ }\textbf {\bibinfo {volume} {622}},\ \bibinfo {pages} {63} (\bibinfo {year} {2023})}\BibitemShut {NoStop}%
		\bibitem [{\citenamefont {Park}\ \emph {et~al.}(2023)\citenamefont {Park}, \citenamefont {Cai}, \citenamefont {Anderson}, \citenamefont {Zhang}, \citenamefont {Zhu}, \citenamefont {Liu}, \citenamefont {Wang}, \citenamefont {Holtzmann}, \citenamefont {Hu}, \citenamefont {Liu} \emph {et~al.}}]{park2023observation}%
		\BibitemOpen
		\bibfield  {author} {\bibinfo {author} {\bibfnamefont {H.}~\bibnamefont {Park}}, \bibinfo {author} {\bibfnamefont {J.}~\bibnamefont {Cai}}, \bibinfo {author} {\bibfnamefont {E.}~\bibnamefont {Anderson}}, \bibinfo {author} {\bibfnamefont {Y.}~\bibnamefont {Zhang}}, \bibinfo {author} {\bibfnamefont {J.}~\bibnamefont {Zhu}}, \bibinfo {author} {\bibfnamefont {X.}~\bibnamefont {Liu}}, \bibinfo {author} {\bibfnamefont {C.}~\bibnamefont {Wang}}, \bibinfo {author} {\bibfnamefont {W.}~\bibnamefont {Holtzmann}}, \bibinfo {author} {\bibfnamefont {C.}~\bibnamefont {Hu}}, \bibinfo {author} {\bibfnamefont {Z.}~\bibnamefont {Liu}}, \emph {et~al.},\ }\bibfield  {title} {\bibinfo {title} {Observation of fractionally quantized anomalous hall effect},\ }\href@noop {} {\bibfield  {journal} {\bibinfo  {journal} {Nature}\ }\textbf {\bibinfo {volume} {622}},\ \bibinfo {pages} {74} (\bibinfo {year} {2023})}\BibitemShut {NoStop}%
		\bibitem [{\citenamefont {Xu}\ \emph {et~al.}(2023)\citenamefont {Xu}, \citenamefont {Sun}, \citenamefont {Jia}, \citenamefont {Liu}, \citenamefont {Xu}, \citenamefont {Li}, \citenamefont {Gu}, \citenamefont {Watanabe}, \citenamefont {Taniguchi}, \citenamefont {Tong} \emph {et~al.}}]{xu2023observation}%
		\BibitemOpen
		\bibfield  {author} {\bibinfo {author} {\bibfnamefont {F.}~\bibnamefont {Xu}}, \bibinfo {author} {\bibfnamefont {Z.}~\bibnamefont {Sun}}, \bibinfo {author} {\bibfnamefont {T.}~\bibnamefont {Jia}}, \bibinfo {author} {\bibfnamefont {C.}~\bibnamefont {Liu}}, \bibinfo {author} {\bibfnamefont {C.}~\bibnamefont {Xu}}, \bibinfo {author} {\bibfnamefont {C.}~\bibnamefont {Li}}, \bibinfo {author} {\bibfnamefont {Y.}~\bibnamefont {Gu}}, \bibinfo {author} {\bibfnamefont {K.}~\bibnamefont {Watanabe}}, \bibinfo {author} {\bibfnamefont {T.}~\bibnamefont {Taniguchi}}, \bibinfo {author} {\bibfnamefont {B.}~\bibnamefont {Tong}}, \emph {et~al.},\ }\bibfield  {title} {\bibinfo {title} {Observation of integer and fractional quantum anomalous hall effects in twisted bilayer mote 2},\ }\href@noop {} {\bibfield  {journal} {\bibinfo  {journal} {Physical Review X}\ }\textbf {\bibinfo {volume} {13}},\ \bibinfo {pages} {031037} (\bibinfo {year} {2023})}\BibitemShut {NoStop}%
		\bibitem [{\citenamefont {Lu}\ \emph {et~al.}(2024)\citenamefont {Lu}, \citenamefont {Han}, \citenamefont {Yao}, \citenamefont {Reddy}, \citenamefont {Yang}, \citenamefont {Seo}, \citenamefont {Watanabe}, \citenamefont {Taniguchi}, \citenamefont {Fu},\ and\ \citenamefont {Ju}}]{lu2024fractional}%
		\BibitemOpen
		\bibfield  {author} {\bibinfo {author} {\bibfnamefont {Z.}~\bibnamefont {Lu}}, \bibinfo {author} {\bibfnamefont {T.}~\bibnamefont {Han}}, \bibinfo {author} {\bibfnamefont {Y.}~\bibnamefont {Yao}}, \bibinfo {author} {\bibfnamefont {A.~P.}\ \bibnamefont {Reddy}}, \bibinfo {author} {\bibfnamefont {J.}~\bibnamefont {Yang}}, \bibinfo {author} {\bibfnamefont {J.}~\bibnamefont {Seo}}, \bibinfo {author} {\bibfnamefont {K.}~\bibnamefont {Watanabe}}, \bibinfo {author} {\bibfnamefont {T.}~\bibnamefont {Taniguchi}}, \bibinfo {author} {\bibfnamefont {L.}~\bibnamefont {Fu}},\ and\ \bibinfo {author} {\bibfnamefont {L.}~\bibnamefont {Ju}},\ }\bibfield  {title} {\bibinfo {title} {Fractional quantum anomalous hall effect in multilayer graphene},\ }\href@noop {} {\bibfield  {journal} {\bibinfo  {journal} {Nature}\ }\textbf {\bibinfo {volume} {626}},\ \bibinfo {pages} {759} (\bibinfo {year} {2024})}\BibitemShut {NoStop}%
		\bibitem [{\citenamefont {Lin}\ \emph {et~al.}(2018)\citenamefont {Lin}, \citenamefont {Choi}, \citenamefont {Zhang}, \citenamefont {Qin}, \citenamefont {Yi}, \citenamefont {Wang}, \citenamefont {Li}, \citenamefont {Wang}, \citenamefont {Zhang}, \citenamefont {Sun} \emph {et~al.}}]{lin2018flatbands}%
		\BibitemOpen
		\bibfield  {author} {\bibinfo {author} {\bibfnamefont {Z.}~\bibnamefont {Lin}}, \bibinfo {author} {\bibfnamefont {J.-H.}\ \bibnamefont {Choi}}, \bibinfo {author} {\bibfnamefont {Q.}~\bibnamefont {Zhang}}, \bibinfo {author} {\bibfnamefont {W.}~\bibnamefont {Qin}}, \bibinfo {author} {\bibfnamefont {S.}~\bibnamefont {Yi}}, \bibinfo {author} {\bibfnamefont {P.}~\bibnamefont {Wang}}, \bibinfo {author} {\bibfnamefont {L.}~\bibnamefont {Li}}, \bibinfo {author} {\bibfnamefont {Y.}~\bibnamefont {Wang}}, \bibinfo {author} {\bibfnamefont {H.}~\bibnamefont {Zhang}}, \bibinfo {author} {\bibfnamefont {Z.}~\bibnamefont {Sun}}, \emph {et~al.},\ }\bibfield  {title} {\bibinfo {title} {Flatbands and emergent ferromagnetic ordering in fe 3 sn 2 kagome lattices},\ }\href@noop {} {\bibfield  {journal} {\bibinfo  {journal} {Physical review letters}\ }\textbf {\bibinfo {volume} {121}},\ \bibinfo {pages} {096401} (\bibinfo {year} {2018})}\BibitemShut {NoStop}%
		\bibitem [{\citenamefont {Hase}\ \emph {et~al.}(2018)\citenamefont {Hase}, \citenamefont {Yanagisawa}, \citenamefont {Aiura},\ and\ \citenamefont {Kawashima}}]{hase2018possibility}%
		\BibitemOpen
		\bibfield  {author} {\bibinfo {author} {\bibfnamefont {I.}~\bibnamefont {Hase}}, \bibinfo {author} {\bibfnamefont {T.}~\bibnamefont {Yanagisawa}}, \bibinfo {author} {\bibfnamefont {Y.}~\bibnamefont {Aiura}},\ and\ \bibinfo {author} {\bibfnamefont {K.}~\bibnamefont {Kawashima}},\ }\bibfield  {title} {\bibinfo {title} {Possibility of flat-band ferromagnetism in hole-doped pyrochlore oxides sn 2 nb 2 o 7 and sn 2 ta 2 o 7},\ }\href@noop {} {\bibfield  {journal} {\bibinfo  {journal} {Physical review letters}\ }\textbf {\bibinfo {volume} {120}},\ \bibinfo {pages} {196401} (\bibinfo {year} {2018})}\BibitemShut {NoStop}%
		\bibitem [{\citenamefont {Arachchige}\ \emph {et~al.}(2022)\citenamefont {Arachchige}, \citenamefont {Meier}, \citenamefont {Marshall}, \citenamefont {Matsuoka}, \citenamefont {Xue}, \citenamefont {McGuire}, \citenamefont {Hermann}, \citenamefont {Cao},\ and\ \citenamefont {Mandrus}}]{arachchige2022charge}%
		\BibitemOpen
		\bibfield  {author} {\bibinfo {author} {\bibfnamefont {H.~W.~S.}\ \bibnamefont {Arachchige}}, \bibinfo {author} {\bibfnamefont {W.~R.}\ \bibnamefont {Meier}}, \bibinfo {author} {\bibfnamefont {M.}~\bibnamefont {Marshall}}, \bibinfo {author} {\bibfnamefont {T.}~\bibnamefont {Matsuoka}}, \bibinfo {author} {\bibfnamefont {R.}~\bibnamefont {Xue}}, \bibinfo {author} {\bibfnamefont {M.~A.}\ \bibnamefont {McGuire}}, \bibinfo {author} {\bibfnamefont {R.~P.}\ \bibnamefont {Hermann}}, \bibinfo {author} {\bibfnamefont {H.}~\bibnamefont {Cao}},\ and\ \bibinfo {author} {\bibfnamefont {D.}~\bibnamefont {Mandrus}},\ }\bibfield  {title} {\bibinfo {title} {Charge density wave in kagome lattice intermetallic scv 6 sn 6},\ }\href@noop {} {\bibfield  {journal} {\bibinfo  {journal} {Physical Review Letters}\ }\textbf {\bibinfo {volume} {129}},\ \bibinfo {pages} {216402} (\bibinfo {year} {2022})}\BibitemShut {NoStop}%
		\bibitem [{\citenamefont {Teng}\ \emph {et~al.}(2023)\citenamefont {Teng}, \citenamefont {Oh}, \citenamefont {Tan}, \citenamefont {Chen}, \citenamefont {Huang}, \citenamefont {Gao}, \citenamefont {Yin}, \citenamefont {Chu}, \citenamefont {Hashimoto}, \citenamefont {Lu} \emph {et~al.}}]{teng2023magnetism}%
		\BibitemOpen
		\bibfield  {author} {\bibinfo {author} {\bibfnamefont {X.}~\bibnamefont {Teng}}, \bibinfo {author} {\bibfnamefont {J.~S.}\ \bibnamefont {Oh}}, \bibinfo {author} {\bibfnamefont {H.}~\bibnamefont {Tan}}, \bibinfo {author} {\bibfnamefont {L.}~\bibnamefont {Chen}}, \bibinfo {author} {\bibfnamefont {J.}~\bibnamefont {Huang}}, \bibinfo {author} {\bibfnamefont {B.}~\bibnamefont {Gao}}, \bibinfo {author} {\bibfnamefont {J.-X.}\ \bibnamefont {Yin}}, \bibinfo {author} {\bibfnamefont {J.-H.}\ \bibnamefont {Chu}}, \bibinfo {author} {\bibfnamefont {M.}~\bibnamefont {Hashimoto}}, \bibinfo {author} {\bibfnamefont {D.}~\bibnamefont {Lu}}, \emph {et~al.},\ }\bibfield  {title} {\bibinfo {title} {Magnetism and charge density wave order in kagome fege},\ }\href@noop {} {\bibfield  {journal} {\bibinfo  {journal} {Nature physics}\ }\textbf {\bibinfo {volume} {19}},\ \bibinfo {pages} {814} (\bibinfo {year} {2023})}\BibitemShut {NoStop}%
		\bibitem [{\citenamefont {Cao}\ \emph {et~al.}(2023)\citenamefont {Cao}, \citenamefont {Xu}, \citenamefont {Fukui}, \citenamefont {Manjo}, \citenamefont {Dong}, \citenamefont {Shi}, \citenamefont {Liu}, \citenamefont {Cao},\ and\ \citenamefont {Song}}]{cao2023competing}%
		\BibitemOpen
		\bibfield  {author} {\bibinfo {author} {\bibfnamefont {S.}~\bibnamefont {Cao}}, \bibinfo {author} {\bibfnamefont {C.}~\bibnamefont {Xu}}, \bibinfo {author} {\bibfnamefont {H.}~\bibnamefont {Fukui}}, \bibinfo {author} {\bibfnamefont {T.}~\bibnamefont {Manjo}}, \bibinfo {author} {\bibfnamefont {Y.}~\bibnamefont {Dong}}, \bibinfo {author} {\bibfnamefont {M.}~\bibnamefont {Shi}}, \bibinfo {author} {\bibfnamefont {Y.}~\bibnamefont {Liu}}, \bibinfo {author} {\bibfnamefont {C.}~\bibnamefont {Cao}},\ and\ \bibinfo {author} {\bibfnamefont {Y.}~\bibnamefont {Song}},\ }\bibfield  {title} {\bibinfo {title} {Competing charge-density wave instabilities in the kagome metal scv6sn6},\ }\href@noop {} {\bibfield  {journal} {\bibinfo  {journal} {Nature Communications}\ }\textbf {\bibinfo {volume} {14}},\ \bibinfo {pages} {7671} (\bibinfo {year} {2023})}\BibitemShut {NoStop}%
		\bibitem [{\citenamefont {Wu}\ \emph {et~al.}(2007)\citenamefont {Wu}, \citenamefont {Bergman}, \citenamefont {Balents},\ and\ \citenamefont {Das~Sarma}}]{wu2007flat}%
		\BibitemOpen
		\bibfield  {author} {\bibinfo {author} {\bibfnamefont {C.}~\bibnamefont {Wu}}, \bibinfo {author} {\bibfnamefont {D.}~\bibnamefont {Bergman}}, \bibinfo {author} {\bibfnamefont {L.}~\bibnamefont {Balents}},\ and\ \bibinfo {author} {\bibfnamefont {S.}~\bibnamefont {Das~Sarma}},\ }\bibfield  {title} {\bibinfo {title} {Flat bands and wigner crystallization in the honeycomb optical lattice},\ }\href@noop {} {\bibfield  {journal} {\bibinfo  {journal} {Physical review letters}\ }\textbf {\bibinfo {volume} {99}},\ \bibinfo {pages} {070401} (\bibinfo {year} {2007})}\BibitemShut {NoStop}%
		\bibitem [{\citenamefont {Bergman}\ \emph {et~al.}(2008)\citenamefont {Bergman}, \citenamefont {Wu},\ and\ \citenamefont {Balents}}]{bergman2008band}%
		\BibitemOpen
		\bibfield  {author} {\bibinfo {author} {\bibfnamefont {D.~L.}\ \bibnamefont {Bergman}}, \bibinfo {author} {\bibfnamefont {C.}~\bibnamefont {Wu}},\ and\ \bibinfo {author} {\bibfnamefont {L.}~\bibnamefont {Balents}},\ }\bibfield  {title} {\bibinfo {title} {Band touching from real-space topology in frustrated hopping models},\ }\href@noop {} {\bibfield  {journal} {\bibinfo  {journal} {Physical Review B}\ }\textbf {\bibinfo {volume} {78}},\ \bibinfo {pages} {125104} (\bibinfo {year} {2008})}\BibitemShut {NoStop}%
		\bibitem [{\citenamefont {Morales-Inostroza}\ and\ \citenamefont {Vicencio}(2016)}]{morales2016simple}%
		\BibitemOpen
		\bibfield  {author} {\bibinfo {author} {\bibfnamefont {L.}~\bibnamefont {Morales-Inostroza}}\ and\ \bibinfo {author} {\bibfnamefont {R.~A.}\ \bibnamefont {Vicencio}},\ }\bibfield  {title} {\bibinfo {title} {Simple method to construct flat-band lattices},\ }\href@noop {} {\bibfield  {journal} {\bibinfo  {journal} {Physical Review A}\ }\textbf {\bibinfo {volume} {94}},\ \bibinfo {pages} {043831} (\bibinfo {year} {2016})}\BibitemShut {NoStop}%
		\bibitem [{\citenamefont {Read}(2017)}]{read2017compactly}%
		\BibitemOpen
		\bibfield  {author} {\bibinfo {author} {\bibfnamefont {N.}~\bibnamefont {Read}},\ }\bibfield  {title} {\bibinfo {title} {Compactly supported wannier functions and algebraic k-theory},\ }\href@noop {} {\bibfield  {journal} {\bibinfo  {journal} {Physical Review B}\ }\textbf {\bibinfo {volume} {95}},\ \bibinfo {pages} {115309} (\bibinfo {year} {2017})}\BibitemShut {NoStop}%
		\bibitem [{\citenamefont {Maimaiti}\ \emph {et~al.}(2017)\citenamefont {Maimaiti}, \citenamefont {Andreanov}, \citenamefont {Park}, \citenamefont {Gendelman},\ and\ \citenamefont {Flach}}]{maimaiti2017compact}%
		\BibitemOpen
		\bibfield  {author} {\bibinfo {author} {\bibfnamefont {W.}~\bibnamefont {Maimaiti}}, \bibinfo {author} {\bibfnamefont {A.}~\bibnamefont {Andreanov}}, \bibinfo {author} {\bibfnamefont {H.~C.}\ \bibnamefont {Park}}, \bibinfo {author} {\bibfnamefont {O.}~\bibnamefont {Gendelman}},\ and\ \bibinfo {author} {\bibfnamefont {S.}~\bibnamefont {Flach}},\ }\bibfield  {title} {\bibinfo {title} {Compact localized states and flat-band generators in one dimension},\ }\href@noop {} {\bibfield  {journal} {\bibinfo  {journal} {Physical Review B}\ }\textbf {\bibinfo {volume} {95}},\ \bibinfo {pages} {115135} (\bibinfo {year} {2017})}\BibitemShut {NoStop}%
		\bibitem [{\citenamefont {R{\"o}ntgen}\ \emph {et~al.}(2018)\citenamefont {R{\"o}ntgen}, \citenamefont {Morfonios},\ and\ \citenamefont {Schmelcher}}]{rontgen2018compact}%
		\BibitemOpen
		\bibfield  {author} {\bibinfo {author} {\bibfnamefont {M.}~\bibnamefont {R{\"o}ntgen}}, \bibinfo {author} {\bibfnamefont {C.}~\bibnamefont {Morfonios}},\ and\ \bibinfo {author} {\bibfnamefont {P.}~\bibnamefont {Schmelcher}},\ }\bibfield  {title} {\bibinfo {title} {Compact localized states and flat bands from local symmetry partitioning},\ }\href@noop {} {\bibfield  {journal} {\bibinfo  {journal} {Physical Review B}\ }\textbf {\bibinfo {volume} {97}},\ \bibinfo {pages} {035161} (\bibinfo {year} {2018})}\BibitemShut {NoStop}%
		\bibitem [{\citenamefont {Rhim}\ and\ \citenamefont {Yang}(2019)}]{rhim2019classification}%
		\BibitemOpen
		\bibfield  {author} {\bibinfo {author} {\bibfnamefont {J.-W.}\ \bibnamefont {Rhim}}\ and\ \bibinfo {author} {\bibfnamefont {B.-J.}\ \bibnamefont {Yang}},\ }\bibfield  {title} {\bibinfo {title} {Classification of flat bands according to the band-crossing singularity of bloch wave functions},\ }\href@noop {} {\bibfield  {journal} {\bibinfo  {journal} {Physical Review B}\ }\textbf {\bibinfo {volume} {99}},\ \bibinfo {pages} {045107} (\bibinfo {year} {2019})}\BibitemShut {NoStop}%
		\bibitem [{\citenamefont {Maimaiti}\ \emph {et~al.}(2019)\citenamefont {Maimaiti}, \citenamefont {Flach},\ and\ \citenamefont {Andreanov}}]{maimaiti2019universal}%
		\BibitemOpen
		\bibfield  {author} {\bibinfo {author} {\bibfnamefont {W.}~\bibnamefont {Maimaiti}}, \bibinfo {author} {\bibfnamefont {S.}~\bibnamefont {Flach}},\ and\ \bibinfo {author} {\bibfnamefont {A.}~\bibnamefont {Andreanov}},\ }\bibfield  {title} {\bibinfo {title} {Universal d= 1 flat band generator from compact localized states},\ }\href@noop {} {\bibfield  {journal} {\bibinfo  {journal} {Physical Review B}\ }\textbf {\bibinfo {volume} {99}},\ \bibinfo {pages} {125129} (\bibinfo {year} {2019})}\BibitemShut {NoStop}%
		\bibitem [{\citenamefont {Maimaiti}\ \emph {et~al.}(2021)\citenamefont {Maimaiti}, \citenamefont {Andreanov},\ and\ \citenamefont {Flach}}]{maimaiti2021flat}%
		\BibitemOpen
		\bibfield  {author} {\bibinfo {author} {\bibfnamefont {W.}~\bibnamefont {Maimaiti}}, \bibinfo {author} {\bibfnamefont {A.}~\bibnamefont {Andreanov}},\ and\ \bibinfo {author} {\bibfnamefont {S.}~\bibnamefont {Flach}},\ }\bibfield  {title} {\bibinfo {title} {Flat-band generator in two dimensions},\ }\href@noop {} {\bibfield  {journal} {\bibinfo  {journal} {Physical Review B}\ }\textbf {\bibinfo {volume} {103}},\ \bibinfo {pages} {165116} (\bibinfo {year} {2021})}\BibitemShut {NoStop}%
		\bibitem [{\citenamefont {Hwang}\ \emph {et~al.}(2021)\citenamefont {Hwang}, \citenamefont {Rhim},\ and\ \citenamefont {Yang}}]{hwang2021general}%
		\BibitemOpen
		\bibfield  {author} {\bibinfo {author} {\bibfnamefont {Y.}~\bibnamefont {Hwang}}, \bibinfo {author} {\bibfnamefont {J.-W.}\ \bibnamefont {Rhim}},\ and\ \bibinfo {author} {\bibfnamefont {B.-J.}\ \bibnamefont {Yang}},\ }\bibfield  {title} {\bibinfo {title} {General construction of flat bands with and without band crossings based on wave function singularity},\ }\href@noop {} {\bibfield  {journal} {\bibinfo  {journal} {Physical Review B}\ }\textbf {\bibinfo {volume} {104}},\ \bibinfo {pages} {085144} (\bibinfo {year} {2021})}\BibitemShut {NoStop}%
		\bibitem [{\citenamefont {Hwang}\ \emph {et~al.}(2021)\citenamefont {Hwang}, \citenamefont {Rhim},\ and\ \citenamefont {Yang}}]{hwang2021flat}%
		\BibitemOpen
		\bibfield  {author} {\bibinfo {author} {\bibfnamefont {Y.}~\bibnamefont {Hwang}}, \bibinfo {author} {\bibfnamefont {J.-W.}\ \bibnamefont {Rhim}},\ and\ \bibinfo {author} {\bibfnamefont {B.-J.}\ \bibnamefont {Yang}},\ }\bibfield  {title} {\bibinfo {title} {Flat bands with band crossings enforced by symmetry representation},\ }\href@noop {} {\bibfield  {journal} {\bibinfo  {journal} {Physical Review B}\ }\textbf {\bibinfo {volume} {104}},\ \bibinfo {pages} {L081104} (\bibinfo {year} {2021})}\BibitemShut {NoStop}%
		\bibitem [{\citenamefont {Graf}\ and\ \citenamefont {Pi{\'e}chon}(2021)}]{graf2021designing}%
		\BibitemOpen
		\bibfield  {author} {\bibinfo {author} {\bibfnamefont {A.}~\bibnamefont {Graf}}\ and\ \bibinfo {author} {\bibfnamefont {F.}~\bibnamefont {Pi{\'e}chon}},\ }\bibfield  {title} {\bibinfo {title} {Designing flat-band tight-binding models with tunable multifold band touching points},\ }\href@noop {} {\bibfield  {journal} {\bibinfo  {journal} {Physical Review B}\ }\textbf {\bibinfo {volume} {104}},\ \bibinfo {pages} {195128} (\bibinfo {year} {2021})}\BibitemShut {NoStop}%
		\bibitem [{\citenamefont {Chen}\ \emph {et~al.}(2023)\citenamefont {Chen}, \citenamefont {Huang}, \citenamefont {Jiang},\ and\ \citenamefont {Hu}}]{chen2023decoding}%
		\BibitemOpen
		\bibfield  {author} {\bibinfo {author} {\bibfnamefont {Y.}~\bibnamefont {Chen}}, \bibinfo {author} {\bibfnamefont {J.}~\bibnamefont {Huang}}, \bibinfo {author} {\bibfnamefont {K.}~\bibnamefont {Jiang}},\ and\ \bibinfo {author} {\bibfnamefont {J.}~\bibnamefont {Hu}},\ }\bibfield  {title} {\bibinfo {title} {Decoding flat bands from compact localized states},\ }\href@noop {} {\bibfield  {journal} {\bibinfo  {journal} {Science Bulletin}\ }\textbf {\bibinfo {volume} {68}},\ \bibinfo {pages} {3165} (\bibinfo {year} {2023})}\BibitemShut {NoStop}%
		\bibitem [{\citenamefont {Ara}\ \emph {et~al.}(2025)\citenamefont {Ara}, \citenamefont {Banerjee}, \citenamefont {Basu},\ and\ \citenamefont {Krishnan}}]{ara2025flat}%
		\BibitemOpen
		\bibfield  {author} {\bibinfo {author} {\bibfnamefont {N.}~\bibnamefont {Ara}}, \bibinfo {author} {\bibfnamefont {A.}~\bibnamefont {Banerjee}}, \bibinfo {author} {\bibfnamefont {R.}~\bibnamefont {Basu}},\ and\ \bibinfo {author} {\bibfnamefont {B.}~\bibnamefont {Krishnan}},\ }\bibfield  {title} {\bibinfo {title} {Flat bands and compact localised states: A carrollian roadmap},\ }\href@noop {} {\bibfield  {journal} {\bibinfo  {journal} {SciPost Physics}\ }\textbf {\bibinfo {volume} {19}},\ \bibinfo {pages} {046} (\bibinfo {year} {2025})}\BibitemShut {NoStop}%
		\bibitem [{\citenamefont {Liu}\ and\ \citenamefont {Liu}(2026)}]{liu2026symmetry}%
		\BibitemOpen
		\bibfield  {author} {\bibinfo {author} {\bibfnamefont {R.-H.}\ \bibnamefont {Liu}}\ and\ \bibinfo {author} {\bibfnamefont {X.}~\bibnamefont {Liu}},\ }\bibfield  {title} {\bibinfo {title} {Symmetry-based real-space framework for realizing flat bands and discovering nodal-line touchings},\ }\href@noop {} {\bibfield  {journal} {\bibinfo  {journal} {Physical Review B}\ }\textbf {\bibinfo {volume} {113}},\ \bibinfo {pages} {035140} (\bibinfo {year} {2026})}\BibitemShut {NoStop}%
		\bibitem [{\citenamefont {Zong}\ \emph {et~al.}(2016)\citenamefont {Zong}, \citenamefont {Xia}, \citenamefont {Tang}, \citenamefont {Song}, \citenamefont {Hu}, \citenamefont {Pei}, \citenamefont {Su}, \citenamefont {Li},\ and\ \citenamefont {Chen}}]{zong2016observation}%
		\BibitemOpen
		\bibfield  {author} {\bibinfo {author} {\bibfnamefont {Y.}~\bibnamefont {Zong}}, \bibinfo {author} {\bibfnamefont {S.}~\bibnamefont {Xia}}, \bibinfo {author} {\bibfnamefont {L.}~\bibnamefont {Tang}}, \bibinfo {author} {\bibfnamefont {D.}~\bibnamefont {Song}}, \bibinfo {author} {\bibfnamefont {Y.}~\bibnamefont {Hu}}, \bibinfo {author} {\bibfnamefont {Y.}~\bibnamefont {Pei}}, \bibinfo {author} {\bibfnamefont {J.}~\bibnamefont {Su}}, \bibinfo {author} {\bibfnamefont {Y.}~\bibnamefont {Li}},\ and\ \bibinfo {author} {\bibfnamefont {Z.}~\bibnamefont {Chen}},\ }\bibfield  {title} {\bibinfo {title} {Observation of localized flat-band states in kagome photonic lattices},\ }\href@noop {} {\bibfield  {journal} {\bibinfo  {journal} {Optics Express}\ }\textbf {\bibinfo {volume} {24}},\ \bibinfo {pages} {8877} (\bibinfo {year} {2016})}\BibitemShut {NoStop}%
		\bibitem [{\citenamefont {Xia}\ \emph {et~al.}(2018)\citenamefont {Xia}, \citenamefont {Ramachandran}, \citenamefont {Xia}, \citenamefont {Li}, \citenamefont {Liu}, \citenamefont {Tang}, \citenamefont {Hu}, \citenamefont {Song}, \citenamefont {Xu}, \citenamefont {Leykam} \emph {et~al.}}]{xia2018unconventional}%
		\BibitemOpen
		\bibfield  {author} {\bibinfo {author} {\bibfnamefont {S.}~\bibnamefont {Xia}}, \bibinfo {author} {\bibfnamefont {A.}~\bibnamefont {Ramachandran}}, \bibinfo {author} {\bibfnamefont {S.}~\bibnamefont {Xia}}, \bibinfo {author} {\bibfnamefont {D.}~\bibnamefont {Li}}, \bibinfo {author} {\bibfnamefont {X.}~\bibnamefont {Liu}}, \bibinfo {author} {\bibfnamefont {L.}~\bibnamefont {Tang}}, \bibinfo {author} {\bibfnamefont {Y.}~\bibnamefont {Hu}}, \bibinfo {author} {\bibfnamefont {D.}~\bibnamefont {Song}}, \bibinfo {author} {\bibfnamefont {J.}~\bibnamefont {Xu}}, \bibinfo {author} {\bibfnamefont {D.}~\bibnamefont {Leykam}}, \emph {et~al.},\ }\bibfield  {title} {\bibinfo {title} {Unconventional flatband line states in photonic lieb lattices},\ }\href@noop {} {\bibfield  {journal} {\bibinfo  {journal} {Physical review letters}\ }\textbf {\bibinfo {volume} {121}},\ \bibinfo {pages} {263902} (\bibinfo {year} {2018})}\BibitemShut {NoStop}%
		\bibitem [{\citenamefont {Rhim}\ \emph {et~al.}(2020)\citenamefont {Rhim}, \citenamefont {Kim},\ and\ \citenamefont {Yang}}]{rhim2020quantum}%
		\BibitemOpen
		\bibfield  {author} {\bibinfo {author} {\bibfnamefont {J.-W.}\ \bibnamefont {Rhim}}, \bibinfo {author} {\bibfnamefont {K.}~\bibnamefont {Kim}},\ and\ \bibinfo {author} {\bibfnamefont {B.-J.}\ \bibnamefont {Yang}},\ }\bibfield  {title} {\bibinfo {title} {Quantum distance and anomalous landau levels of flat bands},\ }\href@noop {} {\bibfield  {journal} {\bibinfo  {journal} {Nature}\ }\textbf {\bibinfo {volume} {584}},\ \bibinfo {pages} {59} (\bibinfo {year} {2020})}\BibitemShut {NoStop}%
		\bibitem [{\citenamefont {Rhim}\ and\ \citenamefont {Yang}(2021)}]{rhim2021singular}%
		\BibitemOpen
		\bibfield  {author} {\bibinfo {author} {\bibfnamefont {J.-W.}\ \bibnamefont {Rhim}}\ and\ \bibinfo {author} {\bibfnamefont {B.-J.}\ \bibnamefont {Yang}},\ }\bibfield  {title} {\bibinfo {title} {Singular flat bands},\ }\href@noop {} {\bibfield  {journal} {\bibinfo  {journal} {Advances in Physics: X}\ }\textbf {\bibinfo {volume} {6}},\ \bibinfo {pages} {1901606} (\bibinfo {year} {2021})}\BibitemShut {NoStop}%
		\bibitem [{\citenamefont {Ma}\ \emph {et~al.}(2020)\citenamefont {Ma}, \citenamefont {Xu}, \citenamefont {Chiu}, \citenamefont {Regnault}, \citenamefont {Houck}, \citenamefont {Song},\ and\ \citenamefont {Bernevig}}]{ma2020spinorbit}%
		\BibitemOpen
		\bibfield  {author} {\bibinfo {author} {\bibfnamefont {D.-S.}\ \bibnamefont {Ma}}, \bibinfo {author} {\bibfnamefont {Y.}~\bibnamefont {Xu}}, \bibinfo {author} {\bibfnamefont {C.~S.}\ \bibnamefont {Chiu}}, \bibinfo {author} {\bibfnamefont {N.}~\bibnamefont {Regnault}}, \bibinfo {author} {\bibfnamefont {A.~A.}\ \bibnamefont {Houck}}, \bibinfo {author} {\bibfnamefont {Z.-D.}\ \bibnamefont {Song}},\ and\ \bibinfo {author} {\bibfnamefont {B.~A.}\ \bibnamefont {Bernevig}},\ }\bibfield  {title} {\bibinfo {title} {Spin-orbit-induced topological flat bands in line and split graphs of bipartite lattices},\ }\href@noop {} {\bibfield  {journal} {\bibinfo  {journal} {Physical Review Letters}\ }\textbf {\bibinfo {volume} {125}},\ \bibinfo {pages} {266403} (\bibinfo {year} {2020})}\BibitemShut {NoStop}%
		\bibitem [{\citenamefont {Sun}\ \emph {et~al.}(2009)\citenamefont {Sun}, \citenamefont {Yao}, \citenamefont {Fradkin},\ and\ \citenamefont {Kivelson}}]{sun2009topological}%
		\BibitemOpen
		\bibfield  {author} {\bibinfo {author} {\bibfnamefont {K.}~\bibnamefont {Sun}}, \bibinfo {author} {\bibfnamefont {H.}~\bibnamefont {Yao}}, \bibinfo {author} {\bibfnamefont {E.}~\bibnamefont {Fradkin}},\ and\ \bibinfo {author} {\bibfnamefont {S.~A.}\ \bibnamefont {Kivelson}},\ }\bibfield  {title} {\bibinfo {title} {Topological insulators and nematic phases from spontaneous symmetry breaking in 2d fermi systems with a quadratic band crossing},\ }\href@noop {} {\bibfield  {journal} {\bibinfo  {journal} {Physical review letters}\ }\textbf {\bibinfo {volume} {103}},\ \bibinfo {pages} {046811} (\bibinfo {year} {2009})}\BibitemShut {NoStop}%
		\bibitem [{\citenamefont {Wen}\ \emph {et~al.}(2010)\citenamefont {Wen}, \citenamefont {R{\"u}egg}, \citenamefont {Wang},\ and\ \citenamefont {Fiete}}]{wen2010interaction}%
		\BibitemOpen
		\bibfield  {author} {\bibinfo {author} {\bibfnamefont {J.}~\bibnamefont {Wen}}, \bibinfo {author} {\bibfnamefont {A.}~\bibnamefont {R{\"u}egg}}, \bibinfo {author} {\bibfnamefont {C.-C.~J.}\ \bibnamefont {Wang}},\ and\ \bibinfo {author} {\bibfnamefont {G.~A.}\ \bibnamefont {Fiete}},\ }\bibfield  {title} {\bibinfo {title} {Interaction-driven topological insulators on the kagome and the decorated honeycomb lattices},\ }\href@noop {} {\bibfield  {journal} {\bibinfo  {journal} {Physical Review B}\ }\textbf {\bibinfo {volume} {82}},\ \bibinfo {pages} {075125} (\bibinfo {year} {2010})}\BibitemShut {NoStop}%
		\bibitem [{\citenamefont {Tsai}\ \emph {et~al.}(2015)\citenamefont {Tsai}, \citenamefont {Fang}, \citenamefont {Yao},\ and\ \citenamefont {Hu}}]{tsai2015interaction}%
		\BibitemOpen
		\bibfield  {author} {\bibinfo {author} {\bibfnamefont {W.-F.}\ \bibnamefont {Tsai}}, \bibinfo {author} {\bibfnamefont {C.}~\bibnamefont {Fang}}, \bibinfo {author} {\bibfnamefont {H.}~\bibnamefont {Yao}},\ and\ \bibinfo {author} {\bibfnamefont {J.}~\bibnamefont {Hu}},\ }\bibfield  {title} {\bibinfo {title} {Interaction-driven topological and nematic phases on the lieb lattice},\ }\href@noop {} {\bibfield  {journal} {\bibinfo  {journal} {New Journal of Physics}\ }\textbf {\bibinfo {volume} {17}},\ \bibinfo {pages} {055016} (\bibinfo {year} {2015})}\BibitemShut {NoStop}%
		\bibitem [{\citenamefont {Zhu}\ \emph {et~al.}(2016)\citenamefont {Zhu}, \citenamefont {Gong}, \citenamefont {Zeng}, \citenamefont {Fu},\ and\ \citenamefont {Sheng}}]{zhu2016interaction}%
		\BibitemOpen
		\bibfield  {author} {\bibinfo {author} {\bibfnamefont {W.}~\bibnamefont {Zhu}}, \bibinfo {author} {\bibfnamefont {S.-S.}\ \bibnamefont {Gong}}, \bibinfo {author} {\bibfnamefont {T.-S.}\ \bibnamefont {Zeng}}, \bibinfo {author} {\bibfnamefont {L.}~\bibnamefont {Fu}},\ and\ \bibinfo {author} {\bibfnamefont {D.}~\bibnamefont {Sheng}},\ }\bibfield  {title} {\bibinfo {title} {Interaction-driven spontaneous quantum hall effect on a kagome lattice},\ }\href@noop {} {\bibfield  {journal} {\bibinfo  {journal} {Physical review letters}\ }\textbf {\bibinfo {volume} {117}},\ \bibinfo {pages} {096402} (\bibinfo {year} {2016})}\BibitemShut {NoStop}%
		\bibitem [{\citenamefont {Zeng}\ \emph {et~al.}(2018)\citenamefont {Zeng}, \citenamefont {Zhu},\ and\ \citenamefont {Sheng}}]{zeng2018tuning}%
		\BibitemOpen
		\bibfield  {author} {\bibinfo {author} {\bibfnamefont {T.-S.}\ \bibnamefont {Zeng}}, \bibinfo {author} {\bibfnamefont {W.}~\bibnamefont {Zhu}},\ and\ \bibinfo {author} {\bibfnamefont {D.}~\bibnamefont {Sheng}},\ }\bibfield  {title} {\bibinfo {title} {Tuning topological phase and quantum anomalous hall effect by interaction in quadratic band touching systems},\ }\href@noop {} {\bibfield  {journal} {\bibinfo  {journal} {npj Quantum Materials}\ }\textbf {\bibinfo {volume} {3}},\ \bibinfo {pages} {49} (\bibinfo {year} {2018})}\BibitemShut {NoStop}%
		\bibitem [{\citenamefont {Chiu}\ \emph {et~al.}(2020)\citenamefont {Chiu}, \citenamefont {Ma}, \citenamefont {Song}, \citenamefont {Bernevig},\ and\ \citenamefont {Houck}}]{chiu2020fragile}%
		\BibitemOpen
		\bibfield  {author} {\bibinfo {author} {\bibfnamefont {C.~S.}\ \bibnamefont {Chiu}}, \bibinfo {author} {\bibfnamefont {D.-S.}\ \bibnamefont {Ma}}, \bibinfo {author} {\bibfnamefont {Z.-D.}\ \bibnamefont {Song}}, \bibinfo {author} {\bibfnamefont {B.~A.}\ \bibnamefont {Bernevig}},\ and\ \bibinfo {author} {\bibfnamefont {A.~A.}\ \bibnamefont {Houck}},\ }\bibfield  {title} {\bibinfo {title} {Fragile topology in line-graph lattices with two, three, or four gapped flat bands},\ }\href@noop {} {\bibfield  {journal} {\bibinfo  {journal} {Physical Review Research}\ }\textbf {\bibinfo {volume} {2}},\ \bibinfo {pages} {043414} (\bibinfo {year} {2020})}\BibitemShut {NoStop}%
		\bibitem [{\citenamefont {C{\u a}lug{\u a}ru}\ \emph {et~al.}(2022)\citenamefont {C{\u a}lug{\u a}ru}, \citenamefont {Chew}, \citenamefont {Elcoro}, \citenamefont {Xu}, \citenamefont {Regnault}, \citenamefont {Song},\ and\ \citenamefont {Bernevig}}]{calugaru2022general}%
		\BibitemOpen
		\bibfield  {author} {\bibinfo {author} {\bibfnamefont {D.}~\bibnamefont {C{\u a}lug{\u a}ru}}, \bibinfo {author} {\bibfnamefont {A.}~\bibnamefont {Chew}}, \bibinfo {author} {\bibfnamefont {L.}~\bibnamefont {Elcoro}}, \bibinfo {author} {\bibfnamefont {Y.}~\bibnamefont {Xu}}, \bibinfo {author} {\bibfnamefont {N.}~\bibnamefont {Regnault}}, \bibinfo {author} {\bibfnamefont {Z.-D.}\ \bibnamefont {Song}},\ and\ \bibinfo {author} {\bibfnamefont {B.~A.}\ \bibnamefont {Bernevig}},\ }\bibfield  {title} {\bibinfo {title} {General construction and topological classification of crystalline flat bands},\ }\href@noop {} {\bibfield  {journal} {\bibinfo  {journal} {Nature Physics}\ }\textbf {\bibinfo {volume} {18}},\ \bibinfo {pages} {185} (\bibinfo {year} {2022})}\BibitemShut {NoStop}%
		\bibitem [{\citenamefont {Chen}\ \emph {et~al.}(2014)\citenamefont {Chen}, \citenamefont {Mazaheri}, \citenamefont {Seidel},\ and\ \citenamefont {Tang}}]{chen2014impossibility}%
		\BibitemOpen
		\bibfield  {author} {\bibinfo {author} {\bibfnamefont {L.}~\bibnamefont {Chen}}, \bibinfo {author} {\bibfnamefont {T.}~\bibnamefont {Mazaheri}}, \bibinfo {author} {\bibfnamefont {A.}~\bibnamefont {Seidel}},\ and\ \bibinfo {author} {\bibfnamefont {X.}~\bibnamefont {Tang}},\ }\bibfield  {title} {\bibinfo {title} {The impossibility of exactly flat non-trivial chern bands in strictly local periodic tight binding models},\ }\href@noop {} {\bibfield  {journal} {\bibinfo  {journal} {Journal of Physics A: Mathematical and Theoretical}\ }\textbf {\bibinfo {volume} {47}},\ \bibinfo {pages} {152001} (\bibinfo {year} {2014})}\BibitemShut {NoStop}%
		\bibitem [{\citenamefont {Dubail}\ and\ \citenamefont {Read}(2015)}]{dubail2015tensor}%
		\BibitemOpen
		\bibfield  {author} {\bibinfo {author} {\bibfnamefont {J.}~\bibnamefont {Dubail}}\ and\ \bibinfo {author} {\bibfnamefont {N.}~\bibnamefont {Read}},\ }\bibfield  {title} {\bibinfo {title} {Tensor network trial states for chiral topological phases in two dimensions and a no-go theorem in any dimension},\ }\href@noop {} {\bibfield  {journal} {\bibinfo  {journal} {Physical Review B}\ }\textbf {\bibinfo {volume} {92}},\ \bibinfo {pages} {205307} (\bibinfo {year} {2015})}\BibitemShut {NoStop}%
		\bibitem [{\citenamefont {Thouless}\ \emph {et~al.}(1982)\citenamefont {Thouless}, \citenamefont {Kohmoto}, \citenamefont {Nightingale},\ and\ \citenamefont {den Nijs}}]{thouless1982quantized}%
		\BibitemOpen
		\bibfield  {author} {\bibinfo {author} {\bibfnamefont {D.~J.}\ \bibnamefont {Thouless}}, \bibinfo {author} {\bibfnamefont {M.}~\bibnamefont {Kohmoto}}, \bibinfo {author} {\bibfnamefont {M.~P.}\ \bibnamefont {Nightingale}},\ and\ \bibinfo {author} {\bibfnamefont {M.}~\bibnamefont {den Nijs}},\ }\bibfield  {title} {\bibinfo {title} {Quantized hall conductance in a two-dimensional periodic potential},\ }\href@noop {} {\bibfield  {journal} {\bibinfo  {journal} {Physical review letters}\ }\textbf {\bibinfo {volume} {49}},\ \bibinfo {pages} {405} (\bibinfo {year} {1982})}\BibitemShut {NoStop}%
		\bibitem [{\citenamefont {Haldane}(1988)}]{haldane1988model}%
		\BibitemOpen
		\bibfield  {author} {\bibinfo {author} {\bibfnamefont {F.~D.~M.}\ \bibnamefont {Haldane}},\ }\bibfield  {title} {\bibinfo {title} {Model for a quantum hall effect without landau levels: Condensed-matter realization of the ``parity anomaly"},\ }\href@noop {} {\bibfield  {journal} {\bibinfo  {journal} {Physical review letters}\ }\textbf {\bibinfo {volume} {61}},\ \bibinfo {pages} {2015} (\bibinfo {year} {1988})}\BibitemShut {NoStop}%
		\bibitem [{\citenamefont {Kane}\ and\ \citenamefont {Mele}(2005{\natexlab{a}})}]{kane2005quantum}%
		\BibitemOpen
		\bibfield  {author} {\bibinfo {author} {\bibfnamefont {C.~L.}\ \bibnamefont {Kane}}\ and\ \bibinfo {author} {\bibfnamefont {E.~J.}\ \bibnamefont {Mele}},\ }\bibfield  {title} {\bibinfo {title} {Quantum spin hall effect in graphene},\ }\href@noop {} {\bibfield  {journal} {\bibinfo  {journal} {Physical review letters}\ }\textbf {\bibinfo {volume} {95}},\ \bibinfo {pages} {226801} (\bibinfo {year} {2005}{\natexlab{a}})}\BibitemShut {NoStop}%
		\bibitem [{\citenamefont {Kane}\ and\ \citenamefont {Mele}(2005{\natexlab{b}})}]{kane2005z}%
		\BibitemOpen
		\bibfield  {author} {\bibinfo {author} {\bibfnamefont {C.~L.}\ \bibnamefont {Kane}}\ and\ \bibinfo {author} {\bibfnamefont {E.~J.}\ \bibnamefont {Mele}},\ }\bibfield  {title} {\bibinfo {title} {Z 2 topological order and the quantum spin hall effect},\ }\href@noop {} {\bibfield  {journal} {\bibinfo  {journal} {Physical review letters}\ }\textbf {\bibinfo {volume} {95}},\ \bibinfo {pages} {146802} (\bibinfo {year} {2005}{\natexlab{b}})}\BibitemShut {NoStop}%
		\bibitem [{\citenamefont {Bernevig}\ \emph {et~al.}(2006)\citenamefont {Bernevig}, \citenamefont {Hughes},\ and\ \citenamefont {Zhang}}]{bernevig2006quantum}%
		\BibitemOpen
		\bibfield  {author} {\bibinfo {author} {\bibfnamefont {B.~A.}\ \bibnamefont {Bernevig}}, \bibinfo {author} {\bibfnamefont {T.~L.}\ \bibnamefont {Hughes}},\ and\ \bibinfo {author} {\bibfnamefont {S.-C.}\ \bibnamefont {Zhang}},\ }\bibfield  {title} {\bibinfo {title} {Quantum spin hall effect and topological phase transition in hgte quantum wells},\ }\href@noop {} {\bibfield  {journal} {\bibinfo  {journal} {science}\ }\textbf {\bibinfo {volume} {314}},\ \bibinfo {pages} {1757} (\bibinfo {year} {2006})}\BibitemShut {NoStop}%
		\bibitem [{\citenamefont {Fu}\ \emph {et~al.}(2007)\citenamefont {Fu}, \citenamefont {Kane},\ and\ \citenamefont {Mele}}]{fu2007topological}%
		\BibitemOpen
		\bibfield  {author} {\bibinfo {author} {\bibfnamefont {L.}~\bibnamefont {Fu}}, \bibinfo {author} {\bibfnamefont {C.~L.}\ \bibnamefont {Kane}},\ and\ \bibinfo {author} {\bibfnamefont {E.~J.}\ \bibnamefont {Mele}},\ }\bibfield  {title} {\bibinfo {title} {Topological insulators in three dimensions},\ }\href@noop {} {\bibfield  {journal} {\bibinfo  {journal} {Physical review letters}\ }\textbf {\bibinfo {volume} {98}},\ \bibinfo {pages} {106803} (\bibinfo {year} {2007})}\BibitemShut {NoStop}%
		\bibitem [{\citenamefont {Moore}\ and\ \citenamefont {Balents}(2007)}]{moore2007topological}%
		\BibitemOpen
		\bibfield  {author} {\bibinfo {author} {\bibfnamefont {J.~E.}\ \bibnamefont {Moore}}\ and\ \bibinfo {author} {\bibfnamefont {L.}~\bibnamefont {Balents}},\ }\bibfield  {title} {\bibinfo {title} {Topological invariants of time-reversal-invariant band structures},\ }\href@noop {} {\bibfield  {journal} {\bibinfo  {journal} {Physical Review B}\ }\textbf {\bibinfo {volume} {75}},\ \bibinfo {pages} {121306} (\bibinfo {year} {2007})}\BibitemShut {NoStop}%
		\bibitem [{\citenamefont {Roy}(2009)}]{roy2009topological}%
		\BibitemOpen
		\bibfield  {author} {\bibinfo {author} {\bibfnamefont {R.}~\bibnamefont {Roy}},\ }\bibfield  {title} {\bibinfo {title} {Topological phases and the quantum spin hall effect in three dimensions},\ }\href@noop {} {\bibfield  {journal} {\bibinfo  {journal} {Physical Review B}\ }\textbf {\bibinfo {volume} {79}},\ \bibinfo {pages} {195322} (\bibinfo {year} {2009})}\BibitemShut {NoStop}%
		\bibitem [{\citenamefont {Isobe}\ and\ \citenamefont {Fu}(2015)}]{isobe2015theory}%
		\BibitemOpen
		\bibfield  {author} {\bibinfo {author} {\bibfnamefont {H.}~\bibnamefont {Isobe}}\ and\ \bibinfo {author} {\bibfnamefont {L.}~\bibnamefont {Fu}},\ }\bibfield  {title} {\bibinfo {title} {Theory of interacting topological crystalline insulators},\ }\href@noop {} {\bibfield  {journal} {\bibinfo  {journal} {Physical Review B}\ }\textbf {\bibinfo {volume} {92}},\ \bibinfo {pages} {081304} (\bibinfo {year} {2015})}\BibitemShut {NoStop}%
		\bibitem [{\citenamefont {Fulga}\ \emph {et~al.}(2016)\citenamefont {Fulga}, \citenamefont {Avraham}, \citenamefont {Beidenkopf},\ and\ \citenamefont {Stern}}]{fulga2016coupled}%
		\BibitemOpen
		\bibfield  {author} {\bibinfo {author} {\bibfnamefont {I.}~\bibnamefont {Fulga}}, \bibinfo {author} {\bibfnamefont {N.}~\bibnamefont {Avraham}}, \bibinfo {author} {\bibfnamefont {H.}~\bibnamefont {Beidenkopf}},\ and\ \bibinfo {author} {\bibfnamefont {A.}~\bibnamefont {Stern}},\ }\bibfield  {title} {\bibinfo {title} {Coupled-layer description of topological crystalline insulators},\ }\href@noop {} {\bibfield  {journal} {\bibinfo  {journal} {Physical Review B}\ }\textbf {\bibinfo {volume} {94}},\ \bibinfo {pages} {125405} (\bibinfo {year} {2016})}\BibitemShut {NoStop}%
		\bibitem [{\citenamefont {Ezawa}(2016)}]{ezawa2016hourglass}%
		\BibitemOpen
		\bibfield  {author} {\bibinfo {author} {\bibfnamefont {M.}~\bibnamefont {Ezawa}},\ }\bibfield  {title} {\bibinfo {title} {Hourglass fermion surface states in stacked topological insulators with nonsymmorphic symmetry},\ }\href@noop {} {\bibfield  {journal} {\bibinfo  {journal} {Physical Review B}\ }\textbf {\bibinfo {volume} {94}},\ \bibinfo {pages} {155148} (\bibinfo {year} {2016})}\BibitemShut {NoStop}%
		\bibitem [{\citenamefont {Song}\ \emph {et~al.}(2017{\natexlab{a}})\citenamefont {Song}, \citenamefont {Huang}, \citenamefont {Fu},\ and\ \citenamefont {Hermele}}]{song2017topological}%
		\BibitemOpen
		\bibfield  {author} {\bibinfo {author} {\bibfnamefont {H.}~\bibnamefont {Song}}, \bibinfo {author} {\bibfnamefont {S.-J.}\ \bibnamefont {Huang}}, \bibinfo {author} {\bibfnamefont {L.}~\bibnamefont {Fu}},\ and\ \bibinfo {author} {\bibfnamefont {M.}~\bibnamefont {Hermele}},\ }\bibfield  {title} {\bibinfo {title} {Topological phases protected by point group symmetry},\ }\href@noop {} {\bibfield  {journal} {\bibinfo  {journal} {Physical Review X}\ }\textbf {\bibinfo {volume} {7}},\ \bibinfo {pages} {011020} (\bibinfo {year} {2017}{\natexlab{a}})}\BibitemShut {NoStop}%
		\bibitem [{\citenamefont {Huang}\ \emph {et~al.}(2017)\citenamefont {Huang}, \citenamefont {Song}, \citenamefont {Huang},\ and\ \citenamefont {Hermele}}]{huang2017building}%
		\BibitemOpen
		\bibfield  {author} {\bibinfo {author} {\bibfnamefont {S.-J.}\ \bibnamefont {Huang}}, \bibinfo {author} {\bibfnamefont {H.}~\bibnamefont {Song}}, \bibinfo {author} {\bibfnamefont {Y.-P.}\ \bibnamefont {Huang}},\ and\ \bibinfo {author} {\bibfnamefont {M.}~\bibnamefont {Hermele}},\ }\bibfield  {title} {\bibinfo {title} {Building crystalline topological phases from lower-dimensional states},\ }\href@noop {} {\bibfield  {journal} {\bibinfo  {journal} {Physical Review B}\ }\textbf {\bibinfo {volume} {96}},\ \bibinfo {pages} {205106} (\bibinfo {year} {2017})}\BibitemShut {NoStop}%
		\bibitem [{\citenamefont {Song}\ \emph {et~al.}(2018)\citenamefont {Song}, \citenamefont {Zhang}, \citenamefont {Fang},\ and\ \citenamefont {Fang}}]{song2018quantitative}%
		\BibitemOpen
		\bibfield  {author} {\bibinfo {author} {\bibfnamefont {Z.}~\bibnamefont {Song}}, \bibinfo {author} {\bibfnamefont {T.}~\bibnamefont {Zhang}}, \bibinfo {author} {\bibfnamefont {Z.}~\bibnamefont {Fang}},\ and\ \bibinfo {author} {\bibfnamefont {C.}~\bibnamefont {Fang}},\ }\bibfield  {title} {\bibinfo {title} {Quantitative mappings between symmetry and topology in solids},\ }\href@noop {} {\bibfield  {journal} {\bibinfo  {journal} {Nature communications}\ }\textbf {\bibinfo {volume} {9}},\ \bibinfo {pages} {3530} (\bibinfo {year} {2018})}\BibitemShut {NoStop}%
		\bibitem [{\citenamefont {Song}\ \emph {et~al.}(2019)\citenamefont {Song}, \citenamefont {Huang}, \citenamefont {Qi}, \citenamefont {Fang},\ and\ \citenamefont {Hermele}}]{song2019topological}%
		\BibitemOpen
		\bibfield  {author} {\bibinfo {author} {\bibfnamefont {Z.}~\bibnamefont {Song}}, \bibinfo {author} {\bibfnamefont {S.-J.}\ \bibnamefont {Huang}}, \bibinfo {author} {\bibfnamefont {Y.}~\bibnamefont {Qi}}, \bibinfo {author} {\bibfnamefont {C.}~\bibnamefont {Fang}},\ and\ \bibinfo {author} {\bibfnamefont {M.}~\bibnamefont {Hermele}},\ }\bibfield  {title} {\bibinfo {title} {Topological states from topological crystals},\ }\href@noop {} {\bibfield  {journal} {\bibinfo  {journal} {Science advances}\ }\textbf {\bibinfo {volume} {5}},\ \bibinfo {pages} {eaax2007} (\bibinfo {year} {2019})}\BibitemShut {NoStop}%
		\bibitem [{\citenamefont {Song}\ \emph {et~al.}(2020)\citenamefont {Song}, \citenamefont {Fang},\ and\ \citenamefont {Qi}}]{song2020real}%
		\BibitemOpen
		\bibfield  {author} {\bibinfo {author} {\bibfnamefont {Z.}~\bibnamefont {Song}}, \bibinfo {author} {\bibfnamefont {C.}~\bibnamefont {Fang}},\ and\ \bibinfo {author} {\bibfnamefont {Y.}~\bibnamefont {Qi}},\ }\bibfield  {title} {\bibinfo {title} {Real-space recipes for general topological crystalline states},\ }\href@noop {} {\bibfield  {journal} {\bibinfo  {journal} {Nature communications}\ }\textbf {\bibinfo {volume} {11}},\ \bibinfo {pages} {4197} (\bibinfo {year} {2020})}\BibitemShut {NoStop}%
		\bibitem [{\citenamefont {Wahl}\ \emph {et~al.}(2013)\citenamefont {Wahl}, \citenamefont {Tu}, \citenamefont {Schuch},\ and\ \citenamefont {Cirac}}]{wahl2013projected}%
		\BibitemOpen
		\bibfield  {author} {\bibinfo {author} {\bibfnamefont {T.~B.}\ \bibnamefont {Wahl}}, \bibinfo {author} {\bibfnamefont {H.-H.}\ \bibnamefont {Tu}}, \bibinfo {author} {\bibfnamefont {N.}~\bibnamefont {Schuch}},\ and\ \bibinfo {author} {\bibfnamefont {J.~I.}\ \bibnamefont {Cirac}},\ }\bibfield  {title} {\bibinfo {title} {Projected entangled-pair states can describe chiral topological states},\ }\href@noop {} {\bibfield  {journal} {\bibinfo  {journal} {Physical review letters}\ }\textbf {\bibinfo {volume} {111}},\ \bibinfo {pages} {236805} (\bibinfo {year} {2013})}\BibitemShut {NoStop}%
		\bibitem [{\citenamefont {Wahl}\ \emph {et~al.}(2014)\citenamefont {Wahl}, \citenamefont {Ha{\ss}ler}, \citenamefont {Tu}, \citenamefont {Cirac},\ and\ \citenamefont {Schuch}}]{wahl2014symmetries}%
		\BibitemOpen
		\bibfield  {author} {\bibinfo {author} {\bibfnamefont {T.~B.}\ \bibnamefont {Wahl}}, \bibinfo {author} {\bibfnamefont {S.~T.}\ \bibnamefont {Ha{\ss}ler}}, \bibinfo {author} {\bibfnamefont {H.-H.}\ \bibnamefont {Tu}}, \bibinfo {author} {\bibfnamefont {J.~I.}\ \bibnamefont {Cirac}},\ and\ \bibinfo {author} {\bibfnamefont {N.}~\bibnamefont {Schuch}},\ }\bibfield  {title} {\bibinfo {title} {Symmetries and boundary theories for chiral projected entangled pair states},\ }\href@noop {} {\bibfield  {journal} {\bibinfo  {journal} {Physical Review B}\ }\textbf {\bibinfo {volume} {90}},\ \bibinfo {pages} {115133} (\bibinfo {year} {2014})}\BibitemShut {NoStop}%
		\bibitem [{\citenamefont {Yang}\ \emph {et~al.}(2015)\citenamefont {Yang}, \citenamefont {Wahl}, \citenamefont {Tu}, \citenamefont {Schuch},\ and\ \citenamefont {Cirac}}]{yang2015chiral}%
		\BibitemOpen
		\bibfield  {author} {\bibinfo {author} {\bibfnamefont {S.}~\bibnamefont {Yang}}, \bibinfo {author} {\bibfnamefont {T.~B.}\ \bibnamefont {Wahl}}, \bibinfo {author} {\bibfnamefont {H.-H.}\ \bibnamefont {Tu}}, \bibinfo {author} {\bibfnamefont {N.}~\bibnamefont {Schuch}},\ and\ \bibinfo {author} {\bibfnamefont {J.~I.}\ \bibnamefont {Cirac}},\ }\bibfield  {title} {\bibinfo {title} {Chiral projected entangled-pair state with topological order},\ }\href@noop {} {\bibfield  {journal} {\bibinfo  {journal} {Physical review letters}\ }\textbf {\bibinfo {volume} {114}},\ \bibinfo {pages} {106803} (\bibinfo {year} {2015})}\BibitemShut {NoStop}%
		\bibitem [{\citenamefont {Yang}\ \emph {et~al.}(2025)\citenamefont {Yang}, \citenamefont {Zhai}, \citenamefont {Tan}, \citenamefont {Fan}, \citenamefont {Lin},\ and\ \citenamefont {Yao}}]{yang2025fractional}%
		\BibitemOpen
		\bibfield  {author} {\bibinfo {author} {\bibfnamefont {W.}~\bibnamefont {Yang}}, \bibinfo {author} {\bibfnamefont {D.}~\bibnamefont {Zhai}}, \bibinfo {author} {\bibfnamefont {T.}~\bibnamefont {Tan}}, \bibinfo {author} {\bibfnamefont {F.-R.}\ \bibnamefont {Fan}}, \bibinfo {author} {\bibfnamefont {Z.}~\bibnamefont {Lin}},\ and\ \bibinfo {author} {\bibfnamefont {W.}~\bibnamefont {Yao}},\ }\bibfield  {title} {\bibinfo {title} {Fractional quantum anomalous hall effect in a singular flat band},\ }\href@noop {} {\bibfield  {journal} {\bibinfo  {journal} {Physical Review Letters}\ }\textbf {\bibinfo {volume} {134}},\ \bibinfo {pages} {196501} (\bibinfo {year} {2025})}\BibitemShut {NoStop}%
		\bibitem [{sup()}]{supp}%
		\BibitemOpen
		\href@noop {} {\bibinfo {title} {See Supplemental Material at [URL will be inserted by publisher] for: A. Details of the two topological conditions; B. Calculation of $\mathbb{Z}_2$ invariants of top$^2$-flat bands; C. Parent Hamiltonian of top$^2$-flat bands; D. Construction of top$^2$-flat bands for topological crystals; E. Perturbative interaction on top$^2$-flat bands; F. Detailed counting on local interacting Hamiltonian and three-operator terms, which includes Refs.~[96--101]}}\BibitemShut {NoStop}%
		\bibitem [{\citenamefont {Roy}(2014)}]{roy2014band}%
		\BibitemOpen
		\bibfield  {author} {\bibinfo {author} {\bibfnamefont {R.}~\bibnamefont {Roy}},\ }\bibfield  {title} {\bibinfo {title} {Band geometry of fractional topological insulators},\ }\href@noop {} {\bibfield  {journal} {\bibinfo  {journal} {Physical Review B}\ }\textbf {\bibinfo {volume} {90}},\ \bibinfo {pages} {165139} (\bibinfo {year} {2014})}\BibitemShut {NoStop}%
		\bibitem [{\citenamefont {Yu}\ \emph {et~al.}(2011)\citenamefont {Yu}, \citenamefont {Qi}, \citenamefont {Bernevig}, \citenamefont {Fang},\ and\ \citenamefont {Dai}}]{yu2011equivalent}%
		\BibitemOpen
		\bibfield  {author} {\bibinfo {author} {\bibfnamefont {R.}~\bibnamefont {Yu}}, \bibinfo {author} {\bibfnamefont {X.~L.}\ \bibnamefont {Qi}}, \bibinfo {author} {\bibfnamefont {A.}~\bibnamefont {Bernevig}}, \bibinfo {author} {\bibfnamefont {Z.}~\bibnamefont {Fang}},\ and\ \bibinfo {author} {\bibfnamefont {X.}~\bibnamefont {Dai}},\ }\bibfield  {title} {\bibinfo {title} {Equivalent expression of $\mathbb{Z}_2$ topological invariant for band insulators using the non-Abelian Berry connection},\ }\href@noop {} {\bibfield  {journal} {\bibinfo  {journal} {Physical Review B}\ }\textbf {\bibinfo {volume} {84}},\ \bibinfo {pages} {075119} (\bibinfo {year} {2011})}\BibitemShut {NoStop}%
		\bibitem [{\citenamefont {Passman}(2004)}]{passman2004course}%
		\BibitemOpen
		\bibfield  {author} {\bibinfo {author} {\bibfnamefont {D.~S.}\ \bibnamefont {Passman}},\ }\href@noop {} {\emph {\bibinfo {title} {A Course in Ring Theory}}}\ (\bibinfo {publisher} {American Mathematical Society},\ \bibinfo {year} {2004})\BibitemShut {NoStop}%
		\bibitem [{\citenamefont {West}(2001)}]{west2001introduction}%
		\BibitemOpen
		\bibfield  {author} {\bibinfo {author} {\bibfnamefont {D.~B.}\ \bibnamefont {West}},\ }\href@noop {} {\emph {\bibinfo {title} {Introduction to Graph Theory}}},\ 2nd ed.\ (\bibinfo {publisher} {Prentice Hall},\ \bibinfo {year} {2001})\BibitemShut {NoStop}%
		\bibitem [{\citenamefont {Kohmoto}(1985)}]{kohmoto1985topological}%
		\BibitemOpen
		\bibfield  {author} {\bibinfo {author} {\bibfnamefont {M.}~\bibnamefont {Kohmoto}},\ }\bibfield  {title} {\bibinfo {title} {Topological invariant and the quantization of the Hall conductance},\ }\href@noop {} {\bibfield  {journal} {\bibinfo  {journal} {Annals of Physics}\ }\textbf {\bibinfo {volume} {160}},\ \bibinfo {pages} {343} (\bibinfo {year} {1985})}\BibitemShut {NoStop}%
		\bibitem [{\citenamefont {Fukui}\ \emph {et~al.}(2005)\citenamefont {Fukui}, \citenamefont {Hatsugai},\ and\ \citenamefont {Suzuki}}]{fukui2005chern}%
		\BibitemOpen
		\bibfield  {author} {\bibinfo {author} {\bibfnamefont {T.}~\bibnamefont {Fukui}}, \bibinfo {author} {\bibfnamefont {Y.}~\bibnamefont {Hatsugai}},\ and\ \bibinfo {author} {\bibfnamefont {H.}~\bibnamefont {Suzuki}},\ }\bibfield  {title} {\bibinfo {title} {Chern numbers in discretized Brillouin zone: Efficient method of computing (spin) Hall conductances},\ }\href@noop {} {\bibfield  {journal} {\bibinfo  {journal} {Journal of the Physical Society of Japan}\ }\textbf {\bibinfo {volume} {74}},\ \bibinfo {pages} {1674} (\bibinfo {year} {2005})}\BibitemShut {NoStop}%
		\bibitem [{\citenamefont {Hwang}\ \emph {et~al.}(2021)\citenamefont {Hwang}, \citenamefont {Jung}, \citenamefont {Rhim},\ and\ \citenamefont {Yang}}]{hwang2021geometry}%
		\BibitemOpen
		\bibfield  {author} {\bibinfo {author} {\bibfnamefont {Y.}~\bibnamefont {Hwang}}, \bibinfo {author} {\bibfnamefont {J.}~\bibnamefont {Jung}}, \bibinfo {author} {\bibfnamefont {J.-W.}\ \bibnamefont {Rhim}},\ and\ \bibinfo {author} {\bibfnamefont {B.-J.}\ \bibnamefont {Yang}},\ }\bibfield  {title} {\bibinfo {title} {Wave-function geometry of band crossing points in two dimensions},\ }\href@noop {} {\bibfield  {journal} {\bibinfo  {journal} {Physical Review B}\ }\textbf {\bibinfo {volume} {103}},\ \bibinfo {pages} {L241102} (\bibinfo {year} {2021})}\BibitemShut {NoStop}%
		\bibitem [{\citenamefont {Udagawa}\ \emph {et~al.}(2024)\citenamefont {Udagawa}, \citenamefont {Nakai},\ and\ \citenamefont {Hotta}}]{udagawa2024pinch}%
		\BibitemOpen
		\bibfield  {author} {\bibinfo {author} {\bibfnamefont {M.}~\bibnamefont {Udagawa}}, \bibinfo {author} {\bibfnamefont {H.}~\bibnamefont {Nakai}},\ and\ \bibinfo {author} {\bibfnamefont {C.}~\bibnamefont {Hotta}},\ }\bibfield  {title} {\bibinfo {title} {Pinch-point spectral singularity from the interference of topological loop states},\ }\href@noop {} {\bibfield  {journal} {\bibinfo  {journal} {Physical Review B}\ }\textbf {\bibinfo {volume} {110}},\ \bibinfo {pages} {L201109} (\bibinfo {year} {2024})}\BibitemShut {NoStop}%
		\bibitem [{\citenamefont {Fu}(2011)}]{fu2011topological}%
		\BibitemOpen
		\bibfield  {author} {\bibinfo {author} {\bibfnamefont {L.}~\bibnamefont {Fu}},\ }\bibfield  {title} {\bibinfo {title} {Topological crystalline insulators},\ }\href@noop {} {\bibfield  {journal} {\bibinfo  {journal} {Physical review letters}\ }\textbf {\bibinfo {volume} {106}},\ \bibinfo {pages} {106802} (\bibinfo {year} {2011})}\BibitemShut {NoStop}%
		\bibitem [{\citenamefont {Benalcazar}\ \emph {et~al.}(2017{\natexlab{a}})\citenamefont {Benalcazar}, \citenamefont {Bernevig},\ and\ \citenamefont {Hughes}}]{benalcazar2017quantized}%
		\BibitemOpen
		\bibfield  {author} {\bibinfo {author} {\bibfnamefont {W.~A.}\ \bibnamefont {Benalcazar}}, \bibinfo {author} {\bibfnamefont {B.~A.}\ \bibnamefont {Bernevig}},\ and\ \bibinfo {author} {\bibfnamefont {T.~L.}\ \bibnamefont {Hughes}},\ }\bibfield  {title} {\bibinfo {title} {Quantized electric multipole insulators},\ }\href@noop {} {\bibfield  {journal} {\bibinfo  {journal} {Science}\ }\textbf {\bibinfo {volume} {357}},\ \bibinfo {pages} {61} (\bibinfo {year} {2017}{\natexlab{a}})}\BibitemShut {NoStop}%
		\bibitem [{\citenamefont {Benalcazar}\ \emph {et~al.}(2017{\natexlab{b}})\citenamefont {Benalcazar}, \citenamefont {Bernevig},\ and\ \citenamefont {Hughes}}]{benalcazar2017electric}%
		\BibitemOpen
		\bibfield  {author} {\bibinfo {author} {\bibfnamefont {W.~A.}\ \bibnamefont {Benalcazar}}, \bibinfo {author} {\bibfnamefont {B.~A.}\ \bibnamefont {Bernevig}},\ and\ \bibinfo {author} {\bibfnamefont {T.~L.}\ \bibnamefont {Hughes}},\ }\bibfield  {title} {\bibinfo {title} {Electric multipole moments, topological multipole moment pumping, and chiral hinge states in crystalline insulators},\ }\href@noop {} {\bibfield  {journal} {\bibinfo  {journal} {Physical Review B}\ }\textbf {\bibinfo {volume} {96}},\ \bibinfo {pages} {245115} (\bibinfo {year} {2017}{\natexlab{b}})}\BibitemShut {NoStop}%
		\bibitem [{\citenamefont {Langbehn}\ \emph {et~al.}(2017)\citenamefont {Langbehn}, \citenamefont {Peng}, \citenamefont {Trifunovic}, \citenamefont {von Oppen},\ and\ \citenamefont {Brouwer}}]{Langbehn2017reflection}%
		\BibitemOpen
		\bibfield  {author} {\bibinfo {author} {\bibfnamefont {J.}~\bibnamefont {Langbehn}}, \bibinfo {author} {\bibfnamefont {Y.}~\bibnamefont {Peng}}, \bibinfo {author} {\bibfnamefont {L.}~\bibnamefont {Trifunovic}}, \bibinfo {author} {\bibfnamefont {F.}~\bibnamefont {von Oppen}},\ and\ \bibinfo {author} {\bibfnamefont {P.~W.}\ \bibnamefont {Brouwer}},\ }\bibfield  {title} {\bibinfo {title} {Reflection-symmetric second-order topological insulators and superconductors},\ }\href@noop {} {\bibfield  {journal} {\bibinfo  {journal} {Physical review letters}\ }\textbf {\bibinfo {volume} {119}},\ \bibinfo {pages} {246401} (\bibinfo {year} {2017})}\BibitemShut {NoStop}%
		\bibitem [{\citenamefont {Song}\ \emph {et~al.}(2017{\natexlab{b}})\citenamefont {Song}, \citenamefont {Fang},\ and\ \citenamefont {Fang}}]{song2017d}%
		\BibitemOpen
		\bibfield  {author} {\bibinfo {author} {\bibfnamefont {Z.}~\bibnamefont {Song}}, \bibinfo {author} {\bibfnamefont {Z.}~\bibnamefont {Fang}},\ and\ \bibinfo {author} {\bibfnamefont {C.}~\bibnamefont {Fang}},\ }\bibfield  {title} {\bibinfo {title} {(d-2)-dimensional edge states of rotation symmetry protected topological states},\ }\href@noop {} {\bibfield  {journal} {\bibinfo  {journal} {Physical review letters}\ }\textbf {\bibinfo {volume} {119}},\ \bibinfo {pages} {246402} (\bibinfo {year} {2017}{\natexlab{b}})}\BibitemShut {NoStop}%
		\bibitem [{\citenamefont {Schindler}\ \emph {et~al.}(2018)\citenamefont {Schindler}, \citenamefont {Cook}, \citenamefont {Vergniory}, \citenamefont {Wang}, \citenamefont {Parkin}, \citenamefont {Bernevig},\ and\ \citenamefont {Neupert}}]{schindler2018higher}%
		\BibitemOpen
		\bibfield  {author} {\bibinfo {author} {\bibfnamefont {F.}~\bibnamefont {Schindler}}, \bibinfo {author} {\bibfnamefont {A.~M.}\ \bibnamefont {Cook}}, \bibinfo {author} {\bibfnamefont {M.~G.}\ \bibnamefont {Vergniory}}, \bibinfo {author} {\bibfnamefont {Z.}~\bibnamefont {Wang}}, \bibinfo {author} {\bibfnamefont {S.~S.}\ \bibnamefont {Parkin}}, \bibinfo {author} {\bibfnamefont {B.~A.}\ \bibnamefont {Bernevig}},\ and\ \bibinfo {author} {\bibfnamefont {T.}~\bibnamefont {Neupert}},\ }\bibfield  {title} {\bibinfo {title} {Higher-order topological insulators},\ }\href@noop {} {\bibfield  {journal} {\bibinfo  {journal} {Science advances}\ }\textbf {\bibinfo {volume} {4}},\ \bibinfo {pages} {eaat0346} (\bibinfo {year} {2018})}\BibitemShut {NoStop}%
		\bibitem [{\citenamefont {Hughes}\ \emph {et~al.}(2011)\citenamefont {Hughes}, \citenamefont {Prodan},\ and\ \citenamefont {Bernevig}}]{hughes2011inversion}%
		\BibitemOpen
		\bibfield  {author} {\bibinfo {author} {\bibfnamefont {T.~L.}\ \bibnamefont {Hughes}}, \bibinfo {author} {\bibfnamefont {E.}~\bibnamefont {Prodan}},\ and\ \bibinfo {author} {\bibfnamefont {B.~A.}\ \bibnamefont {Bernevig}},\ }\bibfield  {title} {\bibinfo {title} {Inversion-symmetric topological insulators},\ }\href@noop {} {\bibfield  {journal} {\bibinfo  {journal} {Physical Review B}\ }\textbf {\bibinfo {volume} {83}},\ \bibinfo {pages} {245132} (\bibinfo {year} {2011})}\BibitemShut {NoStop}%
		\bibitem [{\citenamefont {Peng}\ \emph {et~al.}(2022)\citenamefont {Peng}, \citenamefont {Jiang}, \citenamefont {Fang}, \citenamefont {Weng},\ and\ \citenamefont {Fang}}]{peng2022topological}%
		\BibitemOpen
		\bibfield  {author} {\bibinfo {author} {\bibfnamefont {B.}~\bibnamefont {Peng}}, \bibinfo {author} {\bibfnamefont {Y.}~\bibnamefont {Jiang}}, \bibinfo {author} {\bibfnamefont {Z.}~\bibnamefont {Fang}}, \bibinfo {author} {\bibfnamefont {H.}~\bibnamefont {Weng}},\ and\ \bibinfo {author} {\bibfnamefont {C.}~\bibnamefont {Fang}},\ }\bibfield  {title} {\bibinfo {title} {Topological classification and diagnosis in magnetically ordered electronic materials},\ }\href@noop {} {\bibfield  {journal} {\bibinfo  {journal} {Physical Review B}\ }\textbf {\bibinfo {volume} {105}},\ \bibinfo {pages} {235138} (\bibinfo {year} {2022})}\BibitemShut {NoStop}%
		\bibitem [{\citenamefont {Li}\ \emph {et~al.}(2026)\citenamefont {Li}, \citenamefont {Wang}, \citenamefont {Lin}, \citenamefont {Wang},\ and\ \citenamefont {Song}}]{li2026stable}%
		\BibitemOpen
		\bibfield  {author} {\bibinfo {author} {\bibfnamefont {Y.-Q.}\ \bibnamefont {Li}}, \bibinfo {author} {\bibfnamefont {Y.-J.}\ \bibnamefont {Wang}}, \bibinfo {author} {\bibfnamefont {P.-H.}\ \bibnamefont {Lin}}, \bibinfo {author} {\bibfnamefont {B.}~\bibnamefont {Wang}},\ and\ \bibinfo {author} {\bibfnamefont {Z.-D.}\ \bibnamefont {Song}},\ }\bibfield  {title} {\bibinfo {title} {Stable topology in exactly flat bands},\ }\href@noop {} {\bibfield  {journal} {\bibinfo  {journal} {arXiv preprint arXiv:2603.12258}\ } (\bibinfo {year} {2026})}\BibitemShut {NoStop}%
	\end{thebibliography}
\end{document}